\documentclass[twocolumn,english,aps,pra,longbibliography,superscriptaddress, floatfix]{revtex4-2}

\usepackage[utf8]{inputenc}
\usepackage{xcolor}
\usepackage{amssymb}
\usepackage{amsmath}
\usepackage{braket}
\usepackage{txfonts,stmaryrd}
\usepackage{graphicx}
\usepackage{enumerate}
\usepackage{algorithm}
\usepackage{algpseudocode}
\usepackage{makecell}
\usepackage{soul}

\usepackage{amssymb}\usepackage[colorlinks=true,urlcolor=blue,citecolor=blue,linkcolor=blue]{hyperref}

\usepackage{comment}
\usepackage{natbib}
\usepackage{lipsum}

\newcommand*{\rom}[1]{\expandafter\@slowromancap\romannumeral #1@}

\begin{document}

\title{Role of Nonstabilizerness in Quantum Optimization}

\date{\today}

\author{Chiara Capecci}
\email{chiara.capecci@unitn.it}
\affiliation{Pitaevskii BEC Center, CNR-INO and Department of Physics, University  of  Trento,  Via Sommarive 14, I-38123 Trento, Italy}
\affiliation{INFN-TIFPA, Trento Institute for Fundamental Physics and Applications, Via Sommarive 14, I-38123 Trento, Italy}
\author{Gopal Chandra Santra}
\affiliation{Pitaevskii BEC Center, CNR-INO and Department of Physics, University  of  Trento,  Via Sommarive 14, I-38123 Trento, Italy}
\affiliation{INFN-TIFPA, Trento Institute for Fundamental Physics and Applications, Via Sommarive 14, I-38123 Trento, Italy}
\affiliation{Kirchhoff-Institut f\"ur Physik, Universit\"at Heidelberg, Im Neuenheimer Feld 227, 69120 Heidelberg, Germany}

\author{Alberto Bottarelli}
\affiliation{Pitaevskii BEC Center, CNR-INO and Department of Physics, University  of  Trento,  Via Sommarive 14, I-38123 Trento, Italy}
\affiliation{INFN-TIFPA, Trento Institute for Fundamental Physics and Applications, Via Sommarive 14, I-38123 Trento, Italy}
\affiliation{Honda Research Institute Europe GmbH, Carl-Legien-Str.\ 30, 63073 Offenbach, Germany}
\author{Emanuele Tirrito}
\affiliation{The Abdus Salam International Centre for Theoretical Physics (ICTP), Strada Costiera 11, 34151 Trieste, Italy}
\affiliation{Dipartimento di Fisica ``E. Pancini", Universit\`a di Napoli ``Federico II'', Monte S. Angelo, 80126 Napoli, Italy}
\author{Philipp Hauke}
\email{philipp.hauke@unitn.it}
\affiliation{Pitaevskii BEC Center, CNR-INO and Department of Physics, University  of  Trento,  Via Sommarive 14, I-38123 Trento, Italy}
\affiliation{INFN-TIFPA, Trento Institute for Fundamental Physics and Applications, Via Sommarive 14, I-38123 Trento, Italy}

\begin{abstract}
    Quantum optimization has emerged as a promising approach for tackling complicated classical optimization problems using quantum devices. 
    However, the extent to which such algorithms harness genuine quantum resources and the role of these resources in their success remain open questions. 
    In this work, we investigate the resource requirements of the Quantum Approximate Optimization Algorithm (QAOA) through the lens of the resource theory of nonstabilizerness. We demonstrate that the nonstabilizerness in
    QAOA increases with circuit depth before it reaches a maximum, to fall again during the approach to the final solution state---creating a barrier that limits the algorithm's capability for shallow circuits. 
    We find curves corresponding to different depths to collapse under a simple rescaling, and we reveal a nontrivial relationship between the final nonstabilizerness and the success probability. 
    Finally, we identify a similar nonstabilizerness barrier also in adiabatic quantum annealing.   
    Our results provide deeper insights into how quantum resources influence quantum optimization.

\end{abstract}

\maketitle

\textbf{\textit{Introduction.---}}Combinatorial optimization problems are ubiquitous, spanning from science to industry~\cite{lucas2014ising, mannhold2009combinatorial,yu2013industrial, odili2017combinatorial,yarkoni2022quantum,10082989}. 
Hence, achieving any improvement over state-of-the-art classical algorithms through quantum computing could yield a substantial impact, making quantum optimization a leading candidate for practical uses of quantum technologies~\cite{abbas2023quantum}. 
In particular, the quantum approximate optimization algorithm (QAOA)~\cite{farhi2014quantum} offers a promising strategy in the era of noisy quantum devices~\cite{preskill2018quantum, weidenfeller2022scaling, lotshaw2022scaling,pelofske2023quantum,bottarelli2024inequality}. 
Despite many successful applications~\cite{shaydulin2024evidence,zhou2020quantum,farhi2016quantum, boulebnane2025evidence, omanakuttan2025threshold, brandhofer2022benchmarking, guerreschi2019qaoa, deller2023quantum, ekstrom2025variational, bravyi2022},  it remains unclear to what extent such algorithms truly leverage non-classical quantum resources~\cite{chitambar2019quantum}. 
In this context, entanglement---a defining feature and a key resource in quantum information processing~\cite{RevModPhys.80.517,RevModPhys.82.277,cirac2012goals}---has been extensively studied, both in the form of bipartite~\cite{lanting2014entanglement, hauke2015probing, valle2021quantum, dupont2022calibrating, sreedhar2022quantum, chen2022much} and multipartite entanglement~\cite{hauke2016measuring,santra2024squeezing,vitale2024robust,santra2025genuine}.
However, entanglement alone does not ensure a quantum advantage. Stabilizer states can exhibit large entanglement yet remain efficiently simulable classically via Clifford circuits, as guaranteed by the Gottesman--Knill theorem~\cite{gottesman1997stabilizer,gottesman1998heisenberg,gottesman1998theory}. 
Nonstabilizerness~\cite{PhysRevA.71.022316}---the degree to which a nonstabilizer state, also called a magic state, deviates from the set of stabilizer states---is thus a crucial property to unlock a potential quantum advantage~\cite{PhysRevA.71.022316,PhysRevA.86.052329,campbell2017roads,harrow2017quantum,gross2006hudson,wootters1987wigner}. 
Based on the discrete Wigner function~\cite{PhysRevA.45.6570}, many measures for this quantum resource have been introduced~\cite{veitch2012negative,PhysRevLett.116.250501,bravyi2016trading, PhysRevLett.118.090501,heinrich2019robustness,PhysRevLett.124.090505,heimendahl2021stabilizer} among which Mana~\cite{veitch2012negative, veitch2014resource} and the Stabilizer Rényi Entropy (SRE)~\cite{leone2022stabilizer, haug2023stabilizer,PhysRevA.110.L040403} are notable for being efficiently computable and experimentally accessible. 
In recent years, the study of nonstabilizerness has seen many applications from many-body physics~\cite{sarkar2020characterization,oliviero2022magic,liu2022manybody, viscardi2025interplay,falcao2025magic, santra2025complexity,korbany2025long} over random circuits~\cite{turkeshi2025magic,haug2024probing,dileepNonstab2025} to conformal field theory~\cite{white2021conformal,tarabunga2024critical, hoshino2025stabilizer} and neural network~\cite{Spriggs_2025, sinibaldi2025}. Yet, its role in variational quantum optimization remains largely unexplored. 

In this work, we address this gap by analyzing nonstabilizerness in QAOA applied to combinatorial problems, focusing on both SRE and Mana.
We study the performance of QAOA on the paradigmatic Sherrington–Kirkpatrick (SK) model~\cite{panchenko2013sherrington}, which combines computational hardness with statistical complexity. Our simulations reveal the presence of a “magic barrier”—a transient build-up of nonstabilizerness that occurs during the QAOA run, akin to previously observed entanglement barriers~\cite{hauke2015probing, dupont2022entanglement, santra2025genuine}.
We find that while QAOA begins and ends with low-magic (stabilizer) states, it must pass through a regime of increased magic in order to achieve high fidelity with the target solution. This behavior appears in both qubit and qutrit versions of the algorithm. 
We find that magic curves corresponding to different depths collapse under a simple rescaling, and we identify a similar magic barrier in adiabatic quantum annealing.  
Furthermore, we analyze how the success probability of QAOA correlates with the final nonstabilizerness, uncovering characteristic trends and structure in the fidelity-magic plane. Using analytical calculations on \textit{ansatz} states, made of few-component superpositions, we explain the observed features and clarify under what conditions low-magic solutions may or may not correspond to high-fidelity outcomes.

Through this work, we illuminate the role and significance of nonstabilizerness in quantum optimization, providing deeper insight into its influence on algorithmic performance and complementing existing studies on entanglement and coherence~\cite{dupont2022entanglement,chen2022much,santra2025genuine}.  
The results also have practical relevance, as measurements of magic can provide a figure of merit for estimating the performance of variational or optimization algorithms performed on concrete quantum hardware, such as superconducting qubits~\cite{oliviero2022measuring}, trapped ions~\cite{Niroula2024}, and Rydberg atoms~\cite{bluvstein2024logical}.

\begin{figure*}[hbt!]
    \centering
    {\includegraphics[width=\textwidth]{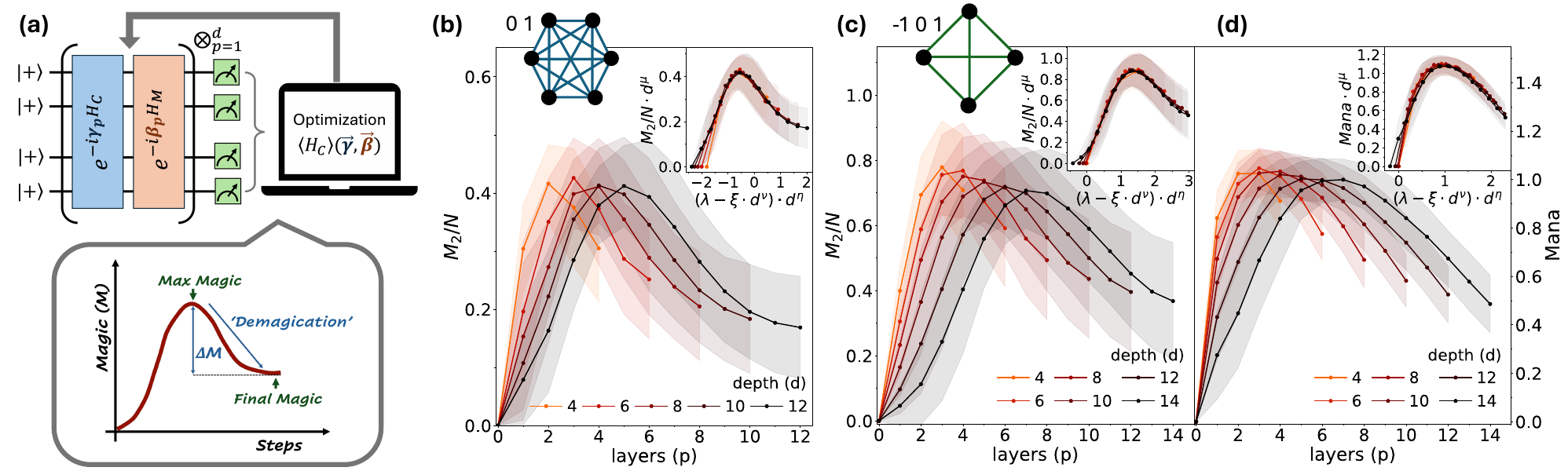} }
    \caption{(a) Quantum Approximate Optimization Algorithm (QAOA) scheme and a pictorial representation of the magic barrier: initially, magic rises up to a maximal value, after which it decreases, an effect we call ``demagication.''
    (b) SRE density as a function of the layer number during a QAOA protocol, for a system of 6 qubits and different depths (i.e., total number of layers). Inset: Using a simple scaling function, the magic barrier collapses onto a universal curve. 
    (c,d) Same plot of panel b for a system of 4 qutrits, showing SRE density (c) and Mana (d) as a function of the layer number. The universal collapses in the insets in (b-d) use the scaling function 
    $\mathcal{M}=d^{-\mu} f[(\lambda-\xi \cdot d^\nu)\cdot d^\eta]$,  with $(\mu, \xi, \nu, \eta)$ equal 
    $(5.11 \cdot10^{-3}, 1.28, -0.34, 0.60)$, $(-8.20 \cdot10^{-2}, 1.19 \cdot10^{-5}, 3.49, 0.46)$, $( - 2.86 \cdot10^{-2}, 3.69\cdot10^{-8}, 5.50, 0.34)$,
    respectively.
    }
    \label{fig:fig1}
\end{figure*}

\textbf{\textit{Quantifying nonstabilizerness.---}}We employ two measures of nonstabilizerness: the stabilizer Rényi entropy (SRE)~\cite{leone2022stabilizer, haug2023stabilizer} and Mana~\cite{veitch2012negative, veitch2014resource}.
The SRE characterizes how a pure state $|\psi\rangle$ of $N$ qudits spreads over the basis of Pauli strings, defined as $\mathcal{P}_N \equiv \big\{P_{\vec{v}^{(1)}} \otimes P_{\vec{v}^{(2)}} \otimes \cdots \otimes P_{\vec{v}^{(N)}}\big\}$. 
For qudits with dimension $D$, each of the $D^2$ possible generalized Pauli operators $ P_{\vec{v}} = X^{v_1}Z^{v_2} $ is specified by $\vec{v} = (v_1, v_2) \in [0, D{-}1]^2 $, where $X$ and $Z$ are shift and phase operator respectively (see SM). The SRE is defined as~\cite{leone2022stabilizer}:
\begin{equation}\label{eq: sre definition}
    M_n(|\psi\rangle) = \frac{1}{1-n} \log_2 \left[ \sum_{P \in \mathcal{P}_N} \frac{|\langle \psi | P | \psi \rangle|^{2n}}{D^N} \right].
\end{equation}
$M_n$ is non-negative and equals zero if and only if $|\psi\rangle$ is a stabilizer state~\cite{haug2023stabilizer, gross2021schurweylduality}. An important feature of Eq.~\eqref{eq: sre definition} is that it is experimentally computable~\cite{Niroula2024}. Throughout this work, we use $M_2$ as our measure.

For qudit systems with odd prime local dimension $D$, the negativity of the discrete Wigner function provides another quantifier of nonstabilizerness~\cite{gross2006hudson, WOOTTERS19871}. Specifically, the log-negativity of the Wigner function $W_\rho$ associated with a quantum state $\rho$ defines a quantity known as Mana~\cite{veitch2014resource}:
\begin{equation}\label{eq: Mana definition}
    {\rm Mana} = \log \sum_{\vec{V}} |W_\rho(\vec{V})|\,.
\end{equation}
Here, $\vec{V}$ ranges over the $D^{2N}$ discrete phase-space points $\vec{V} = \vec{v}^{(1)} \oplus \dots \oplus \vec{v}^{(N)}$, which also label the Pauli strings. For pure states $\rho = |\psi\rangle\langle\psi|$, as considered here, the set of states with non-negative Wigner representation coincides exactly with the set of pure stabilizer states~\cite{gross2006hudson}, in which case the Mana vanishes.

\textbf{\textit{QAOA on SK model.---}}In QAOA, classical optimization problems like MaxCut, MaxSat, or the knapsack problem are mapped to the task of finding the ground state of an Ising Hamiltonian, $\hat{H}_C$~\cite{lucas2014ising}.
As illustrated in Fig.~\ref{fig:fig1}(a), QAOA prepares a trial state $\ket{\psi(\boldsymbol\beta,\boldsymbol\gamma)}$ by applying $d$ layers of alternating unitaries, $\exp(-i\beta_p \hat{H}_M)\exp(-i\gamma_p \hat{H}_C)$, starting from the ground state of the mixer Hamiltonian $\hat{H}_M$~\cite{streif2019comparison}.
When we consider $\hat{H}_M= -\sum_i\hat{X}_i$, the initial state is $\ket{+}^{\otimes N}$. 
A classical optimizer is then tasked to minimize the measured energy $\braket{\psi|\hat{H}_C|\psi}$ by tuning $\boldsymbol\beta=(\beta_1,\dots,\beta_p)^\mathrm{T}$ and $\boldsymbol\gamma=(\gamma_1,\dots,\gamma_p)^\mathrm{T}$.
Originally developed for qubits, the Quantum Approximate Optimization Algorithm (QAOA) can be generalized to qudit systems with any local dimension \( D \)~\cite{deller2023quantum}, thereby expanding its range of potential applications~\cite{bravyi2022,deller2023quantum,bottrillQutrit2023,karacsony2024efficient}.
Given that the initial state is non-magical~\footnote{The $\ket{+}=\frac{1}{\sqrt{D}}\sum_{i=0}^{D-1}\ket{i}$ qudit state is a stabilizer state because it is a $(+1)$ eigenstate of a $d$-element subgroup of the Pauli group; for example, it is $\{I, X\}$ for $d=2$. Using the additive property, $\mathcal{M}(\ket{+}^{\otimes N}) = N\mathcal{M}(\ket{+}) = 0$.} and the solution state of a classical problem with a unique optimum is also non-magical, QAOA offers a natural setting to test the buildup and role of magic during the quantum optimization procedure. 

The model of our interest is the Sherrington–Kirkpatrick (SK) model, a variant of the Ising model with all-to-all interactions randomly drawn from independent Gaussian distributions. Initially introduced as a solvable mean-field model to describe the spin-glass phase~\cite{sherrington1975solvable, panchenko2013sherrington}, it serves as a prototype for real-world optimization problems, which often involve a diverse set of random variables~\cite{venturelli2015quantum, mugel2022dynamic}. 
We use the SK model generalized to qudits (specifically, we study qubits and qutrits), 
\begin{equation}\label{eq: sk hamiltonian}
    \hat{H}_{\rm SK} = \sum_{i\neq j}J_{ij}(Z_iZ_j^{\dagger}+Z_i^{\dagger}Z_j)+\hat{H}_{\rm bias}\,,
\end{equation} 
where the coefficients $J_{ij}$ 
are chosen randomly from a normal distribution with unit variance, and $Z_i$ is the generalized phase operator. The Hermitian conjugate term is necessary to ensure the Hermiticity of the Hamiltonian in the qutrit case. 
To remove degeneracies in the ground state, we add bias magnetic field terms
$\hat{H}_{{\rm bias}}$, see SM. For each random realization of the SK model, we simulate a QAOA protocol using exact numerics, where we select the best outcome from 20 independent QAOA runs initialized using Trotterized quantum annealing~\cite{sack2021quantum}, and average the results over 50 such realizations.

\textbf{\textit{Magic barrier.---}}When targeting the non-degenerate ground state of the SK model, QAOA starts—and ideally ends—in stabilizer states with zero magic. The relevant performance metrics---such as the relative energy $(E_{\rm QAOA}(p)-E_{\rm exact})/ E_{\rm exact}$ and the fidelity with the ground state $\mathcal{F}(p)= |\langle \psi_{\rm exact}|\psi_{\rm QAOA}(p)\rangle|^2$ ---improve monotonically with the number of QAOA layers 
$p$, see SM.
In contrast, the evolution of magic, quantified by SRE and Mana, follows a non-monotonic behavior: it initially rises rapidly as the first QAOA layers are applied, reaches a maximum, and then decreases as the algorithm approaches the final state. We define this structure as the \textit{magic barrier}, marking the necessity of traversing highly nonstabilizer states during the QAOA algorithm.
A pictorial representation of the magic barrier is shown in Fig.~\ref{fig:fig1}(a). Remarkably, this phenomenon is observed consistently across different systems and for both SRE and Mana. For instance, Fig.~\ref{fig:fig1}(b) depicts the evolution of the SRE density $(M_2/N)$ for a $6$-qubit system at varying QAOA depths. A distinct peak emerges at approximately half the total depth, indicating the point of maximal nonstabilizerness. 
A comparable scenario arises in the 4-qutrit system, for both SRE and mana [see Fig.~\ref{fig:fig1}(c,d)]. 

Notably, for a fixed system size, the maximum value of magic encountered remains approximately constant across different QAOA circuit depths, suggesting that the magic peak is independent of circuit depth. 
Moreover, even at its peak, the generated magic remains below that of a typical state in the corresponding Hilbert space, such as given by the Haar-random value-- $M_2^{\rm Haar}= -\log_2(\frac{4}{2^N+3})$ for qubit, and $-\log_2(\frac{3}{3^N+2}) $ for qutrit systems~\cite{turkeshi2025pauli, turkeshi2025magic}. 
This suggests that for a device to successfully run a QAOA, it is sufficient if it is able to build up a limited amount of magic.

The systematic and consistent presence of the magic barrier---across different system sizes, circuit depths, and measures of nonstabilizerness---naturally motivates the formulation of a scaling law to capture the universal features of magic evolution in QAOA. Indeed, shown in the insets of Fig.~\ref{fig:fig1}(b–d), using a simple scaling function $f[\cdot]$,

\begin{equation}\begin{split}
    \frac{\mathcal{M}\left( \ket{\psi_{\rm QAOA_d}(p)}\right)}{N} = d^{-\mu} f\left[ (\lambda - \xi \cdot d^\nu )\cdot  d^\eta \right] ; \\ \mu, \xi, \nu, \eta \in \mathbb{R}; \quad  \lambda=\frac{p}{d}.
\label{eq:scaling}
\end{split}\end{equation}
the data for different depths, both for average SRE density and Mana, collapse onto a single curve. 
For qubit systems, we find a scaling exponent $\mu$ close to zero, suggesting that the peak value of SRE remains independent of both the QAOA circuit depth and the number of qubits. For qutrit systems, the exponent $\mu$ is slightly larger, potentially indicating a mild dependence of the peak magic on circuit depth. 
The behavior of the coefficients $\xi$ and $\nu$ appears to be inversely correlated. In the case of qubit systems, we find that the scaling formulation can be simplified by replacing the combined term $\xi \cdot d^{\nu}$ with a constant critical point $\lambda_c$, leading to the alternative scaling form $d^{-\nu} f\left[ (\lambda - \lambda_c) \cdot d^{\eta} \right]$. The critical point $\lambda_c$, which marks the position of the magic barrier, is consistently found around $0.2$ and remains robust across different system sizes (see SM).
The exponent $\eta$, which governs the width 
rescaling, lies within a relatively narrow range for both the 6-qubit and 4-qutrit systems. Moreover, $\eta$ increases with the number of qubits, suggesting a sharper magic barrier in larger systems (see SM).

\begin{figure}[hbt!]
\centering
{\includegraphics[width=0.94\columnwidth]{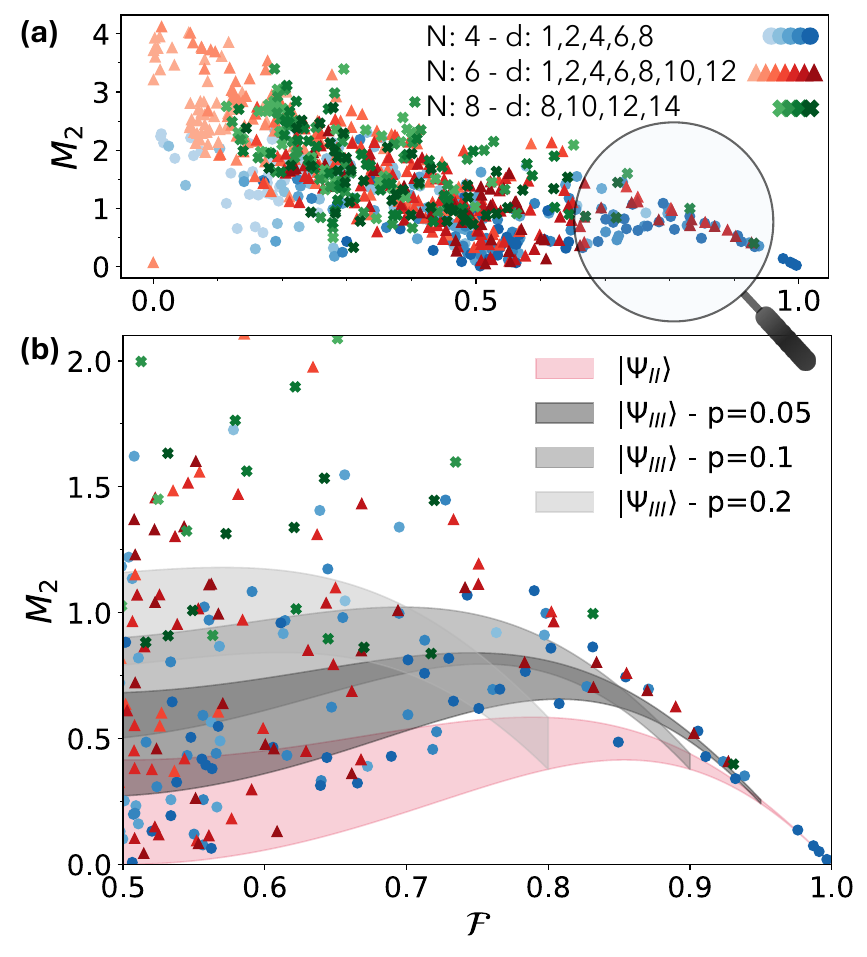}} 
\caption{Magic of the final state as a function of final fidelity. (a) Randomly sampled data for varying numbers of qubits and QAOA depths. (b) Zoom in on the region with $\mathcal{F}\geq 0.5$. The shaded region in red delimits the SRE of any $n$-qubit system described by two computational basis states. The grey regions collect the possible SRE values for a wavefunction spanning three basis states, varying the parameters involved. While higher parts of the figure may be covered by further refined \textit{ansatz} wavefunctions, the empty region towards the bottom is forbidden.}
\label{fig:fig2}
\end{figure}

\textbf{\textit{Final Magic and Fidelity.---}}The ultimate determinant of the success of QAOA lies in the fidelity of the final state with the target solution. 
As the QAOA approaches the final layer, successful instances reach a high overlap with a classical state, implying---in the absence of degenerate solutions---one can associate high-fidelity QAOA instances with low magic.
This behavior is indeed what we observe in randomly sampled instances for different system sizes and depths, shown in Fig.~\ref{fig:fig2}. 
The opposite, however, is not necessarily true: a QAOA sweep can end up in a stabilizer state, including any computational basis state distinct from the solution state, and thus can have small magic while reaching only low fidelity. 

Focusing on instances with $\mathcal{F}\geq0.5$ [Fig.~\ref{fig:fig2}(b)], one observes a void region at the lower axis. 
The pattern can be explained as follows. 
At the end of a rather successful QAOA, we expect the state to consist of the final solution $\ket{\phi_0}$ plus potentially a small contamination, typically from the first excited state, which is another computational basis state $\ket{\phi_1}$. 
We therefore consider an \textit{ansatz} state 
\begin{equation}\label{eq: two state superposition}
    \ket{\psi_{\rm II}}=\sqrt{\mathcal{F}} \ket{\phi_0} +e^{i \theta} \sqrt{1-\mathcal{F}}\ket{\phi_1},\quad  \theta\in [0,2\pi]\,,
\end{equation}
where the probability amplitude to reach the desired solution is $\sqrt{\mathcal{F}}$, and where we allowed for a relative phase $e^{i \theta}$. 
Using the additivity of $M_2$ and permutational invariance, one can show that for 
any number of qubits (see SM), 
\begin{equation} \label{eq: magic superposition}
    {M}_2\left(\ket{\psi_{\rm II}}\right)= - \log_2 \left[1-4\mathcal{F}(1-\mathcal{F}) +2\mathcal{F}^2(1-\mathcal{F})^2
    (7+\cos{4\theta} ) \right] .
\end{equation}
The result is symmetric around $\mathcal{F}=0.5$. The minima (maxima) at $\theta=0 \ (\theta=\pi/4)$ describe a bounded region in the fidelity--magic plane, within which all instances that reach the above \textit{ansatz} are confined, independent of peculiarities of the protocol such as qubit number or circuit depth. 
One can refine this \textit{ansatz} by considering a superposition of three computational basis states, 
\begin{align}\label{eq: three state superposition}
    \ket{\psi_{\rm III}}=\sqrt{\mathcal{F}} \ket{\phi_0} +e^{i \theta_1}& \sqrt{p} \ket{\phi_1}+e^{i \theta_2} \sqrt{1-\mathcal{F}-p}\ket{\phi_2}, \nonumber
    \\ & \quad p\in[0,1-\mathcal{F}], \ \theta_{1/2}\in [0, 2\pi]\,,  
\end{align}
for which we can again calculate $M_2$ analytically (see SM). 
The shaded regions in Fig.~\ref{fig:fig2}(b) correspond to $M_2$ contained in the \textit{ansatz} states \(\ket{\psi_{\rm II, III}}\) for varying $\mathcal{F}, \theta, p$. 
Since \textit{ansätze} with a further increased number of basis states will lead to higher magic, these analytic considerations explain why no instances are found numerically in the empty region at the bottom of the figure. 
Moreover, they suggest a reduced likelihood of obtaining a medium-to-high-fidelity state $\mathcal{F}\sim (0.6-0.9)$ with low magic, further indicating that QAOA has to go through a high-magic state to reach a good solution.

\begin{figure}[hbt!]
    \centering
    \includegraphics[width=1\columnwidth]{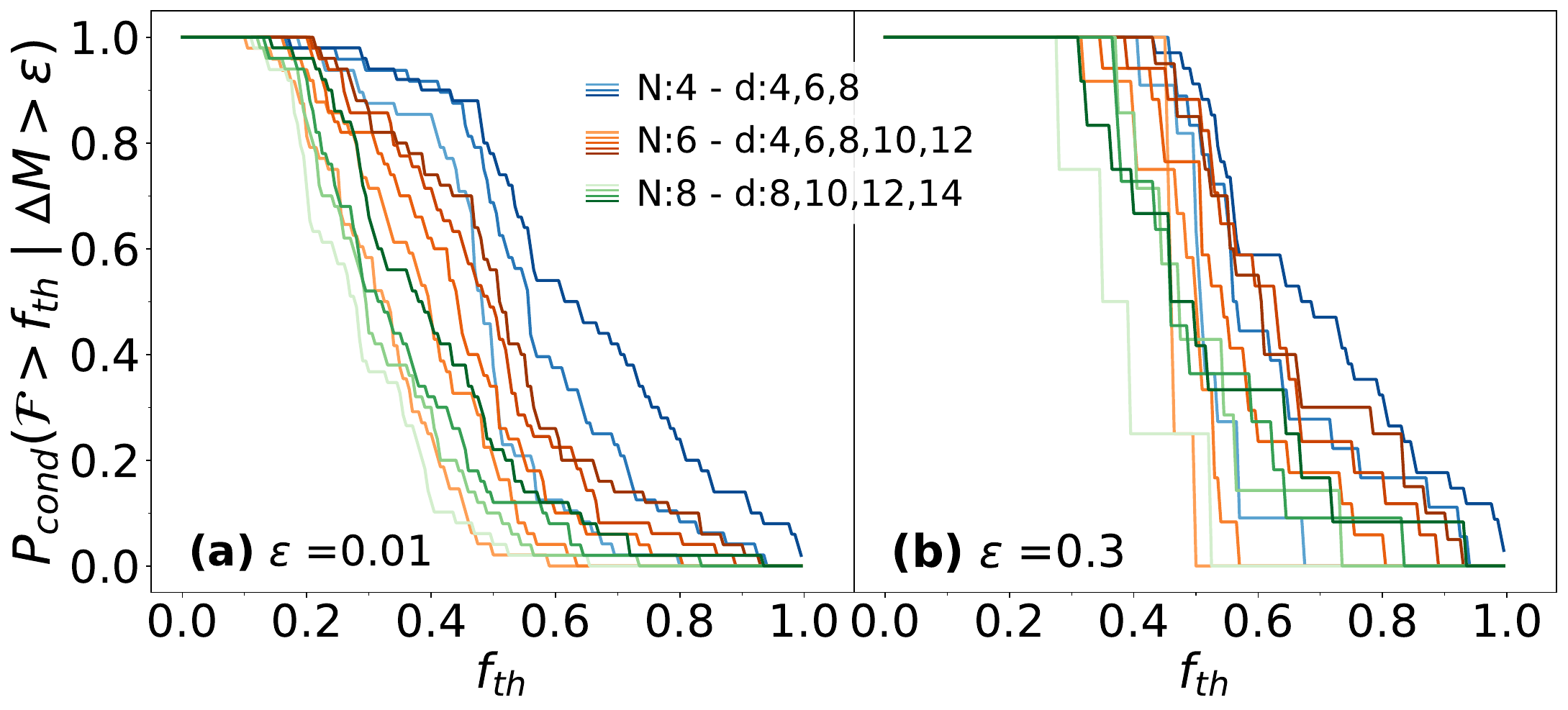}
    \caption{
    Conditional probability of reaching a high final fidelity $\geq f_\mathrm{th}$ given a demagication (difference between peak and final value) $\Delta M$ larger than a given fixed value $\epsilon$, for various system sizes and QAOA depths. 
    (a) $\epsilon=0.01$ (b) $\epsilon=0.3$. The rightward shift of panel b vs.\ a indicates that stronger demagication leads to a higher probability of achieving high fidelity. Darker shades denote higher depths.}
    \label{fig:fig3}
\end{figure}

\textbf{\textit{Demagication and success of QAOA.---}}One may wonder whether the effect of demagication, i.e., the amount by which SRE decreases after the barrier, $\Delta M= M_2^{\rm max}- {M}_2^{\rm final}$, is related to the success fidelity. 
To analyze this question, we calculate the conditional probability ($\mathcal{P}_{\rm cond}[f_{\rm th}, \epsilon]$) of obtaining a fidelity greater than some $f_{\rm th}$, provided that the QAOA has shown some minimum amount of demagication, $\Delta M > \epsilon$:
\begin{equation}
    \mathcal{P}_{\rm cond}\Big[F > f_{\rm th} \big| \Delta M > \epsilon\Big]= \frac{\mathcal{P}\Big[(F > f_{\rm th}) \cap (\Delta M > \epsilon)\Big]}{ \mathcal{P} \Big[ \Delta M > \epsilon\Big]}.
\label{eq: cond_prob}
\end{equation}
As we see in Fig.~\ref{fig:fig3}, 
a larger demagication (larger $\epsilon$), appears to result in a higher probability to solve the QAOA with high fidelity. Additionally, the probability to reach or surpass a given fidelity also increases (shifts to the right) as the depth increases, which is to be expected. Similar results hold for the qutrit case (see SM). 

\textbf{\textit{Magic barrier in quantum annealing.---}}To assess the generality of the magic barrier beyond QAOA, we analyze its emergence in continuous-time quantum annealing protocols \cite{hauke2020perspectives, Santoro_2002_QA, morita2008mathematical}. 
Quantum annealing aims to solve combinatorial optimization problems by initializing the system in the ground state of a simple Hamiltonian, here chosen as the transverse-field mixer $H_M$, and gradually transforming it into the problem-specific cost Hamiltonian $H_C$, whose ground state encodes the solution~\cite{Santoro_2002_QA,hauke2020perspectives,rajak2023quantum}.
In this setting, the evolution is modeled by a time-dependent Hamiltonian of the form
\begin{equation}
 H(\lambda)=(1-\lambda)H_M+\lambda H_C,   
\end{equation}
where the interpolation parameter $\lambda \in [0,1]$ represents the progress along the annealing path. In the adiabatic limit, where the evolution is sufficiently slow, the system remains in the instantaneous ground state of $H(\lambda)$ throughout the protocol~\cite{hauke2020perspectives}.  

\begin{figure}
    \centering
    \includegraphics[width=1\linewidth]{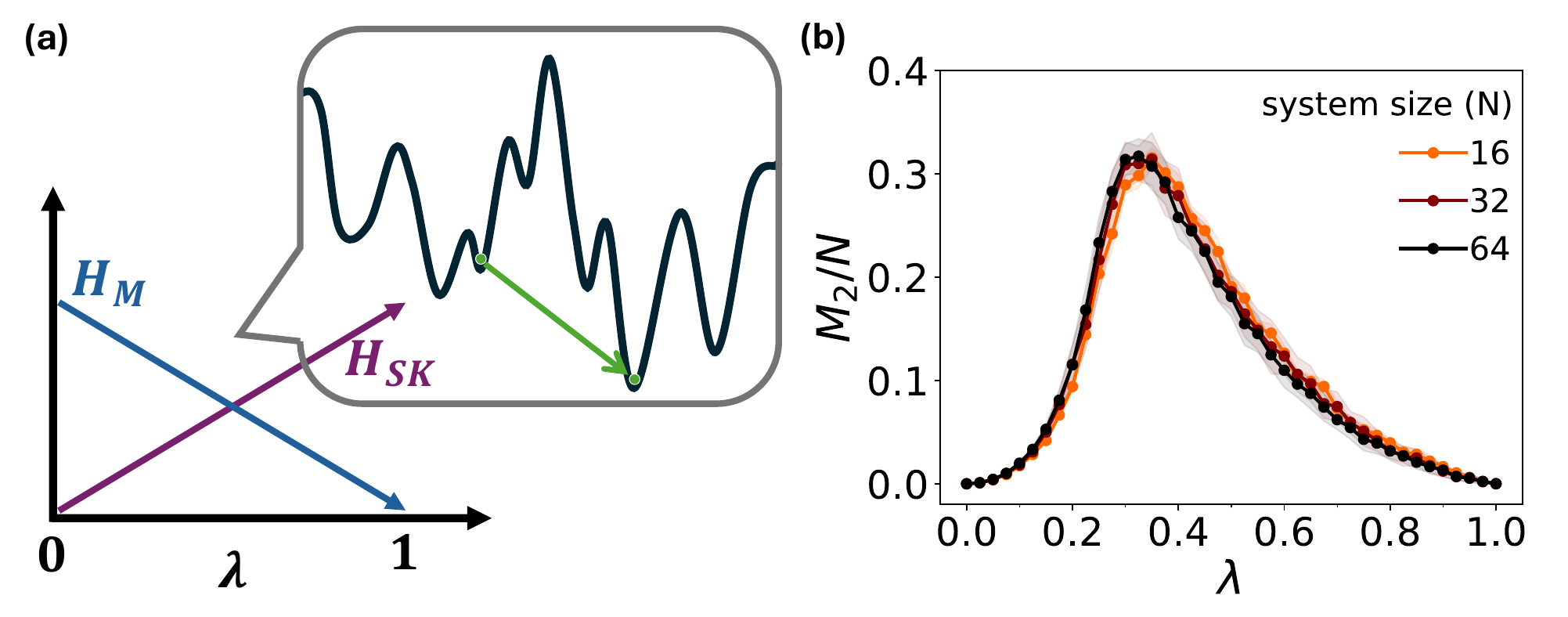}
    \caption{(a) Scheme of the quantum annealing protocol: as the annealing parameter $\lambda$ increases, the strength of the cost Hamiltonian $H_\mathrm{SK}$ rises while that of the mixing Hamiltonian $H_\mathrm{M}$. Pictorially, quantum annealing reaches the solution state by tunneling between local minima in the energy landscape. (b) The SRE density, computed on the ground state of the instantaneous Hamiltonian---corresponding to ideal adiabatic quantum annealing with normalized annealing time $\lambda \in [0,1]$---exhibits a magic barrier around $\lambda \sim 0.35$. The almost-overlapping curves for $N = 16, 32, 64$ indicate that the SRE density traces a barrier that is largely independent of system size, consistent with observations in QAOA.
    }
    \label{fig:annealing}
\end{figure}

To simulate the stationary regime of the annealing protocol, we employ tensor network techniques,  
specifically matrix product state (MPS) methods~\cite{schollwock2011density,orus2014practical,orus2019tensor,ran2020tensor}, which allow us to compute numerically exact ground 
the ground state $|\psi(\lambda)\rangle$ of the instantaneous Hamiltonian $H(\lambda)$, at arbitrary points along the interpolation, with small bond dimension $\chi$ (we fixed $\chi=60$ in our simulation, at which value the result is converged).  
This framework also enables efficient computation of SRE using the Pauli Matrix Product State (Pauli MPS) formalism~\cite{tarabunga2024nonstabilizerness}, with bond dimension up to $\chi_P=1024$. To make the computation tractable, we truncate the fully connected spin-glass interactions in the cost Hamiltonian to fifth-neighbor couplings. We verified that this truncation does not qualitatively affect the key features of the observed magic dynamics.

The resulting SRE profile, shown in Fig.~\ref{fig:annealing}, reveals a clear magic barrier: magic rises during the early stages of the sweep, peaks around intermediate $\lambda$, and decreases as the system approaches the classical solution. This behavior closely mirrors the magic evolution observed in QAOA. Moreover, based on earlier studies of entanglement growth in non-adiabatic annealing protocols~\cite{hauke2015probing}, we expect that deviation from perfect adiabacity would generally enhance the amount of magic generating during the sweep.

\textit{\textbf{Conclusions.---}}To summarize, magic arises in QAOA even when the target problem has a non-magical solution. Irrespective of the depth of the QAOA, magic rises to a peak and, in a well-performing run, falls towards the end of the protocol, incurring an extra resource cost to overcome, a phenomenon we call the magic barrier. This peak consistently appears near the middle of the QAOA, and upon rescaling, the average magic collapses onto a universal curve across different depths, for both qubits and qutrits, and for SRE and Mana. The fidelity of the final state has a characteristic relation to the final magic, whose qualitative behavior can be explained analytically through the dependence on probability amplitude and local phase of a simple \textit{ansatz} state. 

While in current noisy devices non-Clifford gates are not necessarily more expensive than Clifford gates, the situation is different in early fault-tolerant machines, where non-Clifford operations are a major bottleneck~\cite{PhysRevA.86.052329,ogorman2017quantum,souza2011experimental,beverland2020lower,campbell2021early}. In such cases, the emergence of a magic barrier in QAOA becomes a crucial factor. This motivates the question of whether QAOA can be parameterized to reach the solution state while remaining close to the convex hull of stabilizer states, thus minimizing magic generation.

A natural next step is to test these findings on a quantum processor such as one based on superconducting qubits, cold atoms, or trapped ions, which would also provide a figure of merit for the performance characteristics of a given device. Trapped ions, e.g., appear well suited for a direct test, as they naturally implement long-range interactions as occur in the Sherrington--Kirkpatrick model studied here, and they have recently permitted for measuring nonstabilizerness \cite{Niroula2024}, and for the implementations of quantum optimization protocols~\cite{pagano2020,Huber2021}

\textit{\textbf{Acknowledgments.---}}We thank Matteo M.~Wauters, and Andrea Legramandi for useful discussions. We acknowledge the MAGIC App project, funded by the German Federal Ministry for Education and Research under the funding reference number 13N16437.
A.B.\ acknowledges funding from the Honda Research Institute Europe.
This project has received funding by the European Union under Horizon Europe Programme - Grant Agreement 101080086 - NeQST. This project was supported by the Provincia Autonoma di Trento, and by Q@TN, the joint lab between the University of Trento, FBK-Fondazione Bruno Kessler, INFN-National Institute for Nuclear Physics and CNR-National Research Council.
E.\,T. acknowledges support from  ERC under grant agreement n.101053159 (RAVE), and
CINECA (Consorzio Interuniversitario per il Calcolo Automatico) award, under the ISCRA 
initiative and Leonardo early access program, for the availability of high-performance computing resources and support.
Views and opinions expressed are however those of the author(s) only and do not necessarily reflect those of the European Union or the European Commission.
Neither the European Union nor the granting authority can be held responsible for them.

\onecolumngrid
\vspace{3em}
\begin{center}
    \large \textbf{Supplementary Material: \\ 
    Role of Nonstabilizerness in Quantum Optimization}\\[1em]
\end{center}

\vspace{1em}

This supplementary material provides additional details about the results reported in the main text. The content is organized as follows: Sec. \ref{app: qaoa_details} describes the optimization method used in the QAOA algorithm, Sec. \ref{app: Detailed definition} presents a detailed definition of nonstabilizerness for qudit systems, Sec. \ref{app: Degeneracy_breaking_SK} provides a mathematical explanation of the degeneracy breaking in the Sherrington–Kirkpatrick model, Sec. \ref{app: Scaling} includes additional insights and data on the scaling law of nonstabilizerness, Sec. \ref{app: magic-fidelity} contains the analytical proof of the relationship between magic and fidelity; Sec. \ref{app: energy-fidelity},  Sec. \ref{app: magic-in-opt}, Sec. \ref{app: final_QAOA_data} and Sec. \ref{app: add_qutrits}  report, respectively, additional data on energy and fidelity as a function of the layer index, the behavior of nonstabilizerness during the optimization loop, final results as a function of circuit depth and additional data for the qutrit systems studied. 

\section{Details on QAOA optimization method}
\label{app: qaoa_details}

We execute all QAOA simulations using a custom implementation. To find the optimal parameters $\{\boldsymbol{\beta}, \boldsymbol{\gamma}\}$, we employ the Constrained Optimization BY Linear Approximation (COBYLA) algorithm~\cite{powell1994direct}, a gradient-free method that solves the problem by linearly approximating the cost function. The parameters are initialized using the Trotterized Quantum Annealing (TQA) strategy~\cite{sack2021quantum}. 
The classical optimizer stops when the trust-region radius decreases below the specified tolerance of $10^{-4}$.
The SRE is computed exactly, with computational cost scaling as $D^{2N}$ due to the exponential growth in the number of Pauli strings, where $D$ is the qudit dimension and $N$ is the number of qudits.

\section{Nonstabilizerness measures for qudits}\label{app: Detailed definition}

Central to the SRE calculation is the Pauli group on $N$-qudits, which is formed by the strings of qudit Pauli operators. The qudit Pauli operators are also related to the Heisenberg--Weyl operators, which also define the phase space operators for calculating Mana~\cite{veitch2014resource}. 

The local (single-qudit) Pauli operators are generated by the product of $X$ and $Z$~\cite{gheorghiu2014standard},
\begin{equation}
        P_{\vec{v}}= X^{v_1}Z^{v_2}; \quad v_i\in \{0,1,2...,D-1\},
\end{equation}
where $Z$ and $X$ are defined as the phase and shift operators, 
\begin{equation}
    Z= \sum_{j=0}^{D-1} \omega^j |j \rangle \langle j|, \quad X=\sum_{j=0}^{D-1} |j+1\rangle \langle j|\,; \quad \omega=e^{2\pi i/D} \,.
\end{equation}

We consider a system of $N$ qudits with Hilbert space $\mathcal{H}=\otimes^N_{j=1} \mathcal{H}_i$, where $\mathcal{H}_i$ correspond to local Hilbert space for each qubits of size $D^N$
The $N$-qudit Pauli group $\mathcal{P}_N$ encompasses all the possible Pauli strings with overall phases $\pm 1$ or $\pm i$. However, neither SRE nor Mana depends on this phase.
The $N$-qudit Pauli group $\mathcal{P}_N$ is defined as
\begin{equation}
    \mathcal{P}_N = \left \lbrace e^{i\theta \frac{\pi}{2}} P_{\vec{v}^{(1)}} \otimes \cdots P_{\vec{a}^{(N)}} ; \  \theta \in [0,2\pi]  \right \rbrace .
\end{equation}

We define a stabilizer state using Clifford unitaries. The Clifford group $\mathcal{C}_N$ is defined as the normalizer of the $N$-qudit.

Stabilizer gates are defined as the set of unitary matrices 
\begin{equation}
    \mathcal{C}_N=\left \lbrace U \  \mbox{such that} \ UPU^{\dagger} \in \mathcal{P}_N \, \text{for  all} \ P \in \mathcal{P}_N \right \rbrace .
\end{equation}

The Clifford group is generated by the Hadamard gate, the $\pi/4$ phase gate, and CNOT gate. Stabilizer states encompass all the states that can be generated by Clifford operations acting on the computational basis state $|0\rangle^{\otimes N}$. 
The amount of nonstabilizerness, or magic, of any state is measured using magic monotones. The properties required for a good monotone $\mathcal{M}$ of nonstabilizerness are~\cite{chitambar2019quantum, leone2022stabilizer}: (i) $\mathcal{M} (|\psi\rangle)=0$ iff $|\psi\rangle$ is a stabilizer state (ii) nonincreasing under Clifford operations: $\mathcal{M}(\Gamma |\psi \rangle) \leq \mathcal{M}(|\psi \rangle)$, and (iii) $\mathcal{M}(|\psi\rangle \otimes |\phi \rangle) = \mathcal{M}(|\psi \rangle)+\mathcal{M}(|\phi \rangle)$ (in general, sub-additivity is sufficient).

\paragraph*{\textbf{Stabilizer Renyi entropy.}---}The SRE defined in Eq.(1) of main text satisfies all properties (i-iii) necessary to be a valid magic measure \cite{leone2022stabilizer}
For $d>2$, the Pauli operators are no longer Hermitian, and thus their expectation values can be complex. For this reason, the absolute values are taken in definition (1) in the main text.
\paragraph*{\textbf{Mana.}---}Defining Mana requires knowledge of two other concepts: Heisenberg--Weyl operators, and discrete Wigner functions, which we define in the following section in detail. 
Given a point $\vec{v} = v_1+iv_2 $ in the discretized local phase space $\mathbb{Z}_{D}^{\otimes 2}$, local Heisenberg--Weyl operators are defined as~\cite{veitch2014resource}
\begin{equation}
    T_{\vec{v}}= \tau^{-v_1 v_2} X^{v_1} Z^{v_2}, \quad \tau=e^{i(D+1)\pi /D}\,,
\end{equation}
and generalized to many qudits as
\begin{equation}\label{eq: generalized Heisenberg-Weyl}
    T_{\vec{V}} = T_{\vec{v}^{(1)}} \otimes \dots \otimes T_{\vec{v}^{(N)}},
\end{equation}
where $\vec{V} \equiv {\vec{v}^{(1)} \oplus \dots \oplus \vec{v}_{(N)}}\in \mathbb{Z}_{D}^{\otimes 2} \times \dots \times \mathbb{Z}_{D}^{\otimes 2}$.
Given a vector $\vec{V}$, we define the corresponding phase space operators
\begin{align}
    &A_0 = \frac{1}{D^N} \sum_{\vec{V}} T_{\vec{V}} \,, \nonumber \\
    &A_{\vec{V}}= T_{\vec{V}}A_0 T_{\vec{V}}^{\dagger} \,,
\end{align}
from where we arrive at the definition of the discrete Wigner function of a state $\rho$ as 
\begin{equation}
    W_{\rho}(\vec{V}) \coloneqq \frac{1}{D^N} \rm{Tr} [A_{\vec{V}} \rho].
\end{equation}
Mana is defined as the logarithmic negativity of the Wigner function, as described in Eq.(2) of the main text.

\section{Degeneracy breaking in SK model}
\label{app: Degeneracy_breaking_SK}

The Sherrington-Kirkpatrick (SK) model is an Ising-like model with all-to-all interactions. The generalization of the model to qudit systems is represented by the following Hamiltonian:
\begin{equation}
    \hat{H} = \sum_{i\neq j}J_{ij}(Z_iZ_j^{\dagger}+Z_i^{\dagger}Z_j)\,,
    \label{eq: sk_hamiltonian_appendix}
\end{equation} 
where the interactions $J_{ij}$ are sampled from a Gaussian distribution with unit variance, and the Hermitian conjugate term guarantees the Hermiticity of the Hamiltonian in the qutrit case, where the $Z_i$s are not Hermitian.  
Since we aim to understand the magic generated solely through QAOA, it is crucial to remove any degeneracy, as it may contribute additional magic arising from a superposition of equally valid solution states rather than the algorithm itself. The Hamiltonian (Eq.~\eqref{eq: sk_hamiltonian_appendix}) has symmetry under reflection, $Z_i \leftrightarrow -Z_i$, thus creating two degenerate states. 
The reflection degeneracy can be lifted by adding a longitudinal field, 
$\hat{H}_{{\rm bias},2}=\sum_i h_i (Z_i+Z_i^{\dagger})$, which is reduced to $\sum_i 2h_i Z_i$ for the qubit case. 

However, for qutrits, the Hamiltonian is furthermore symmetric under $Z_i \to cZ_i$, where $c \in \{\pm1, \pm \omega, \pm \omega^2\}$, which introduces $4$ additional symmetries. For example, $Z_iZ_j^\dagger \xrightarrow[]{+\omega} (\omega Z_i) (\omega Z_j)^\dagger= (\omega Z_i) (\omega^2 Z_j^\dagger)=Z_iZ_j^\dagger$. Thus, the qutrit Hamiltonian contains $6$ degenerate eigenstates, equivalent to permutation among the local basis of qutrits, which contains $3!=6$ degenerate states. The longitudinal field $\hat{H}_{{\rm bias},2}$ removes the degeneracy between $\ket{0}$ and $\{\ket{1}, \ket{2}\}$ but retains the degeneracy between $\ket{1}$ and $\ket{2}$, as $(Z +Z^\dagger)= \rm{diag}(2,-1,-1)$. To lift this degeneracy, we add another term $\sum_i ih'_i (Z_i-Z_i^{\dagger})$, as $i(Z-Z^\dagger)=\rm{diag}(0, -\sqrt{3}, +\sqrt{3})$. In our numerics, $h_i$ and $h'_i$ are chosen randomly from a normal distribution with variance $0.3$.

\section{Scaling of nonstabilizerness}
\label{app: Scaling}

Although the states generated by QAOA strongly depend on the circuit depth, optimization parameters, and system size, we observe a qualitatively uniform behavior of the magic barrier in Fig.1 of the main text. For a fixed $N$, while the magic curve varies with the QAOA depth, certain patterns emerge consistently: (1) the barrier height remains nearly unchanged, and (2) as the depth increases, the curve appears to stretch horizontally. These observations motivate the search for a scaling collapse of the magic barrier, which could offer valuable insights into the underlying behavior.

We begin by seeking a scaling collapse for fixed $N$, using Eq.(4) of the main text, where $\mu$ and $\eta$ control the stretching along the $y$- and $x$-axes, respectively. The curves only collapse onto each other when centered around a depth ($d$)-dependent parameter, which we achieve through the introduction of $\xi$ and $\nu$.
We use these parameters to rescale the data for different depths and observe whether the curves align.
The collected scaling parameters for different system sizes are reported in Table~\ref{tab:tab-scaling-1}, which includes the values of $\mu$, $\xi$, $\nu$, and $\eta$ obtained from numerical fits for $N=4,6,8$ qubit systems. 

\begin{table}[h]
    \centering
    \begin{tabular}{cccccc}
        N  & $\mu$ &$\xi$ & $\nu$ & $\eta$ \\
        4 & $-6.47 \cdot 10^{-2}$ & $1.64\cdot 10^{-2}$ & $0.76$ & $0.23$ \\
        6 & $5.11 \cdot 10^{-3}$ & $1.28$ & -$0.34$ & $0.60$\\
        8 & $-6.83 \cdot 10^{-2}$ & $1.38\cdot 10^{-2}$ & $0.95$ & $0.86$\\
    \end{tabular}
    \caption{Scaling parameters $\mu$, $\xi$, $\nu$, and $\eta$ used for data collapse at fixed system size $N$. Values obtained from fitting QAOA magic curves for $N=4,6,8$.}
    \label{tab:tab-scaling-1}
\end{table}

In systems exhibiting universality--such as those undergoing phase transitions---the central parameter (around which the scaling occurs) becomes a constant, signaling a universal or critical point. Motivated by this, we reformulate the scaling ansatz into a slightly modified form 
\begin{equation}\begin{split}
    \frac{\mathcal{M}\left( \ket{\psi_{\rm QAOA_d}(p)}\right)}{N} = d^{-\mu} f\left[ (\lambda - \lambda_c)\cdot  d^\eta \right] ; \quad  \mu, \lambda_c, \eta \in \mathbb{R}; \quad  \lambda=\frac{p}{d}.
\label{eq:scaling_critical_point}
\end{split}\end{equation}
This revised form uses the scaling variable $\lambda = p/d$ and centers the collapse around a critical point $\lambda_c$. The goal is to determine whether such a $\lambda_c$ can be found consistently across system sizes, as a signature of universal behavior.
Table~\ref{tab:tab-scaling-2} reports the corresponding values of $\mu$, $\lambda_c$, and $\eta$ obtained from scaling collapses performed using Eq.~\eqref{eq:scaling_critical_point}  for $N=4,6,8$ qubit systems. Remarkably, the extracted values of $\lambda_c$ are rather consistent across system sizes, suggesting the existence of a constant point in the qubit case, even at small system sizes. This behavior is in contrast to the qutrit case, where no such consistent value of $\lambda_c$ is observed.

\begin{table}[h]
    \centering
    \begin{tabular}{cccc}
        N  & $\mu$ &$\lambda_c$ & $\eta$ \\
        4 &$ -6.51 \cdot 10^{-2}$ & $0.27$ &  $0.23$ \\
        6 & $ 6.50 \cdot 10^{-3}$  &  $0.27$ & $0.59$\\
        8 & $-9.99 \cdot 10^{-2}$ & $0.29$ & $0.92$\\
    \end{tabular}
    \caption{Fitted scaling parameters $\mu$, $\lambda_c$, and $\eta$ using the critical-point form in Eq.~\eqref{eq:scaling_critical_point}. Results shown for $N=4,6,8$ qubits.}
    \label{tab:tab-scaling-2}
\end{table}

\section{Proof of the relationship between Magic and Fidelity}
\label{app: magic-fidelity}

In Fig.2 of main text, the observed correlation between fidelity and final SRE exhibits a distinctive behavior, which can be explained by analytical expressions of the SRE for simple \textit{ansatz} states, which we prove in detail in this section.

\textit{\textbf{Qubits case.---}}We show the explicit derivation of Eq.(6) of the main text.
We start by considering a single qubit and insert the state defined in Eq.(5) into the definition (1) of the main text.
Since for qubits we have only $4$ possible Pauli strings, $P_i= \mathbb{I}, X, Y, Z$, the terms contributing to the sum are the following:
\begin{align}
\langle \psi|I \ket{\psi}  =1 , \quad 
\langle \psi|Z \ket{\psi}  =2\mathcal{F}-1,  \quad 
\langle \psi|X \ket{\psi}  = 2 \sqrt{\mathcal{F}(1-\mathcal{F})}\cos{\theta}, \quad
\langle \psi|Y \ket{\psi}  = 2i \sqrt{\mathcal{F}(1-\mathcal{F})}\sin{\theta}\,.
\end{align}
By summing them, we get 
\begin{align}\label{eq: final-magic-two-states}
    = & 2\left(1-4\mathcal{F}(1-\mathcal{F})+ 14\mathcal{F}^2(1-\mathcal{F})^2 + 2\mathcal{F}^2(1-\mathcal{F})^2 \cos{4\theta}\right)\,.
\end{align}
A similar calculation has been recently published in Ref.~\cite{liu2025maximal}. However, this result is more general than $1$ qubit, as we show below.
Consider a system composed of $n$ qubits. Since the states $\ket{\phi_0}$ and  $\ket{\phi_1}$ are in the computational basis, we can assume that they have Hamming weight distance $h$. Assuming $h<n$, we can write the state $\ket{\psi_{II}}$ as $\ket{\psi'_{II}}_h \otimes\ket{d}_{n-h}$. Here, $\ket{d}=\ket{0}^{\otimes (n-h)}$ or $\ket{1}^{\otimes (n-h)}$. By the additive property of SRE, $M_n(\ket{\psi_{II}})=M_n(\ket{\psi'_{II}}_h)+M_n(\ket{d}_{n-h})$. The contribution $M_n(\ket{d}_{n-h})$ is equal to zero, and thus, the non-trivial contribution to SRE comes from $M_n(\ket{\psi'_{II}}_h)$. We can write $\ket{\psi'_{II}}_h$ as  $\sqrt{\mathcal{F}}\ket{a}_n + e^{i\theta} \sqrt{1-\mathcal{F}}\ket{\bar{a}}_n$. 

This state has the same SRE of $\sqrt{\mathcal{F}}\ket{0}^{\otimes n} + e^{i\theta} \sqrt{1-\mathcal{F}}\ket{1}^{\otimes n}$, as they can be obtained from each other by applying  $X_{\vec{j}}$ at the appropriate positions ($\vec{j}$). This is equivalent to using $P'_i= X_{\vec{j}} P_i X_{\vec{j}}$ on the earlier states. As the Pauli group is invariant under such an operation, we can compute the magic for the latter state without any loss of generality.

We can write all four terms appearing in the expectation values as,
\begin{align}
   \mathcal{F} & \langle 0 |^{\otimes n} P_i \ket{0}^{\otimes n},  \quad
   (1-\mathcal{F}) & \langle 1 |^{\otimes n}  P_i \ket{1}^{\otimes n}, \quad
   \sqrt{\mathcal{F}(1-\mathcal{F})} e^{i\theta} & \langle 0 |^{\otimes n} P_i \ket{1}^{\otimes n}, \quad
  \sqrt{\mathcal{F}(1-\mathcal{F})} e^{-i\theta}  & \langle 1|^{\otimes n}P_i \ket{0}^{\otimes n}, \quad \forall i
\end{align}
The first two terms are non-zero only when the involved string $P_i$ is built only from combinations of $\{I,Z\}$ Pauli operators. In that case, the last two terms are zero. When $P_i$ contains a number $k$ of $Z$ operators, there are ${n \choose k}$ terms contributing to the sum. These will  cumulatively contribute to 
\begin{equation}
    \sum_{k=0}^n {n \choose k} \left| \mathcal{F}+(-1)^k(1-\mathcal{F})\right|^4 = 2^{n-1} [1+ (2\mathcal{F}-1)^4]\,.
\end{equation}

Similarly, the last two terms are non-zero when the $P_i= Y^{\otimes k}X^{\otimes n-k}$. Their contribution to the sum is
\begin{align}
    & \sum_{k=0}^n {n \choose k} \mathcal{F}^2(1-\mathcal{F})^2|e^{i\theta} (-i)^k +e^{-i \theta}(+i)^k|^4 
     =  2^{n-1}\mathcal{F}^2(1-\mathcal{F})^2 2^4 [\cos^4{\theta} + \sin^4{\theta}] \,.
\end{align}

By summing these two contributions and performing some simple algebra, one obtains the final value 
\begin{align}
    M_2  
    = & -\log_2\left[1-4\mathcal{F}(1-\mathcal{F})+ 2\mathcal{F}^2(1-\mathcal{F})^2(7+\cos{4\theta})\right] 
\label{eq: magic-two-state}
\end{align}

The formula is $n$-independent because what counts is the number of states in superposition, not the dimension of the system or which two basis states constitute the state.

Motivated by the $n$ independence of the previous result, we also investigate the SRE of a superposition of three computational basis states as in Eq.(7). For different system sizes, we find an expression consistent for different $N$ with the following:
\begin{equation}
\begin{aligned}
M_2 = -\log_2 & \Big[ 1 + 14 f^4 + 28 f^3 (p - 1)  
 + 2(p - 1)p\left(2 + 7p(p - 1)\right) + 6 f^2 \left(3 + p(7p - 6)\right) 
+ 4 f(p - 1)\left(1 + p(7p - 6)\right) \\
&+ 2 f^2 (f + p - 1)^2 \cos(4\phi) 
+ 4 p^2 f^2 \cos(2\phi)\cos(2\phi - 4\theta) 
+ 2 p^2 (p - 1)(2f + p - 1) \cos(4\phi - 4\theta) \Big]\,.
\end{aligned}
\label{eq: magic-threee-state}
\end{equation}

\begin{figure}[h!]
    \centering
    \includegraphics[width=\textwidth]{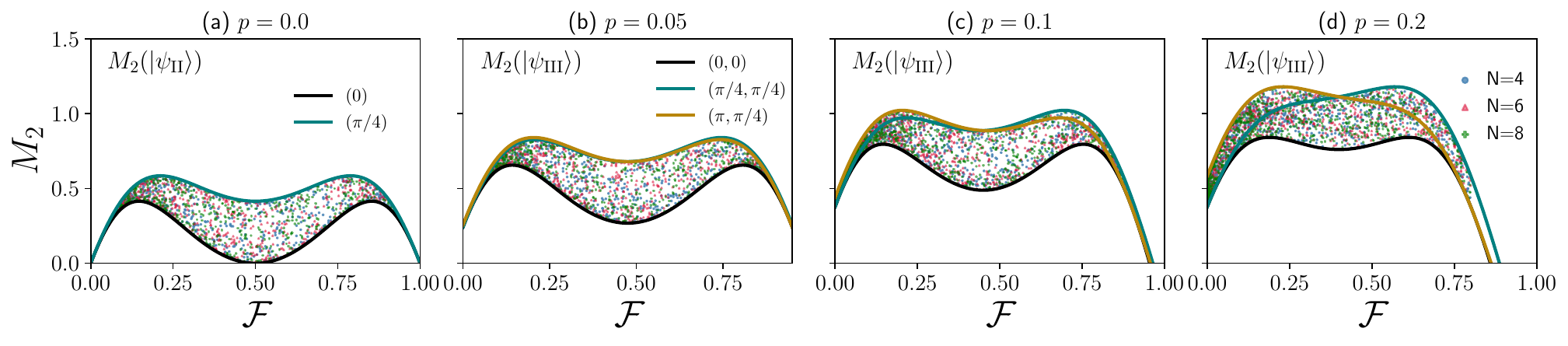}
    \caption{SRE vs.\ fidelity for a superposition of two (a) and three (b,c,d) states at different values of $p$. Continuous lines represent analytic bounds given by Eq.~\eqref{eq: magic-two-state} and \eqref{eq: magic-threee-state}, respectively. Dots represent numerically computed SRE for random states of the form $\ket{\psi_{\rm II}}$ (Eq. 5 of main text) (a) and $\ket{\psi_{\rm III}}$(Eq. 7 of main text) (b,c,d). The points are clearly enclosed within the bounds given by the continuous lines, certifying the validity of \eqref{eq: magic-threee-state}. }
    \label{fig: benchmarks}
\end{figure}

As in the case of magic computed for a superposition of two computational basis states, the analytical expression derived here remains valid for states of arbitrary dimension.
The generalization of the Eq.~\eqref{eq: magic-two-state} and \eqref{eq: magic-threee-state} to any dimension systems is shown in Fig.~\ref{fig: benchmarks}, where random benchmarks are used for this purpose.

\textit{\textbf{Qutrits case.---}} 
The pauli group for qutrits is defined as $P = \{X^rZ^j|r,j \in (0,1,2)\}$. By repeating the calculation of \eqref{eq: magic-two-state} for a qutrit state $\ket{\psi} = \sqrt{\mathcal{F}}\ket{a}+e^{i\theta}\sqrt{1-\mathcal{F}}\ket{b}$, we get 
\begin{align}
    \bra{\psi}P_i\ket{\psi} =&\mathcal{F}\bra{a}P_i\ket{a}+(1-\mathcal{F})\bra{b}P_i\ket{b}+
    e^{i\theta}\sqrt{\mathcal{F}(1-\mathcal{F})}\bra{a}P_i\ket{b}+e^{-i\theta}\sqrt{\mathcal{F}(1-\mathcal{F})}\bra{b}P_i\ket{a}
\end{align}
The first two terms will be different from zero when $a=0$, while the second two terms will be different from zero when $a\neq 0$. Thus, we have
\begin{align}
    e^{-M_2} = &\sum_{=0,1,2} |\mathcal{F}\bra{a}Z^j\ket{a}+(1-\mathcal{F})\bra{b}Z^j\ket{b}|^4
              +\sum_{r=1,2;j = 0,1,2}|e^{i\theta}\sqrt{\mathcal{F}(1-\mathcal{F})}\bra{a}X^{r}Z^{j}\ket{b}
              +e^{-i\theta}\sqrt{\mathcal{F}(1-\mathcal{F})}\bra{b}X^{r}Z^{j}\ket{a}|^4\nonumber\\
        = & 1+|\omega^a \mathcal{F}+(1-\mathcal{F})\omega^b|^4+|\omega^{2a} \mathcal{F}+(1-\mathcal{F})\omega^{2b}|^4
         + \mathcal{F}^2(1-\mathcal{F})^2 \sum_{r=1,2; j=0,1,2} |e^{i \theta} \omega^{jb} \delta_{a,b\oplus r } +e^{-i\theta} \omega^{ja} \delta_{b,a\oplus r} |^4\,.
\end{align}
Notice that for a qutrit, $a= b\oplus r $ and $b= a\oplus r $ cannot be satisfied at the same time. This will remove any dependence from the relative phase $\theta$ between the two basis states ($\oplus$ indicated sum mod $d$).  
Considering $a=0,b=1$ we get
\begin{align*}
    e^{-M_2} =1+|\mathcal{F}+(1-\mathcal{F})\omega|^4+|\mathcal{F}+(1-\mathcal{F})\omega^{2}|^4+3\mathcal{F}^2(1-\mathcal{F})^2\,.
\end{align*}

\section{Energy and Fidelity as a function of layer index}
\label{app: energy-fidelity}
\begin{figure*}[ht!]
    \centering
    \includegraphics[width=0.85 \textwidth]{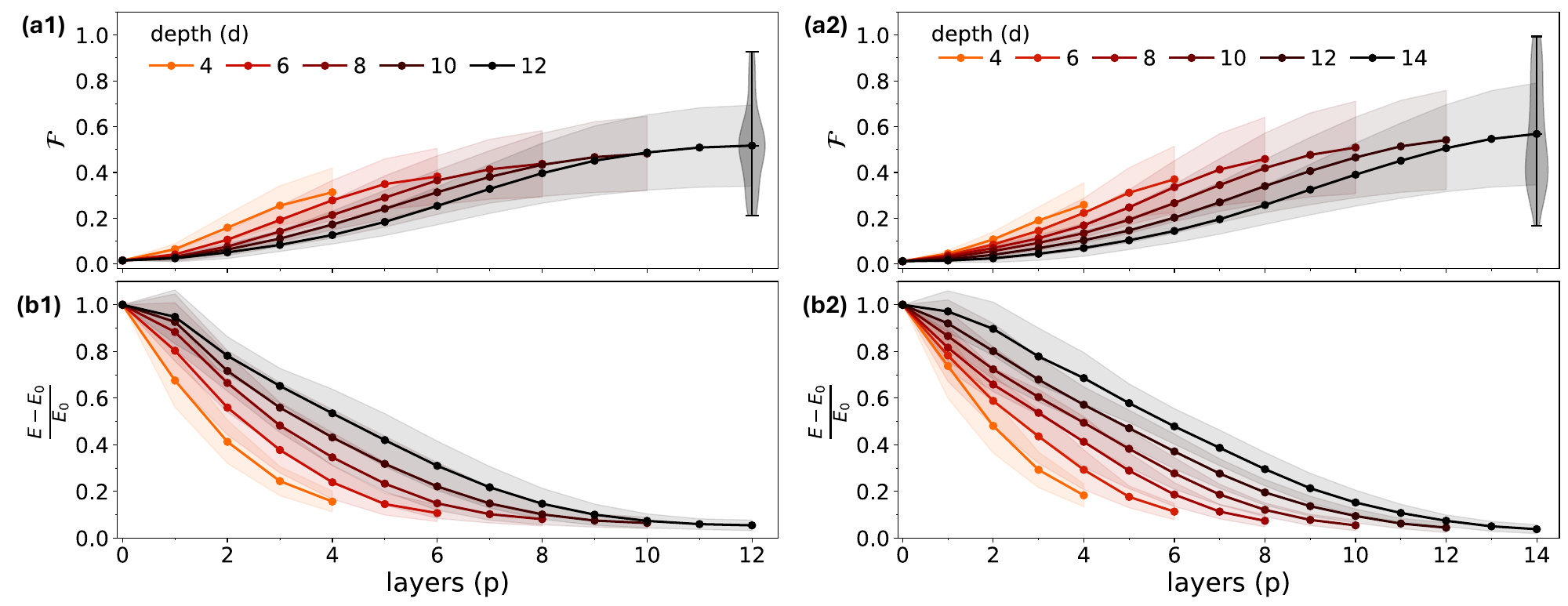}
    \caption{(a1) Mean value of the fidelity over 50 different realizations, with relative standard deviation, for each layer at different depths of the 6-qubit system. The violin plot for the curve at depth 12, using the full 12-layer wave function, highlights the high fidelity achieved by some realizations.(b1) Mean value of the relative energy under the same conditions.(a2) and (b2) show the corresponding results for a 4-qutrit system.}
    \label{fig:energy-fidelity}
\end{figure*}

For completeness, we show the evolution of the approximation ratio $\frac{E-E_0}{E_0}$(Fig.~\ref{fig:energy-fidelity} (a)) and fidelity $\mathcal{F}$(Fig.~\ref{fig:energy-fidelity} (b)) of the optimized QAOA state with the solution of the problem. We see that even though the energy reaches values close to the ground state, the fidelity is far from being equal to one. However, both of these quantities monotonically improve despite the appearance of the magic barrier. 
Consequently, the quantum computational resources required during the intermediate steps of QAOA significantly exceed those needed for direct simulation of the final target state. This additional ``cost'' in nonstabilizerness, $\Delta \mathcal{M}=\mathcal{M}_{\rm max}-\mathcal{M}_{\rm final}$ emerges inherently from the QAOA ansatz structure. It is crucial to account for this elevated resource demand in the design and benchmarking of fault-tolerant quantum devices.

\section{Magic during variational optimization feedback loop}
\label{app: magic-in-opt}

\begin{figure*}[hbt!]
\centering\includegraphics[width=0.85\textwidth]{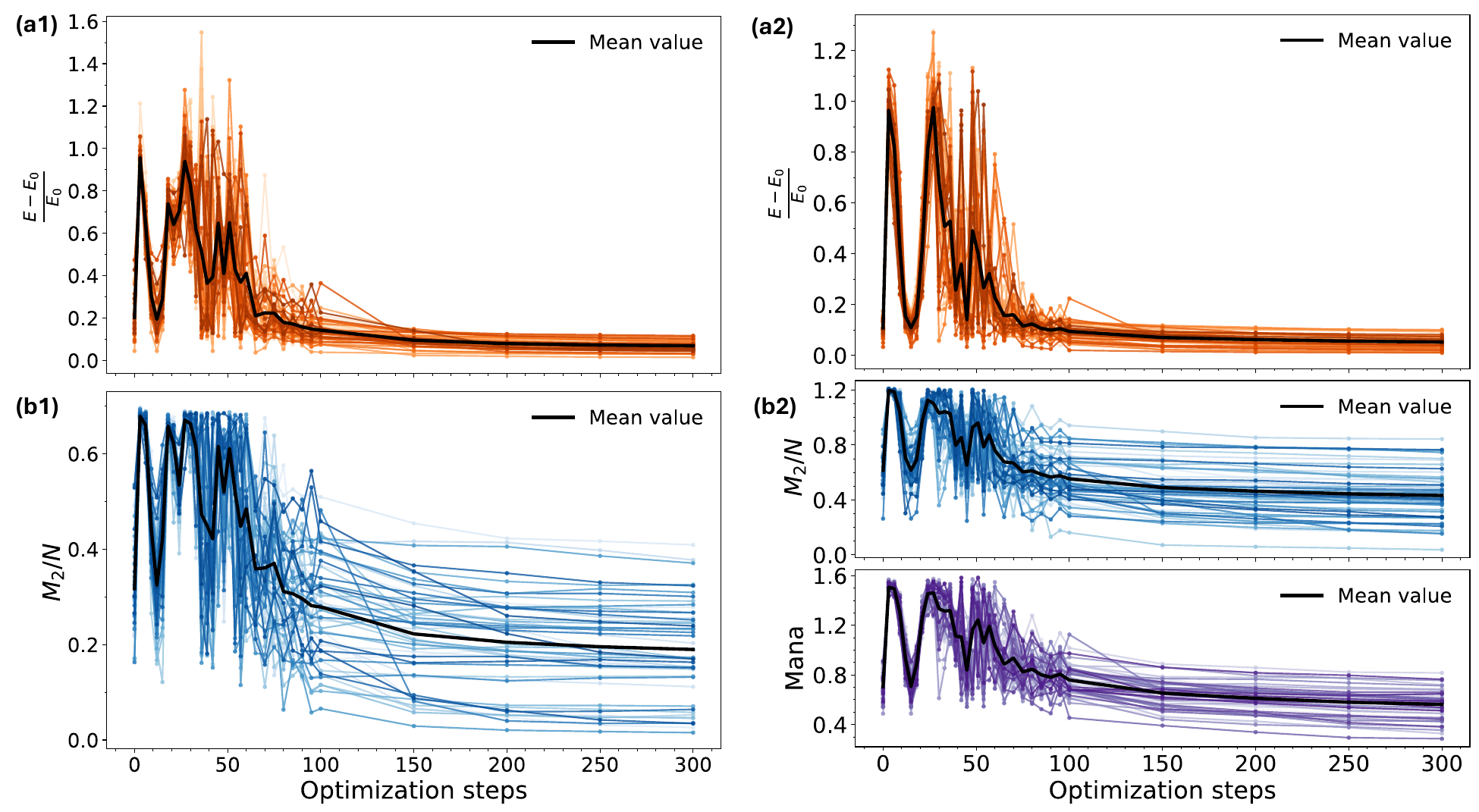} 
    \caption{Energy and Magic barrier during optimization protocol. 
    The black curves represent the mean values calculated over $50$ realizations. (a1), (a2) Energy ratio as a function of the number of optimization steps during a QAOA run of each problem realization for 6-qubit and 4-qudits systems, respectively. (b1) Magic as a function of optimization steps of each QAOA run. (b2) Magic in the upper plot and Mana in the lower plot as a function of optimization steps.}
    \label{fig: barriers_in_optimization}
\end{figure*}

To reach a more complete understanding of the role of nonstabilizerness in solving optimization problems using variational quantum algorithms, we analyze how it evolves throughout the optimization process.
QAOA heavily relies on classical optimization to determine optimal parameters, requiring repeated execution of quantum circuits (Fig.1(a) in the main text) that include non-Clifford gates, making them resource-intensive in terms of quantum magic.
By studying how magic evolves during the process of optimization of the circuit parameters, we extract information about the resources required to achieve convergence.
In particular, we consider $50$ problem instances, each defined by different Hamiltonian coefficients, and run $50$ QAOA optimizations per instance, initialized with annealing-inspired parameters (decreasing $\gamma$, increasing $\beta$) ~\cite{sack2021quantum}. For each instance, we select the run achieving the lowest energy as the best-performing optimization.

Figure ~\ref{fig: barriers_in_optimization} shows energy values and nonstabilizerness measures as a function of the optimization steps for the best-performing QAOA run of each of the $50$ realizations. Panels (a1) and (b1) refer results for a 6-qubits system using a QAOA with circuit depth of 12, while panels (a2) and (b2) report results for a 4-qutrits system using a QAOA circuit with depth 14.
The success of QAOA is measured here using the approximation ratio, defined as $\frac{E-E_0}{E_0}$, where $E$ is the expectation value of the Hamiltonian for the optimized state, and $E_0$ represents the minimum energy obtained through exact diagonalization. During optimization, the energy initially increases, crosses a barrier, and eventually converges close to $E_0$ for nearly all problem instances.
Despite the initial parameterization ensuring a good starting point, the optimization process allows the algorithm to escape local minima and reach a lower-energy minimum, closer to the ground state. 
This energy barrier is accompanied by a peak in magic, suggesting that crossing a high-nonstabilizerness region is required for convergence. A similar pattern is observed for the Mana in the 4-qutrit case, shown in Fig.~\ref{fig: barriers_in_optimization}(b2).

Although magic, like energy, shows a barrier during optimization, it captures a different property: the distance from stabilizer states in the Clifford group. As such, it is not necessarily expected to follow the energy landscape directly. While the energy approximation ratio consistently approaches zero across different instances—indicating convergence toward the ground state \cite{farhi2022quantum}—the final magic values remain widely spread. This highlights a key point: although the ground state has zero magic, optimized QAOA states often retain finite magic, showing that energy convergence does not imply convergence to the exact solution.
Moreover, our Hamiltonians are non-degenerate and thus their ground state are product-states with zero magic. This suggests that even near the exact solution energy, QAOA may converge to a superposition of low-lying eigenstates. This occurs when energy level splittings---which depend on the external field strength---are small, preventing the algorithm from fully collapsing onto the true ground state. We explore this further in the next section through fidelity analysis with the exact solution.

\section{Final results as a function of QAOA depth}
\label{app: final_QAOA_data}

To round off our analysis, we now focus on the final results at the end of the optimization process. Specifically, we analyze the 
 following quantities: final value of the energy ratio, fidelity of the optimized state with the ground state, final SRE, and also Mana in the case of qutrits.
 \begin{figure*}[b]
    \centering
    \includegraphics[width=0.85\textwidth]{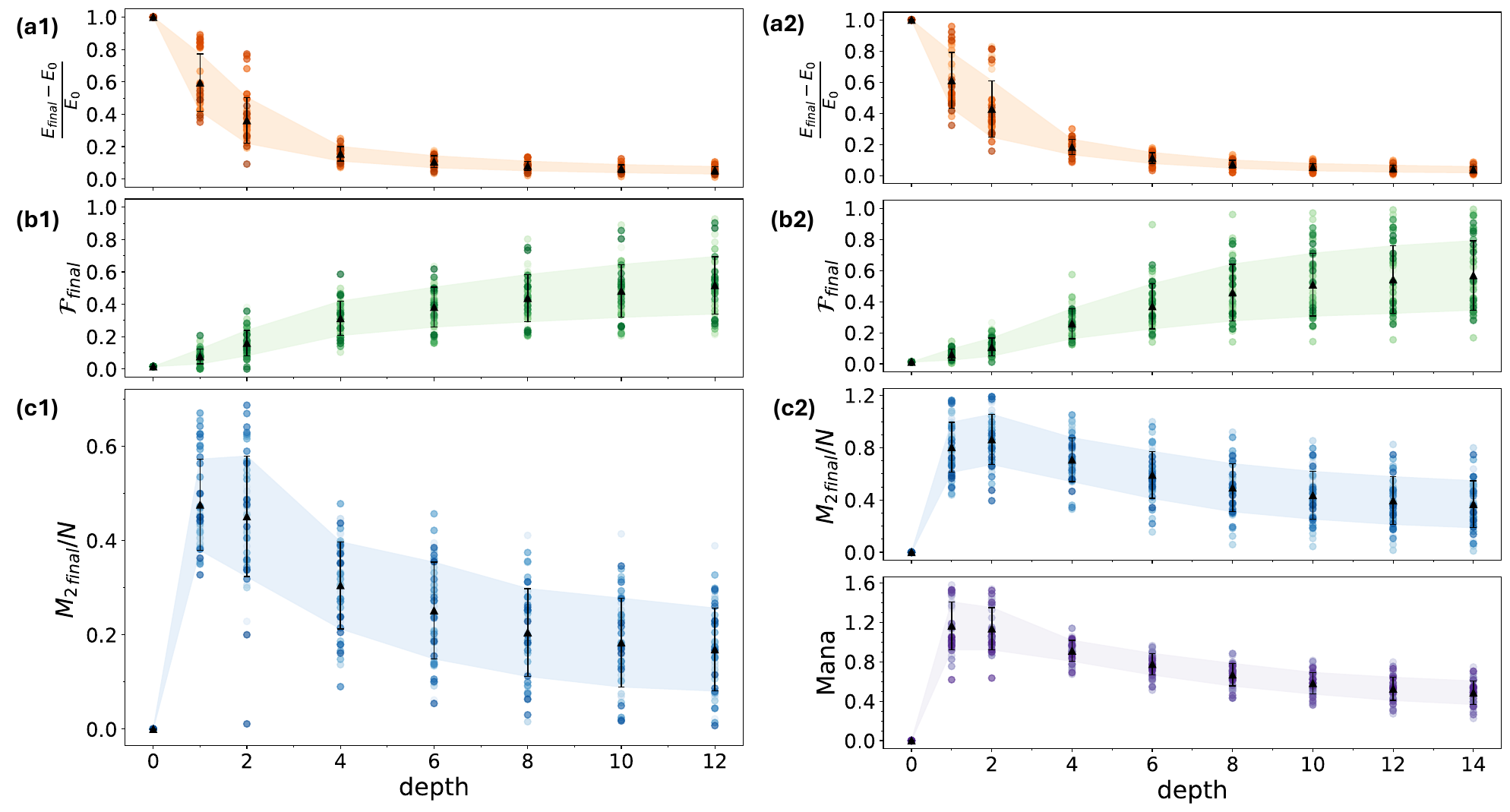}
    \caption{Final results obtained from the optimized outcomes as a function of QAOA optimization depth for a 6-qubit system and a 4-qudit system across 50 random realizations of the SK model. (a1), (b1), and (c1): Energy ratio (orange), Fidelity (green), and SRE density (blue), respectively, for the qubit system. Colored dots represent individual realizations, black triangles indicate the average over all realizations, and the shaded regions correspond to the standard deviation from the mean values. (a2), (b2), and (c2): Energy ratio (orange), Fidelity (green), SRE density (blue), and Mana (violet), respectively, for the qudit system. 
    }
    \label{fig: final_results}
\end{figure*}
As shown in Fig.~\ref{fig: final_results}, which presents results for 6-qubit and 4-qutrit systems optimized using different circuit depths in the QAOA ansatz, the energy converges to values close to the ground state for sufficiently high depth. The fidelity between the optimized and ground states is low for small circuit depths and gradually increases with depth, though it never reaches the ideal value of 1. The value of SRE (and Mana), in contrast, initially rises rapidly, reaches its maximum at small depths, and then decreases without ever reaching zero. 
Both fidelity and magic can vary significantly depending on the specific realization.
The fidelity values different from zero and the related nonzero magic suggest that the algorithm drives the system toward a superposition of states with similar energy values. 
The lower plot of Fig.~\ref{fig: final_results}(c2) shows that Mana computed for the system of qutrits has a similar behavior to the qubit SRE.

\begin{figure*}[hbt!]
    \centering
    \includegraphics[width=1\textwidth]{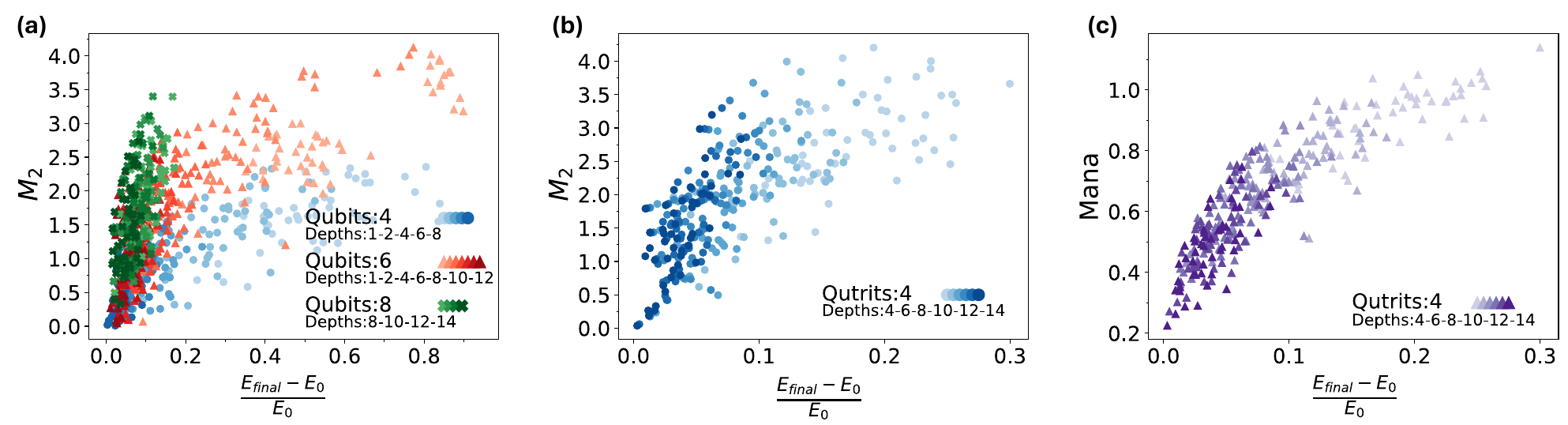}
    \caption{Magic and Mana values calculated for the optimized state as a function of the final optimized energy ratio. (a) Magic for qubit systems. In the scatter plots, all the data related to different numbers of qubit systems and different ansatz depths are collected. (b),(c) Magic and Mana for 4 qutrit system.}
    \label{fig: magic_vs_energy}
\end{figure*}
A similar trend is observed in Fig.~\ref{fig: magic_vs_energy}, which shows the final SRE (and Mana) as a function of the approximation ratio for different systems and varying circuit depth. The plot highlights that many solutions achieve energy values close to the ground state while still retaining nonzero magic.

\section{Additional Results for qutrits}
\label{app: add_qutrits}
For completeness, we also present the results for the 4-qutrit system, which were not reported in the main text. The additional data include the conditional probability, computed as in Eq.(8) of main text, where  `demagication' refers to the difference between the maximum and minimum values of the SRE. We also show the fidelity of the optimized states as a function of nonstabilizerness measures, considering both SRE and Mana.

\begin{figure}[hbt!]
    \centering\includegraphics[width=0.6\columnwidth]{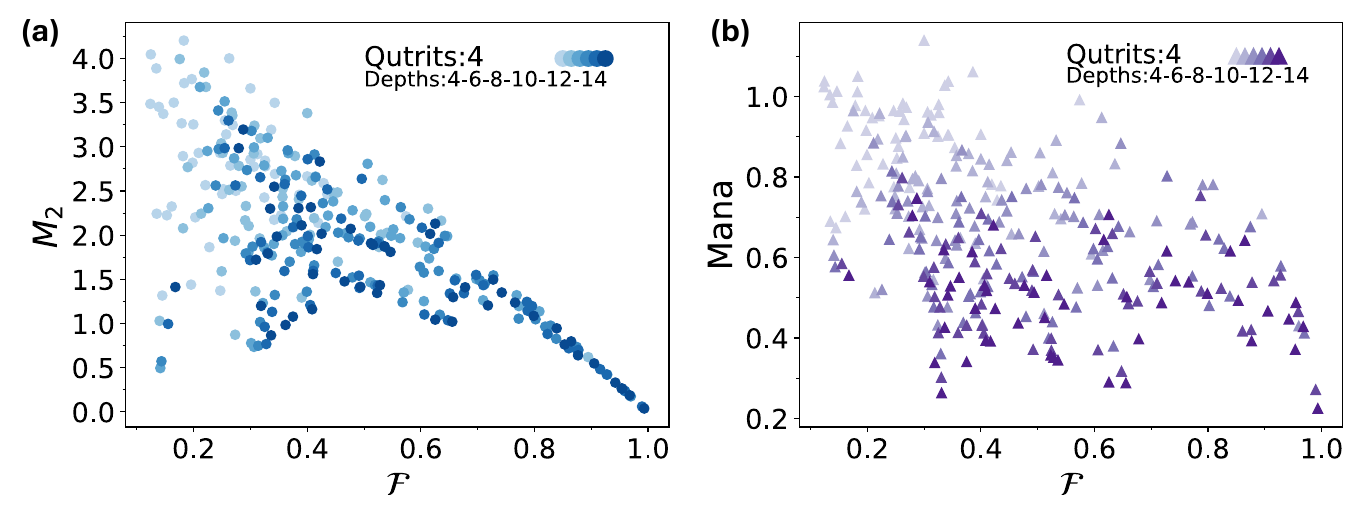} 
    \caption{SRE (a) and Mana (b) vs.\ Fidelity for a system of four qutrits. Different shades represent different depths of the QAOA algorithm.}
    \label{fig:fidelity_vs_magic}
\end{figure}

\textit{\textbf{Final nonstabilizerness and Fidelity}}.
Figure \ref{fig:fidelity_vs_magic} (a-b) presents the values of nonstabilizerness vs.\ fidelity of the optimized QAOA state for 4-qutrit systems,  analogous to the qubits case shown in Fig. 2 of the main text. 

\begin{figure}[hbt!]
    \centering\includegraphics[width=0.5\columnwidth]{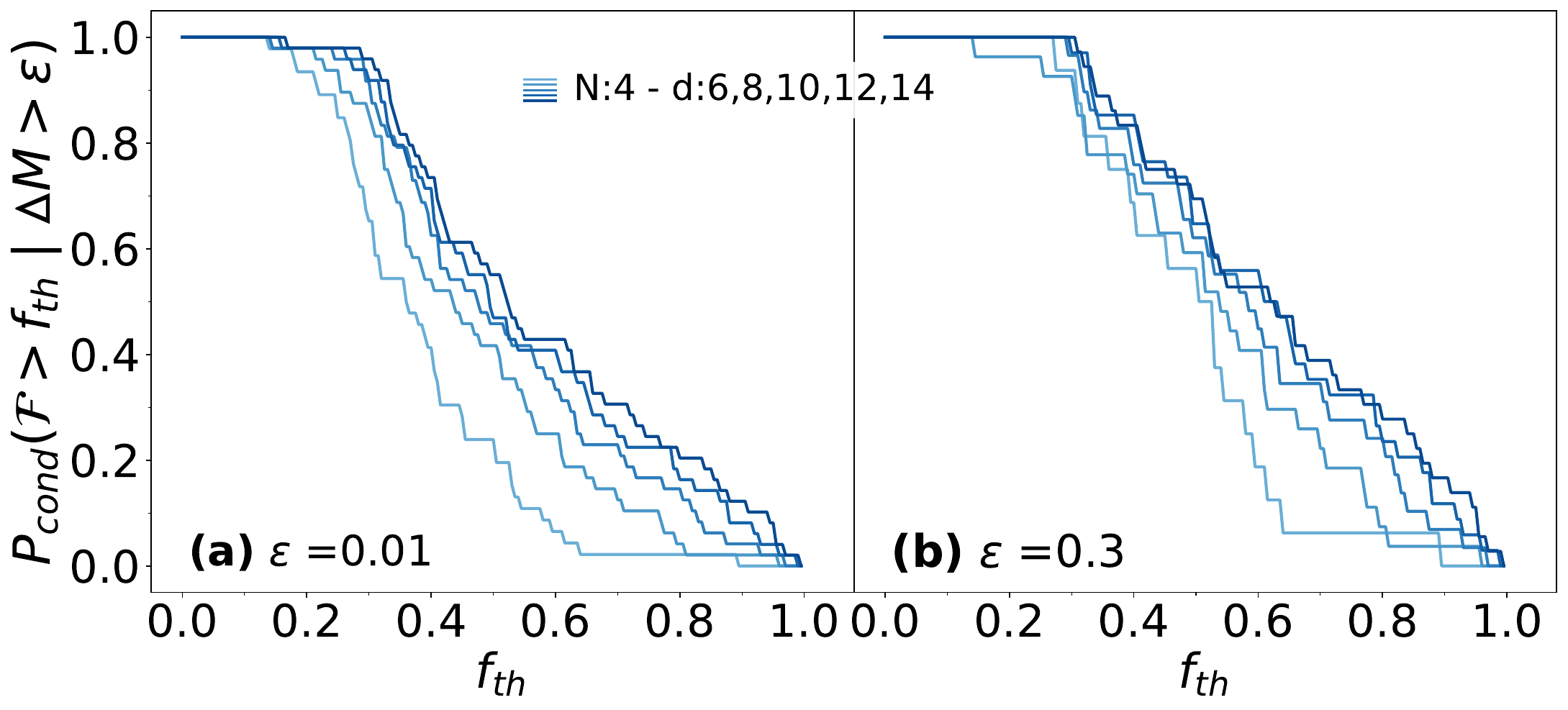} 
    \caption{(a) Conditional probability curve for $\epsilon=0.01$ and (b) $\epsilon=0.03$  obtained for four qutrits at QAOA depths $6-8-10-12-14$. Also in this case, similar to Fig. 3 of the main text, it is possible to see that higher demagication thresholds lead to higher fidelities.}
    \label{fig:demagication_qutrits}
\end{figure}

\textit{\textbf{Demagication}}. Figure \ref{fig:demagication_qutrits} (a-b) shows the conditional probability of achieving a fidelity above a given threshold, provided the QAOA reaches a minimum amount of demagication, for a 4-qutrit system. The results are comparable to those obtained with the qubit-based algorithm reported in Fig. 3 of the main text.

\twocolumngrid


\begin{thebibliography}{108}%
\makeatletter
\providecommand \@ifxundefined [1]{%
 \@ifx{#1\undefined}
}%
\providecommand \@ifnum [1]{%
 \ifnum #1\expandafter \@firstoftwo
 \else \expandafter \@secondoftwo
 \fi
}%
\providecommand \@ifx [1]{%
 \ifx #1\expandafter \@firstoftwo
 \else \expandafter \@secondoftwo
 \fi
}%
\providecommand \natexlab [1]{#1}%
\providecommand \enquote  [1]{``#1''}%
\providecommand \bibnamefont  [1]{#1}%
\providecommand \bibfnamefont [1]{#1}%
\providecommand \citenamefont [1]{#1}%
\providecommand \href@noop [0]{\@secondoftwo}%
\providecommand \href [0]{\begingroup \@sanitize@url \@href}%
\providecommand \@href[1]{\@@startlink{#1}\@@href}%
\providecommand \@@href[1]{\endgroup#1\@@endlink}%
\providecommand \@sanitize@url [0]{\catcode `\\12\catcode `\$12\catcode `\&12\catcode `\#12\catcode `\^12\catcode `\_12\catcode `\%12\relax}%
\providecommand \@@startlink[1]{}%
\providecommand \@@endlink[0]{}%
\providecommand \url  [0]{\begingroup\@sanitize@url \@url }%
\providecommand \@url [1]{\endgroup\@href {#1}{\urlprefix }}%
\providecommand \urlprefix  [0]{URL }%
\providecommand \Eprint [0]{\href }%
\providecommand \doibase [0]{https://doi.org/}%
\providecommand \selectlanguage [0]{\@gobble}%
\providecommand \bibinfo  [0]{\@secondoftwo}%
\providecommand \bibfield  [0]{\@secondoftwo}%
\providecommand \translation [1]{[#1]}%
\providecommand \BibitemOpen [0]{}%
\providecommand \bibitemStop [0]{}%
\providecommand \bibitemNoStop [0]{.\EOS\space}%
\providecommand \EOS [0]{\spacefactor3000\relax}%
\providecommand \BibitemShut  [1]{\csname bibitem#1\endcsname}%
\let\auto@bib@innerbib\@empty
\bibitem [{\citenamefont {Lucas}(2014)}]{lucas2014ising}%
  \BibitemOpen
  \bibfield  {author} {\bibinfo {author} {\bibfnamefont {A.}~\bibnamefont {Lucas}},\ }\bibfield  {title} {\bibinfo {title} {Ising formulations of many {NP} problems},\ }\href {https://www.frontiersin.org/journals/physics/articles/10.3389/fphy.2014.00005/full} {\bibfield  {journal} {\bibinfo  {journal} {Frontiers in physics}\ }\textbf {\bibinfo {volume} {2}},\ \bibinfo {pages} {5} (\bibinfo {year} {2014})}\BibitemShut {NoStop}%
\bibitem [{\citenamefont {Mannhold}\ \emph {et~al.}(2009)\citenamefont {Mannhold}, \citenamefont {Kubinyi},\ and\ \citenamefont {Timmerman}}]{mannhold2009combinatorial}%
  \BibitemOpen
  \bibfield  {author} {\bibinfo {author} {\bibfnamefont {R.}~\bibnamefont {Mannhold}}, \bibinfo {author} {\bibfnamefont {H.}~\bibnamefont {Kubinyi}},\ and\ \bibinfo {author} {\bibfnamefont {H.}~\bibnamefont {Timmerman}},\ }\href {https://onlinelibrary.wiley.com/doi/book/10.1002/9783527614141} {\emph {\bibinfo {title} {Combinatorial Chemistry: A practical approach}}}\ (\bibinfo  {publisher} {John Wiley \& Sons},\ \bibinfo {year} {2009})\BibitemShut {NoStop}%
\bibitem [{\citenamefont {Yu}(2013)}]{yu2013industrial}%
  \BibitemOpen
  \bibfield  {author} {\bibinfo {author} {\bibfnamefont {G.}~\bibnamefont {Yu}},\ }\href {https://link.springer.com/book/10.1007/978-1-4757-2876-7} {\emph {\bibinfo {title} {Industrial applications of combinatorial optimization}}},\ Vol.~\bibinfo {volume} {16}\ (\bibinfo  {publisher} {Springer Science \& Business Media},\ \bibinfo {year} {2013})\BibitemShut {NoStop}%
\bibitem [{\citenamefont {Odili}(2017)}]{odili2017combinatorial}%
  \BibitemOpen
  \bibfield  {author} {\bibinfo {author} {\bibfnamefont {J.~B.}\ \bibnamefont {Odili}},\ }\bibfield  {title} {\bibinfo {title} {Combinatorial optimization in science and engineering},\ }\href {https://www.jstor.org/stable/26493534?seq=1} {\bibfield  {journal} {\bibinfo  {journal} {Current Science}\ ,\ \bibinfo {pages} {2268}} (\bibinfo {year} {2017})}\BibitemShut {NoStop}%
\bibitem [{\citenamefont {Yarkoni}\ \emph {et~al.}(2022)\citenamefont {Yarkoni}, \citenamefont {Raponi}, \citenamefont {B{\"a}ck},\ and\ \citenamefont {Schmitt}}]{yarkoni2022quantum}%
  \BibitemOpen
  \bibfield  {author} {\bibinfo {author} {\bibfnamefont {S.}~\bibnamefont {Yarkoni}}, \bibinfo {author} {\bibfnamefont {E.}~\bibnamefont {Raponi}}, \bibinfo {author} {\bibfnamefont {T.}~\bibnamefont {B{\"a}ck}},\ and\ \bibinfo {author} {\bibfnamefont {S.}~\bibnamefont {Schmitt}},\ }\bibfield  {title} {\bibinfo {title} {Quantum annealing for industry applications: Introduction and review},\ }\href {https://iopscience.iop.org/article/10.1088/1361-6633/ac8c54/meta} {\bibfield  {journal} {\bibinfo  {journal} {Reports on Progress in Physics}\ }\textbf {\bibinfo {volume} {85}},\ \bibinfo {pages} {104001} (\bibinfo {year} {2022})}\BibitemShut {NoStop}%
\bibitem [{\citenamefont {Wang}\ \emph {et~al.}(2023)\citenamefont {Wang}, \citenamefont {Chen}, \citenamefont {Yang}, \citenamefont {Lee},\ and\ \citenamefont {Tseng}}]{10082989}%
  \BibitemOpen
  \bibfield  {author} {\bibinfo {author} {\bibfnamefont {P.-H.}\ \bibnamefont {Wang}}, \bibinfo {author} {\bibfnamefont {J.-H.}\ \bibnamefont {Chen}}, \bibinfo {author} {\bibfnamefont {Y.-Y.}\ \bibnamefont {Yang}}, \bibinfo {author} {\bibfnamefont {C.}~\bibnamefont {Lee}},\ and\ \bibinfo {author} {\bibfnamefont {Y.~J.}\ \bibnamefont {Tseng}},\ }\bibfield  {title} {\bibinfo {title} {Recent advances in quantum computing for drug discovery and development},\ }\href {https://doi.org/10.1109/MNANO.2023.3249499} {\bibfield  {journal} {\bibinfo  {journal} {IEEE Nanotechnology Magazine}\ }\textbf {\bibinfo {volume} {17}},\ \bibinfo {pages} {26} (\bibinfo {year} {2023})}\BibitemShut {NoStop}%
\bibitem [{\citenamefont {Abbas}\ \emph {et~al.}(2023)\citenamefont {Abbas}, \citenamefont {Ambainis}, \citenamefont {Augustino}, \citenamefont {B{\"a}rtschi}, \citenamefont {Buhrman}, \citenamefont {Coffrin}, \citenamefont {Cortiana}, \citenamefont {Dunjko}, \citenamefont {Egger}, \citenamefont {Elmegreen} \emph {et~al.}}]{abbas2023quantum}%
  \BibitemOpen
  \bibfield  {author} {\bibinfo {author} {\bibfnamefont {A.}~\bibnamefont {Abbas}}, \bibinfo {author} {\bibfnamefont {A.}~\bibnamefont {Ambainis}}, \bibinfo {author} {\bibfnamefont {B.}~\bibnamefont {Augustino}}, \bibinfo {author} {\bibfnamefont {A.}~\bibnamefont {B{\"a}rtschi}}, \bibinfo {author} {\bibfnamefont {H.}~\bibnamefont {Buhrman}}, \bibinfo {author} {\bibfnamefont {C.}~\bibnamefont {Coffrin}}, \bibinfo {author} {\bibfnamefont {G.}~\bibnamefont {Cortiana}}, \bibinfo {author} {\bibfnamefont {V.}~\bibnamefont {Dunjko}}, \bibinfo {author} {\bibfnamefont {D.~J.}\ \bibnamefont {Egger}}, \bibinfo {author} {\bibfnamefont {B.~G.}\ \bibnamefont {Elmegreen}}, \emph {et~al.},\ }\bibfield  {title} {\bibinfo {title} {Quantum optimization: Potential, challenges, and the path forward},\ }\href {https://arxiv.org/abs/2312.02279} {\bibfield  {journal} {\bibinfo  {journal} {arXiv:2312.02279}\ } (\bibinfo {year} {2023})}\BibitemShut {NoStop}%
\bibitem [{\citenamefont {Farhi}\ \emph {et~al.}(2014)\citenamefont {Farhi}, \citenamefont {Goldstone},\ and\ \citenamefont {Gutmann}}]{farhi2014quantum}%
  \BibitemOpen
  \bibfield  {author} {\bibinfo {author} {\bibfnamefont {E.}~\bibnamefont {Farhi}}, \bibinfo {author} {\bibfnamefont {J.}~\bibnamefont {Goldstone}},\ and\ \bibinfo {author} {\bibfnamefont {S.}~\bibnamefont {Gutmann}},\ }\bibfield  {title} {\bibinfo {title} {A quantum approximate optimization algorithm},\ }\href {https://arxiv.org/abs/1411.4028} {\bibfield  {journal} {\bibinfo  {journal} {arXiv:1411.4028}\ } (\bibinfo {year} {2014})}\BibitemShut {NoStop}%
\bibitem [{\citenamefont {Preskill}(2018)}]{preskill2018quantum}%
  \BibitemOpen
  \bibfield  {author} {\bibinfo {author} {\bibfnamefont {J.}~\bibnamefont {Preskill}},\ }\bibfield  {title} {\bibinfo {title} {Quantum computing in the nisq era and beyond},\ }\href {https://quantum-journal.org/papers/q-2018-08-06-79/} {\bibfield  {journal} {\bibinfo  {journal} {Quantum}\ }\textbf {\bibinfo {volume} {2}},\ \bibinfo {pages} {79} (\bibinfo {year} {2018})}\BibitemShut {NoStop}%
\bibitem [{\citenamefont {Weidenfeller}\ \emph {et~al.}(2022)\citenamefont {Weidenfeller}, \citenamefont {Valor}, \citenamefont {Gacon}, \citenamefont {Tornow}, \citenamefont {Bello}, \citenamefont {Woerner},\ and\ \citenamefont {Egger}}]{weidenfeller2022scaling}%
  \BibitemOpen
  \bibfield  {author} {\bibinfo {author} {\bibfnamefont {J.}~\bibnamefont {Weidenfeller}}, \bibinfo {author} {\bibfnamefont {L.~C.}\ \bibnamefont {Valor}}, \bibinfo {author} {\bibfnamefont {J.}~\bibnamefont {Gacon}}, \bibinfo {author} {\bibfnamefont {C.}~\bibnamefont {Tornow}}, \bibinfo {author} {\bibfnamefont {L.}~\bibnamefont {Bello}}, \bibinfo {author} {\bibfnamefont {S.}~\bibnamefont {Woerner}},\ and\ \bibinfo {author} {\bibfnamefont {D.~J.}\ \bibnamefont {Egger}},\ }\bibfield  {title} {\bibinfo {title} {Scaling of the quantum approximate optimization algorithm on superconducting qubit based hardware},\ }\href {https://quantum-journal.org/papers/q-2022-12-07-870/} {\bibfield  {journal} {\bibinfo  {journal} {Quantum}\ }\textbf {\bibinfo {volume} {6}},\ \bibinfo {pages} {870} (\bibinfo {year} {2022})}\BibitemShut {NoStop}%
\bibitem [{\citenamefont {Lotshaw}\ \emph {et~al.}(2022)\citenamefont {Lotshaw}, \citenamefont {Nguyen}, \citenamefont {Santana}, \citenamefont {McCaskey}, \citenamefont {Herrman}, \citenamefont {Ostrowski}, \citenamefont {Siopsis},\ and\ \citenamefont {Humble}}]{lotshaw2022scaling}%
  \BibitemOpen
  \bibfield  {author} {\bibinfo {author} {\bibfnamefont {P.~C.}\ \bibnamefont {Lotshaw}}, \bibinfo {author} {\bibfnamefont {T.}~\bibnamefont {Nguyen}}, \bibinfo {author} {\bibfnamefont {A.}~\bibnamefont {Santana}}, \bibinfo {author} {\bibfnamefont {A.}~\bibnamefont {McCaskey}}, \bibinfo {author} {\bibfnamefont {R.}~\bibnamefont {Herrman}}, \bibinfo {author} {\bibfnamefont {J.}~\bibnamefont {Ostrowski}}, \bibinfo {author} {\bibfnamefont {G.}~\bibnamefont {Siopsis}},\ and\ \bibinfo {author} {\bibfnamefont {T.~S.}\ \bibnamefont {Humble}},\ }\bibfield  {title} {\bibinfo {title} {Scaling quantum approximate optimization on near-term hardware},\ }\href {https://www.nature.com/articles/s41598-022-14767-w} {\bibfield  {journal} {\bibinfo  {journal} {Scientific Reports}\ }\textbf {\bibinfo {volume} {12}},\ \bibinfo {pages} {12388} (\bibinfo {year} {2022})}\BibitemShut {NoStop}%
\bibitem [{\citenamefont {Pelofske}\ \emph {et~al.}(2023)\citenamefont {Pelofske}, \citenamefont {B{\"a}rtschi},\ and\ \citenamefont {Eidenbenz}}]{pelofske2023quantum}%
  \BibitemOpen
  \bibfield  {author} {\bibinfo {author} {\bibfnamefont {E.}~\bibnamefont {Pelofske}}, \bibinfo {author} {\bibfnamefont {A.}~\bibnamefont {B{\"a}rtschi}},\ and\ \bibinfo {author} {\bibfnamefont {S.}~\bibnamefont {Eidenbenz}},\ }\bibfield  {title} {\bibinfo {title} {Quantum annealing vs. qaoa: 127 qubit higher-order ising problems on nisq computers},\ }in\ \href {https://link.springer.com/chapter/10.1007/978-3-031-32041-5_13} {\emph {\bibinfo {booktitle} {International Conference on High Performance Computing}}}\ (\bibinfo {organization} {Springer},\ \bibinfo {year} {2023})\ pp.\ \bibinfo {pages} {240--258}\BibitemShut {NoStop}%
\bibitem [{\citenamefont {Bottarelli}\ \emph {et~al.}(2024)\citenamefont {Bottarelli}, \citenamefont {Schmitt},\ and\ \citenamefont {Hauke}}]{bottarelli2024inequality}%
  \BibitemOpen
  \bibfield  {author} {\bibinfo {author} {\bibfnamefont {A.}~\bibnamefont {Bottarelli}}, \bibinfo {author} {\bibfnamefont {S.}~\bibnamefont {Schmitt}},\ and\ \bibinfo {author} {\bibfnamefont {P.}~\bibnamefont {Hauke}},\ }\bibfield  {title} {\bibinfo {title} {Inequality constraints in variational quantum circuits with qudits},\ }\href@noop {} {\bibfield  {journal} {\bibinfo  {journal} {arXiv preprint arXiv:2410.07674}\ } (\bibinfo {year} {2024})}\BibitemShut {NoStop}%
\bibitem [{\citenamefont {Shaydulin}\ \emph {et~al.}(2024)\citenamefont {Shaydulin}, \citenamefont {Li}, \citenamefont {Chakrabarti}, \citenamefont {DeCross}, \citenamefont {Herman}, \citenamefont {Kumar}, \citenamefont {Larson}, \citenamefont {Lykov}, \citenamefont {Minssen}, \citenamefont {Sun} \emph {et~al.}}]{shaydulin2024evidence}%
  \BibitemOpen
  \bibfield  {author} {\bibinfo {author} {\bibfnamefont {R.}~\bibnamefont {Shaydulin}}, \bibinfo {author} {\bibfnamefont {C.}~\bibnamefont {Li}}, \bibinfo {author} {\bibfnamefont {S.}~\bibnamefont {Chakrabarti}}, \bibinfo {author} {\bibfnamefont {M.}~\bibnamefont {DeCross}}, \bibinfo {author} {\bibfnamefont {D.}~\bibnamefont {Herman}}, \bibinfo {author} {\bibfnamefont {N.}~\bibnamefont {Kumar}}, \bibinfo {author} {\bibfnamefont {J.}~\bibnamefont {Larson}}, \bibinfo {author} {\bibfnamefont {D.}~\bibnamefont {Lykov}}, \bibinfo {author} {\bibfnamefont {P.}~\bibnamefont {Minssen}}, \bibinfo {author} {\bibfnamefont {Y.}~\bibnamefont {Sun}}, \emph {et~al.},\ }\bibfield  {title} {\bibinfo {title} {Evidence of scaling advantage for the quantum approximate optimization algorithm on a classically intractable problem},\ }\href {https://www.science.org/doi/10.1126/sciadv.adm6761} {\bibfield  {journal} {\bibinfo  {journal} {Science Advances}\ }\textbf {\bibinfo {volume} {10}},\ \bibinfo {pages} {eadm6761} (\bibinfo {year}
  {2024})}\BibitemShut {NoStop}%
\bibitem [{\citenamefont {Zhou}\ \emph {et~al.}(2020)\citenamefont {Zhou}, \citenamefont {Wang}, \citenamefont {Choi}, \citenamefont {Pichler},\ and\ \citenamefont {Lukin}}]{zhou2020quantum}%
  \BibitemOpen
  \bibfield  {author} {\bibinfo {author} {\bibfnamefont {L.}~\bibnamefont {Zhou}}, \bibinfo {author} {\bibfnamefont {S.-T.}\ \bibnamefont {Wang}}, \bibinfo {author} {\bibfnamefont {S.}~\bibnamefont {Choi}}, \bibinfo {author} {\bibfnamefont {H.}~\bibnamefont {Pichler}},\ and\ \bibinfo {author} {\bibfnamefont {M.~D.}\ \bibnamefont {Lukin}},\ }\bibfield  {title} {\bibinfo {title} {Quantum approximate optimization algorithm: Performance, mechanism, and implementation on near-term devices},\ }\href {https://doi.org/10.1103/PhysRevX.10.021067} {\bibfield  {journal} {\bibinfo  {journal} {Phys. Rev. X}\ }\textbf {\bibinfo {volume} {10}},\ \bibinfo {pages} {021067} (\bibinfo {year} {2020})}\BibitemShut {NoStop}%
\bibitem [{\citenamefont {Farhi}\ and\ \citenamefont {Harrow}(2016)}]{farhi2016quantum}%
  \BibitemOpen
  \bibfield  {author} {\bibinfo {author} {\bibfnamefont {E.}~\bibnamefont {Farhi}}\ and\ \bibinfo {author} {\bibfnamefont {A.~W.}\ \bibnamefont {Harrow}},\ }\bibfield  {title} {\bibinfo {title} {Quantum supremacy through the quantum approximate optimization algorithm},\ }\href {https://arxiv.org/abs/1602.07674} {\bibfield  {journal} {\bibinfo  {journal} {arXiv:1602.07674}\ } (\bibinfo {year} {2016})}\BibitemShut {NoStop}%
\bibitem [{\citenamefont {Boulebnane}\ \emph {et~al.}(2025)\citenamefont {Boulebnane}, \citenamefont {Khan}, \citenamefont {Liu}, \citenamefont {Larson}, \citenamefont {Herman}, \citenamefont {Shaydulin},\ and\ \citenamefont {Pistoia}}]{boulebnane2025evidence}%
  \BibitemOpen
  \bibfield  {author} {\bibinfo {author} {\bibfnamefont {S.}~\bibnamefont {Boulebnane}}, \bibinfo {author} {\bibfnamefont {A.}~\bibnamefont {Khan}}, \bibinfo {author} {\bibfnamefont {M.}~\bibnamefont {Liu}}, \bibinfo {author} {\bibfnamefont {J.}~\bibnamefont {Larson}}, \bibinfo {author} {\bibfnamefont {D.}~\bibnamefont {Herman}}, \bibinfo {author} {\bibfnamefont {R.}~\bibnamefont {Shaydulin}},\ and\ \bibinfo {author} {\bibfnamefont {M.}~\bibnamefont {Pistoia}},\ }\bibfield  {title} {\bibinfo {title} {Evidence that the quantum approximate optimization algorithm optimizes the sherrington-kirkpatrick model efficiently in the average case},\ }\href {https://arxiv.org/abs/2505.07929} {\bibfield  {journal} {\bibinfo  {journal} {arXiv:2505.07929}\ } (\bibinfo {year} {2025})}\BibitemShut {NoStop}%
\bibitem [{\citenamefont {Omanakuttan}\ \emph {et~al.}(2025)\citenamefont {Omanakuttan}, \citenamefont {He}, \citenamefont {Zhang}, \citenamefont {Hao}, \citenamefont {Babakhani}, \citenamefont {Boulebnane}, \citenamefont {Chakrabarti}, \citenamefont {Herman}, \citenamefont {Sullivan}, \citenamefont {Perlin} \emph {et~al.}}]{omanakuttan2025threshold}%
  \BibitemOpen
  \bibfield  {author} {\bibinfo {author} {\bibfnamefont {S.}~\bibnamefont {Omanakuttan}}, \bibinfo {author} {\bibfnamefont {Z.}~\bibnamefont {He}}, \bibinfo {author} {\bibfnamefont {Z.}~\bibnamefont {Zhang}}, \bibinfo {author} {\bibfnamefont {T.}~\bibnamefont {Hao}}, \bibinfo {author} {\bibfnamefont {A.}~\bibnamefont {Babakhani}}, \bibinfo {author} {\bibfnamefont {S.}~\bibnamefont {Boulebnane}}, \bibinfo {author} {\bibfnamefont {S.}~\bibnamefont {Chakrabarti}}, \bibinfo {author} {\bibfnamefont {D.}~\bibnamefont {Herman}}, \bibinfo {author} {\bibfnamefont {J.}~\bibnamefont {Sullivan}}, \bibinfo {author} {\bibfnamefont {M.~A.}\ \bibnamefont {Perlin}}, \emph {et~al.},\ }\bibfield  {title} {\bibinfo {title} {Threshold for fault-tolerant quantum advantage with the quantum approximate optimization algorithm},\ }\href {https://arxiv.org/abs/2504.01897} {\bibfield  {journal} {\bibinfo  {journal} {arXiv:2504.01897}\ } (\bibinfo {year} {2025})}\BibitemShut {NoStop}%
\bibitem [{\citenamefont {Brandhofer}\ \emph {et~al.}(2022)\citenamefont {Brandhofer}, \citenamefont {Braun}, \citenamefont {Dehn}, \citenamefont {Hellstern}, \citenamefont {H{\"u}ls}, \citenamefont {Ji}, \citenamefont {Polian}, \citenamefont {Bhatia},\ and\ \citenamefont {Wellens}}]{brandhofer2022benchmarking}%
  \BibitemOpen
  \bibfield  {author} {\bibinfo {author} {\bibfnamefont {S.}~\bibnamefont {Brandhofer}}, \bibinfo {author} {\bibfnamefont {D.}~\bibnamefont {Braun}}, \bibinfo {author} {\bibfnamefont {V.}~\bibnamefont {Dehn}}, \bibinfo {author} {\bibfnamefont {G.}~\bibnamefont {Hellstern}}, \bibinfo {author} {\bibfnamefont {M.}~\bibnamefont {H{\"u}ls}}, \bibinfo {author} {\bibfnamefont {Y.}~\bibnamefont {Ji}}, \bibinfo {author} {\bibfnamefont {I.}~\bibnamefont {Polian}}, \bibinfo {author} {\bibfnamefont {A.~S.}\ \bibnamefont {Bhatia}},\ and\ \bibinfo {author} {\bibfnamefont {T.}~\bibnamefont {Wellens}},\ }\bibfield  {title} {\bibinfo {title} {Benchmarking the performance of portfolio optimization with qaoa},\ }\href {https://link.springer.com/article/10.1007/s11128-022-03766-5} {\bibfield  {journal} {\bibinfo  {journal} {Quantum Information Processing}\ }\textbf {\bibinfo {volume} {22}},\ \bibinfo {pages} {25} (\bibinfo {year} {2022})}\BibitemShut {NoStop}%
\bibitem [{\citenamefont {Guerreschi}\ and\ \citenamefont {Matsuura}(2019)}]{guerreschi2019qaoa}%
  \BibitemOpen
  \bibfield  {author} {\bibinfo {author} {\bibfnamefont {G.~G.}\ \bibnamefont {Guerreschi}}\ and\ \bibinfo {author} {\bibfnamefont {A.~Y.}\ \bibnamefont {Matsuura}},\ }\bibfield  {title} {\bibinfo {title} {Qaoa for max-cut requires hundreds of qubits for quantum speed-up},\ }\href {https://www.nature.com/articles/s41598-019-43176-9} {\bibfield  {journal} {\bibinfo  {journal} {Scientific reports}\ }\textbf {\bibinfo {volume} {9}},\ \bibinfo {pages} {6903} (\bibinfo {year} {2019})}\BibitemShut {NoStop}%
\bibitem [{\citenamefont {Deller}\ \emph {et~al.}(2023)\citenamefont {Deller}, \citenamefont {Schmitt}, \citenamefont {Lewenstein}, \citenamefont {Lenk}, \citenamefont {Federer}, \citenamefont {Jendrzejewski}, \citenamefont {Hauke},\ and\ \citenamefont {Kasper}}]{deller2023quantum}%
  \BibitemOpen
  \bibfield  {author} {\bibinfo {author} {\bibfnamefont {Y.}~\bibnamefont {Deller}}, \bibinfo {author} {\bibfnamefont {S.}~\bibnamefont {Schmitt}}, \bibinfo {author} {\bibfnamefont {M.}~\bibnamefont {Lewenstein}}, \bibinfo {author} {\bibfnamefont {S.}~\bibnamefont {Lenk}}, \bibinfo {author} {\bibfnamefont {M.}~\bibnamefont {Federer}}, \bibinfo {author} {\bibfnamefont {F.}~\bibnamefont {Jendrzejewski}}, \bibinfo {author} {\bibfnamefont {P.}~\bibnamefont {Hauke}},\ and\ \bibinfo {author} {\bibfnamefont {V.}~\bibnamefont {Kasper}},\ }\bibfield  {title} {\bibinfo {title} {Quantum approximate optimization algorithm for qudit systems},\ }\href {https://doi.org/10.1103/PhysRevA.107.062410} {\bibfield  {journal} {\bibinfo  {journal} {Phys. Rev. A}\ }\textbf {\bibinfo {volume} {107}},\ \bibinfo {pages} {062410} (\bibinfo {year} {2023})}\BibitemShut {NoStop}%
\bibitem [{\citenamefont {Ekstrom}\ \emph {et~al.}(2025)\citenamefont {Ekstrom}, \citenamefont {Wang},\ and\ \citenamefont {Schmitt}}]{ekstrom2025variational}%
  \BibitemOpen
  \bibfield  {author} {\bibinfo {author} {\bibfnamefont {L.}~\bibnamefont {Ekstrom}}, \bibinfo {author} {\bibfnamefont {H.}~\bibnamefont {Wang}},\ and\ \bibinfo {author} {\bibfnamefont {S.}~\bibnamefont {Schmitt}},\ }\bibfield  {title} {\bibinfo {title} {Variational quantum multiobjective optimization},\ }\href {https://doi.org/10.1103/PhysRevResearch.7.023141} {\bibfield  {journal} {\bibinfo  {journal} {Phys. Rev. Res.}\ }\textbf {\bibinfo {volume} {7}},\ \bibinfo {pages} {023141} (\bibinfo {year} {2025})}\BibitemShut {NoStop}%
\bibitem [{\citenamefont {Bravyi}\ \emph {et~al.}(2022)\citenamefont {Bravyi}, \citenamefont {Kliesch}, \citenamefont {Koenig},\ and\ \citenamefont {Tang}}]{bravyi2022}%
  \BibitemOpen
  \bibfield  {author} {\bibinfo {author} {\bibfnamefont {S.}~\bibnamefont {Bravyi}}, \bibinfo {author} {\bibfnamefont {A.}~\bibnamefont {Kliesch}}, \bibinfo {author} {\bibfnamefont {R.}~\bibnamefont {Koenig}},\ and\ \bibinfo {author} {\bibfnamefont {E.}~\bibnamefont {Tang}},\ }\bibfield  {title} {\bibinfo {title} {Hybrid quantum-classical algorithms for approximate graph coloring},\ }\href {https://quantum-journal.org/papers/q-2022-03-30-678/} {\bibfield  {journal} {\bibinfo  {journal} {Quantum}\ }\textbf {\bibinfo {volume} {6}},\ \bibinfo {pages} {678} (\bibinfo {year} {2022})}\BibitemShut {NoStop}%
\bibitem [{\citenamefont {Chitambar}\ and\ \citenamefont {Gour}(2019)}]{chitambar2019quantum}%
  \BibitemOpen
  \bibfield  {author} {\bibinfo {author} {\bibfnamefont {E.}~\bibnamefont {Chitambar}}\ and\ \bibinfo {author} {\bibfnamefont {G.}~\bibnamefont {Gour}},\ }\bibfield  {title} {\bibinfo {title} {Quantum resource theories},\ }\href {https://doi.org/10.1103/RevModPhys.91.025001} {\bibfield  {journal} {\bibinfo  {journal} {Rev. Mod. Phys.}\ }\textbf {\bibinfo {volume} {91}},\ \bibinfo {pages} {025001} (\bibinfo {year} {2019})}\BibitemShut {NoStop}%
\bibitem [{\citenamefont {Amico}\ \emph {et~al.}(2008)\citenamefont {Amico}, \citenamefont {Fazio}, \citenamefont {Osterloh},\ and\ \citenamefont {Vedral}}]{RevModPhys.80.517}%
  \BibitemOpen
  \bibfield  {author} {\bibinfo {author} {\bibfnamefont {L.}~\bibnamefont {Amico}}, \bibinfo {author} {\bibfnamefont {R.}~\bibnamefont {Fazio}}, \bibinfo {author} {\bibfnamefont {A.}~\bibnamefont {Osterloh}},\ and\ \bibinfo {author} {\bibfnamefont {V.}~\bibnamefont {Vedral}},\ }\bibfield  {title} {\bibinfo {title} {Entanglement in many-body systems},\ }\href {https://doi.org/10.1103/RevModPhys.80.517} {\bibfield  {journal} {\bibinfo  {journal} {Rev. Mod. Phys.}\ }\textbf {\bibinfo {volume} {80}},\ \bibinfo {pages} {517} (\bibinfo {year} {2008})}\BibitemShut {NoStop}%
\bibitem [{\citenamefont {Eisert}\ \emph {et~al.}(2010)\citenamefont {Eisert}, \citenamefont {Cramer},\ and\ \citenamefont {Plenio}}]{RevModPhys.82.277}%
  \BibitemOpen
  \bibfield  {author} {\bibinfo {author} {\bibfnamefont {J.}~\bibnamefont {Eisert}}, \bibinfo {author} {\bibfnamefont {M.}~\bibnamefont {Cramer}},\ and\ \bibinfo {author} {\bibfnamefont {M.~B.}\ \bibnamefont {Plenio}},\ }\bibfield  {title} {\bibinfo {title} {Colloquium: Area laws for the entanglement entropy},\ }\href {https://doi.org/10.1103/RevModPhys.82.277} {\bibfield  {journal} {\bibinfo  {journal} {Rev. Mod. Phys.}\ }\textbf {\bibinfo {volume} {82}},\ \bibinfo {pages} {277} (\bibinfo {year} {2010})}\BibitemShut {NoStop}%
\bibitem [{\citenamefont {Cirac}\ and\ \citenamefont {Zoller}(2012)}]{cirac2012goals}%
  \BibitemOpen
  \bibfield  {author} {\bibinfo {author} {\bibfnamefont {J.~I.}\ \bibnamefont {Cirac}}\ and\ \bibinfo {author} {\bibfnamefont {P.}~\bibnamefont {Zoller}},\ }\bibfield  {title} {\bibinfo {title} {Goals and opportunities in quantum simulation},\ }\href {https://www.nature.com/articles/nphys2275} {\bibfield  {journal} {\bibinfo  {journal} {Nature physics}\ }\textbf {\bibinfo {volume} {8}},\ \bibinfo {pages} {264} (\bibinfo {year} {2012})}\BibitemShut {NoStop}%
\bibitem [{\citenamefont {Lanting}\ \emph {et~al.}(2014)\citenamefont {Lanting}, \citenamefont {Przybysz}, \citenamefont {Smirnov}, \citenamefont {Spedalieri}, \citenamefont {Amin}, \citenamefont {Berkley}, \citenamefont {Harris}, \citenamefont {Altomare}, \citenamefont {Boixo}, \citenamefont {Bunyk}, \citenamefont {Dickson}, \citenamefont {Enderud}, \citenamefont {Hilton}, \citenamefont {Hoskinson}, \citenamefont {Johnson}, \citenamefont {Ladizinsky}, \citenamefont {Ladizinsky}, \citenamefont {Neufeld}, \citenamefont {Oh}, \citenamefont {Perminov}, \citenamefont {Rich}, \citenamefont {Thom}, \citenamefont {Tolkacheva}, \citenamefont {Uchaikin}, \citenamefont {Wilson},\ and\ \citenamefont {Rose}}]{lanting2014entanglement}%
  \BibitemOpen
  \bibfield  {author} {\bibinfo {author} {\bibfnamefont {T.}~\bibnamefont {Lanting}}, \bibinfo {author} {\bibfnamefont {A.~J.}\ \bibnamefont {Przybysz}}, \bibinfo {author} {\bibfnamefont {A.~Y.}\ \bibnamefont {Smirnov}}, \bibinfo {author} {\bibfnamefont {F.~M.}\ \bibnamefont {Spedalieri}}, \bibinfo {author} {\bibfnamefont {M.~H.}\ \bibnamefont {Amin}}, \bibinfo {author} {\bibfnamefont {A.~J.}\ \bibnamefont {Berkley}}, \bibinfo {author} {\bibfnamefont {R.}~\bibnamefont {Harris}}, \bibinfo {author} {\bibfnamefont {F.}~\bibnamefont {Altomare}}, \bibinfo {author} {\bibfnamefont {S.}~\bibnamefont {Boixo}}, \bibinfo {author} {\bibfnamefont {P.}~\bibnamefont {Bunyk}}, \bibinfo {author} {\bibfnamefont {N.}~\bibnamefont {Dickson}}, \bibinfo {author} {\bibfnamefont {C.}~\bibnamefont {Enderud}}, \bibinfo {author} {\bibfnamefont {J.~P.}\ \bibnamefont {Hilton}}, \bibinfo {author} {\bibfnamefont {E.}~\bibnamefont {Hoskinson}}, \bibinfo {author} {\bibfnamefont {M.~W.}\ \bibnamefont {Johnson}}, \bibinfo {author}
  {\bibfnamefont {E.}~\bibnamefont {Ladizinsky}}, \bibinfo {author} {\bibfnamefont {N.}~\bibnamefont {Ladizinsky}}, \bibinfo {author} {\bibfnamefont {R.}~\bibnamefont {Neufeld}}, \bibinfo {author} {\bibfnamefont {T.}~\bibnamefont {Oh}}, \bibinfo {author} {\bibfnamefont {I.}~\bibnamefont {Perminov}}, \bibinfo {author} {\bibfnamefont {C.}~\bibnamefont {Rich}}, \bibinfo {author} {\bibfnamefont {M.~C.}\ \bibnamefont {Thom}}, \bibinfo {author} {\bibfnamefont {E.}~\bibnamefont {Tolkacheva}}, \bibinfo {author} {\bibfnamefont {S.}~\bibnamefont {Uchaikin}}, \bibinfo {author} {\bibfnamefont {A.~B.}\ \bibnamefont {Wilson}},\ and\ \bibinfo {author} {\bibfnamefont {G.}~\bibnamefont {Rose}},\ }\bibfield  {title} {\bibinfo {title} {Entanglement in a quantum annealing processor},\ }\href {https://doi.org/10.1103/PhysRevX.4.021041} {\bibfield  {journal} {\bibinfo  {journal} {Phys. Rev. X}\ }\textbf {\bibinfo {volume} {4}},\ \bibinfo {pages} {021041} (\bibinfo {year} {2014})}\BibitemShut {NoStop}%
\bibitem [{\citenamefont {Hauke}\ \emph {et~al.}(2015)\citenamefont {Hauke}, \citenamefont {Bonnes}, \citenamefont {Heyl},\ and\ \citenamefont {Lechner}}]{hauke2015probing}%
  \BibitemOpen
  \bibfield  {author} {\bibinfo {author} {\bibfnamefont {P.}~\bibnamefont {Hauke}}, \bibinfo {author} {\bibfnamefont {L.}~\bibnamefont {Bonnes}}, \bibinfo {author} {\bibfnamefont {M.}~\bibnamefont {Heyl}},\ and\ \bibinfo {author} {\bibfnamefont {W.}~\bibnamefont {Lechner}},\ }\bibfield  {title} {\bibinfo {title} {Probing entanglement in adiabatic quantum optimization with trapped ions},\ }\href {https://www.frontiersin.org/journals/physics/articles/10.3389/fphy.2015.00021/full} {\bibfield  {journal} {\bibinfo  {journal} {Frontiers in Physics}\ }\textbf {\bibinfo {volume} {3}},\ \bibinfo {pages} {21} (\bibinfo {year} {2015})}\BibitemShut {NoStop}%
\bibitem [{\citenamefont {D\'{\i}ez-Valle}\ \emph {et~al.}(2021)\citenamefont {D\'{\i}ez-Valle}, \citenamefont {Porras},\ and\ \citenamefont {Garc\'{\i}a-Ripoll}}]{valle2021quantum}%
  \BibitemOpen
  \bibfield  {author} {\bibinfo {author} {\bibfnamefont {P.}~\bibnamefont {D\'{\i}ez-Valle}}, \bibinfo {author} {\bibfnamefont {D.}~\bibnamefont {Porras}},\ and\ \bibinfo {author} {\bibfnamefont {J.~J.}\ \bibnamefont {Garc\'{\i}a-Ripoll}},\ }\bibfield  {title} {\bibinfo {title} {Quantum variational optimization: The role of entanglement and problem hardness},\ }\href {https://doi.org/10.1103/PhysRevA.104.062426} {\bibfield  {journal} {\bibinfo  {journal} {Phys. Rev. A}\ }\textbf {\bibinfo {volume} {104}},\ \bibinfo {pages} {062426} (\bibinfo {year} {2021})}\BibitemShut {NoStop}%
\bibitem [{\citenamefont {Dupont}\ \emph {et~al.}(2022{\natexlab{a}})\citenamefont {Dupont}, \citenamefont {Didier}, \citenamefont {Hodson}, \citenamefont {Moore},\ and\ \citenamefont {Reagor}}]{dupont2022calibrating}%
  \BibitemOpen
  \bibfield  {author} {\bibinfo {author} {\bibfnamefont {M.}~\bibnamefont {Dupont}}, \bibinfo {author} {\bibfnamefont {N.}~\bibnamefont {Didier}}, \bibinfo {author} {\bibfnamefont {M.~J.}\ \bibnamefont {Hodson}}, \bibinfo {author} {\bibfnamefont {J.~E.}\ \bibnamefont {Moore}},\ and\ \bibinfo {author} {\bibfnamefont {M.~J.}\ \bibnamefont {Reagor}},\ }\bibfield  {title} {\bibinfo {title} {Calibrating the classical hardness of the quantum approximate optimization algorithm},\ }\href {https://doi.org/10.1103/PRXQuantum.3.040339} {\bibfield  {journal} {\bibinfo  {journal} {PRX Quantum}\ }\textbf {\bibinfo {volume} {3}},\ \bibinfo {pages} {040339} (\bibinfo {year} {2022}{\natexlab{a}})}\BibitemShut {NoStop}%
\bibitem [{\citenamefont {Sreedhar}\ \emph {et~al.}(2022)\citenamefont {Sreedhar}, \citenamefont {Vikst{\aa}l}, \citenamefont {Svensson}, \citenamefont {Ask}, \citenamefont {Johansson},\ and\ \citenamefont {Garc{\'\i}a-{\'A}lvarez}}]{sreedhar2022quantum}%
  \BibitemOpen
  \bibfield  {author} {\bibinfo {author} {\bibfnamefont {R.}~\bibnamefont {Sreedhar}}, \bibinfo {author} {\bibfnamefont {P.}~\bibnamefont {Vikst{\aa}l}}, \bibinfo {author} {\bibfnamefont {M.}~\bibnamefont {Svensson}}, \bibinfo {author} {\bibfnamefont {A.}~\bibnamefont {Ask}}, \bibinfo {author} {\bibfnamefont {G.}~\bibnamefont {Johansson}},\ and\ \bibinfo {author} {\bibfnamefont {L.}~\bibnamefont {Garc{\'\i}a-{\'A}lvarez}},\ }\bibfield  {title} {\bibinfo {title} {The quantum approximate optimization algorithm performance with low entanglement and high circuit depth},\ }\href {https://arxiv.org/abs/2207.03404} {\bibfield  {journal} {\bibinfo  {journal} {arXiv:2207.03404}\ } (\bibinfo {year} {2022})}\BibitemShut {NoStop}%
\bibitem [{\citenamefont {Chen}\ \emph {et~al.}(2022)\citenamefont {Chen}, \citenamefont {Zhu}, \citenamefont {Mayhall}, \citenamefont {Barnes},\ and\ \citenamefont {Economou}}]{chen2022much}%
  \BibitemOpen
  \bibfield  {author} {\bibinfo {author} {\bibfnamefont {Y.}~\bibnamefont {Chen}}, \bibinfo {author} {\bibfnamefont {L.}~\bibnamefont {Zhu}}, \bibinfo {author} {\bibfnamefont {N.~J.}\ \bibnamefont {Mayhall}}, \bibinfo {author} {\bibfnamefont {E.}~\bibnamefont {Barnes}},\ and\ \bibinfo {author} {\bibfnamefont {S.~E.}\ \bibnamefont {Economou}},\ }\bibfield  {title} {\bibinfo {title} {How much entanglement do quantum optimization algorithms require?},\ }\href {https://doi.org/10.1364/QUANTUM.2022.QM4A.2} {\bibfield  {journal} {\bibinfo  {journal} {Quantum 2.0 Conference and Exhibition}\ ,\ \bibinfo {pages} {QM4A.2}} (\bibinfo {year} {2022})}\BibitemShut {NoStop}%
\bibitem [{\citenamefont {Hauke}\ \emph {et~al.}(2016)\citenamefont {Hauke}, \citenamefont {Heyl}, \citenamefont {Tagliacozzo},\ and\ \citenamefont {Zoller}}]{hauke2016measuring}%
  \BibitemOpen
  \bibfield  {author} {\bibinfo {author} {\bibfnamefont {P.}~\bibnamefont {Hauke}}, \bibinfo {author} {\bibfnamefont {M.}~\bibnamefont {Heyl}}, \bibinfo {author} {\bibfnamefont {L.}~\bibnamefont {Tagliacozzo}},\ and\ \bibinfo {author} {\bibfnamefont {P.}~\bibnamefont {Zoller}},\ }\bibfield  {title} {\bibinfo {title} {Measuring multipartite entanglement through dynamic susceptibilities},\ }\href {https://doi.org/10.1038/nphys3700} {\bibfield  {journal} {\bibinfo  {journal} {Nat. Phys.}\ }\textbf {\bibinfo {volume} {12}},\ \bibinfo {pages} {778} (\bibinfo {year} {2016})}\BibitemShut {NoStop}%
\bibitem [{\citenamefont {Santra}\ \emph {et~al.}(2024)\citenamefont {Santra}, \citenamefont {Jendrzejewski}, \citenamefont {Hauke},\ and\ \citenamefont {Egger}}]{santra2024squeezing}%
  \BibitemOpen
  \bibfield  {author} {\bibinfo {author} {\bibfnamefont {G.~C.}\ \bibnamefont {Santra}}, \bibinfo {author} {\bibfnamefont {F.}~\bibnamefont {Jendrzejewski}}, \bibinfo {author} {\bibfnamefont {P.}~\bibnamefont {Hauke}},\ and\ \bibinfo {author} {\bibfnamefont {D.~J.}\ \bibnamefont {Egger}},\ }\bibfield  {title} {\bibinfo {title} {Squeezing and quantum approximate optimization},\ }\href {https://doi.org/10.1103/PhysRevA.109.012413} {\bibfield  {journal} {\bibinfo  {journal} {Phys. Rev. A}\ }\textbf {\bibinfo {volume} {109}},\ \bibinfo {pages} {012413} (\bibinfo {year} {2024})}\BibitemShut {NoStop}%
\bibitem [{\citenamefont {Vitale}\ \emph {et~al.}(2024)\citenamefont {Vitale}, \citenamefont {Rath}, \citenamefont {Jurcevic}, \citenamefont {Elben}, \citenamefont {Branciard},\ and\ \citenamefont {Vermersch}}]{vitale2024robust}%
  \BibitemOpen
  \bibfield  {author} {\bibinfo {author} {\bibfnamefont {V.}~\bibnamefont {Vitale}}, \bibinfo {author} {\bibfnamefont {A.}~\bibnamefont {Rath}}, \bibinfo {author} {\bibfnamefont {P.}~\bibnamefont {Jurcevic}}, \bibinfo {author} {\bibfnamefont {A.}~\bibnamefont {Elben}}, \bibinfo {author} {\bibfnamefont {C.}~\bibnamefont {Branciard}},\ and\ \bibinfo {author} {\bibfnamefont {B.}~\bibnamefont {Vermersch}},\ }\bibfield  {title} {\bibinfo {title} {Robust estimation of the quantum fisher information on a quantum processor},\ }\href {https://doi.org/10.1103/PRXQuantum.5.030338} {\bibfield  {journal} {\bibinfo  {journal} {PRX Quantum}\ }\textbf {\bibinfo {volume} {5}},\ \bibinfo {pages} {030338} (\bibinfo {year} {2024})}\BibitemShut {NoStop}%
\bibitem [{\citenamefont {Santra}\ \emph {et~al.}(2025{\natexlab{a}})\citenamefont {Santra}, \citenamefont {Roy}, \citenamefont {Egger},\ and\ \citenamefont {Hauke}}]{santra2025genuine}%
  \BibitemOpen
  \bibfield  {author} {\bibinfo {author} {\bibfnamefont {G.~C.}\ \bibnamefont {Santra}}, \bibinfo {author} {\bibfnamefont {S.~S.}\ \bibnamefont {Roy}}, \bibinfo {author} {\bibfnamefont {D.~J.}\ \bibnamefont {Egger}},\ and\ \bibinfo {author} {\bibfnamefont {P.}~\bibnamefont {Hauke}},\ }\bibfield  {title} {\bibinfo {title} {Genuine multipartite entanglement in quantum optimization},\ }\href {https://doi.org/10.1103/PhysRevA.111.022434} {\bibfield  {journal} {\bibinfo  {journal} {Phys. Rev. A}\ }\textbf {\bibinfo {volume} {111}},\ \bibinfo {pages} {022434} (\bibinfo {year} {2025}{\natexlab{a}})}\BibitemShut {NoStop}%
\bibitem [{\citenamefont {Gottesman}(1997)}]{gottesman1997stabilizer}%
  \BibitemOpen
  \bibfield  {author} {\bibinfo {author} {\bibfnamefont {D.}~\bibnamefont {Gottesman}},\ }\bibfield  {title} {\bibinfo {title} {Stabilizer codes and quantum error correction},\ }\href {https://arxiv.org/abs/quant-ph/9705052} {\bibfield  {journal} {\bibinfo  {journal} {arXiv: quant-ph/9705052}\ } (\bibinfo {year} {1997})}\BibitemShut {NoStop}%
\bibitem [{\citenamefont {Gottesman}(1998{\natexlab{a}})}]{gottesman1998heisenberg}%
  \BibitemOpen
  \bibfield  {author} {\bibinfo {author} {\bibfnamefont {D.}~\bibnamefont {Gottesman}},\ }\bibfield  {title} {\bibinfo {title} {The heisenberg representation of quantum computers},\ }\href {https://arxiv.org/abs/quant-ph/9807006} {\bibfield  {journal} {\bibinfo  {journal} {arXiv quant-ph/9807006}\ } (\bibinfo {year} {1998}{\natexlab{a}})}\BibitemShut {NoStop}%
\bibitem [{\citenamefont {Gottesman}(1998{\natexlab{b}})}]{gottesman1998theory}%
  \BibitemOpen
  \bibfield  {author} {\bibinfo {author} {\bibfnamefont {D.}~\bibnamefont {Gottesman}},\ }\bibfield  {title} {\bibinfo {title} {Theory of fault-tolerant quantum computation},\ }\href {https://doi.org/10.1103/PhysRevA.57.127} {\bibfield  {journal} {\bibinfo  {journal} {Phys. Rev. A}\ }\textbf {\bibinfo {volume} {57}},\ \bibinfo {pages} {127} (\bibinfo {year} {1998}{\natexlab{b}})}\BibitemShut {NoStop}%
\bibitem [{\citenamefont {Bravyi}\ and\ \citenamefont {Kitaev}(2005)}]{PhysRevA.71.022316}%
  \BibitemOpen
  \bibfield  {author} {\bibinfo {author} {\bibfnamefont {S.}~\bibnamefont {Bravyi}}\ and\ \bibinfo {author} {\bibfnamefont {A.}~\bibnamefont {Kitaev}},\ }\bibfield  {title} {\bibinfo {title} {Universal quantum computation with ideal clifford gates and noisy ancillas},\ }\href {https://doi.org/10.1103/PhysRevA.71.022316} {\bibfield  {journal} {\bibinfo  {journal} {Phys. Rev. A}\ }\textbf {\bibinfo {volume} {71}},\ \bibinfo {pages} {022316} (\bibinfo {year} {2005})}\BibitemShut {NoStop}%
\bibitem [{\citenamefont {Bravyi}\ and\ \citenamefont {Haah}(2012)}]{PhysRevA.86.052329}%
  \BibitemOpen
  \bibfield  {author} {\bibinfo {author} {\bibfnamefont {S.}~\bibnamefont {Bravyi}}\ and\ \bibinfo {author} {\bibfnamefont {J.}~\bibnamefont {Haah}},\ }\bibfield  {title} {\bibinfo {title} {Magic-state distillation with low overhead},\ }\href {https://doi.org/10.1103/PhysRevA.86.052329} {\bibfield  {journal} {\bibinfo  {journal} {Phys. Rev. A}\ }\textbf {\bibinfo {volume} {86}},\ \bibinfo {pages} {052329} (\bibinfo {year} {2012})}\BibitemShut {NoStop}%
\bibitem [{\citenamefont {Campbell}\ \emph {et~al.}(2017)\citenamefont {Campbell}, \citenamefont {Terhal},\ and\ \citenamefont {Vuillot}}]{campbell2017roads}%
  \BibitemOpen
  \bibfield  {author} {\bibinfo {author} {\bibfnamefont {E.~T.}\ \bibnamefont {Campbell}}, \bibinfo {author} {\bibfnamefont {B.~M.}\ \bibnamefont {Terhal}},\ and\ \bibinfo {author} {\bibfnamefont {C.}~\bibnamefont {Vuillot}},\ }\bibfield  {title} {\bibinfo {title} {Roads towards fault-tolerant universal quantum computation},\ }\href {https://www.nature.com/articles/nature23460} {\bibfield  {journal} {\bibinfo  {journal} {Nature}\ }\textbf {\bibinfo {volume} {549}},\ \bibinfo {pages} {172} (\bibinfo {year} {2017})}\BibitemShut {NoStop}%
\bibitem [{\citenamefont {Harrow}\ and\ \citenamefont {Montanaro}(2017)}]{harrow2017quantum}%
  \BibitemOpen
  \bibfield  {author} {\bibinfo {author} {\bibfnamefont {A.~W.}\ \bibnamefont {Harrow}}\ and\ \bibinfo {author} {\bibfnamefont {A.}~\bibnamefont {Montanaro}},\ }\bibfield  {title} {\bibinfo {title} {Quantum computational supremacy},\ }\href {https://www.nature.com/articles/nature23458} {\bibfield  {journal} {\bibinfo  {journal} {Nature}\ }\textbf {\bibinfo {volume} {549}},\ \bibinfo {pages} {203} (\bibinfo {year} {2017})}\BibitemShut {NoStop}%
\bibitem [{\citenamefont {Gross}(2006)}]{gross2006hudson}%
  \BibitemOpen
  \bibfield  {author} {\bibinfo {author} {\bibfnamefont {D.}~\bibnamefont {Gross}},\ }\bibfield  {title} {\bibinfo {title} {Hudson’s theorem for finite-dimensional quantum systems},\ }\href {https://pubs.aip.org/aip/jmp/article/47/12/122107/919571/Hudson-s-theorem-for-finite-dimensional-quantum} {\bibfield  {journal} {\bibinfo  {journal} {Journal of mathematical physics}\ }\textbf {\bibinfo {volume} {47}} (\bibinfo {year} {2006})}\BibitemShut {NoStop}%
\bibitem [{\citenamefont {Wootters}(1987{\natexlab{a}})}]{wootters1987wigner}%
  \BibitemOpen
  \bibfield  {author} {\bibinfo {author} {\bibfnamefont {W.~K.}\ \bibnamefont {Wootters}},\ }\bibfield  {title} {\bibinfo {title} {A wigner-function formulation of finite-state quantum mechanics},\ }\href {https://www.sciencedirect.com/science/article/abs/pii/000349168790176X} {\bibfield  {journal} {\bibinfo  {journal} {Annals of Physics}\ }\textbf {\bibinfo {volume} {176}},\ \bibinfo {pages} {1} (\bibinfo {year} {1987}{\natexlab{a}})}\BibitemShut {NoStop}%
\bibitem [{\citenamefont {Bu\ifmmode~\check{z}\else \v{z}\fi{}ek}\ \emph {et~al.}(1992)\citenamefont {Bu\ifmmode~\check{z}\else \v{z}\fi{}ek}, \citenamefont {Vidiella-Barranco},\ and\ \citenamefont {Knight}}]{PhysRevA.45.6570}%
  \BibitemOpen
  \bibfield  {author} {\bibinfo {author} {\bibfnamefont {V.}~\bibnamefont {Bu\ifmmode~\check{z}\else \v{z}\fi{}ek}}, \bibinfo {author} {\bibfnamefont {A.}~\bibnamefont {Vidiella-Barranco}},\ and\ \bibinfo {author} {\bibfnamefont {P.~L.}\ \bibnamefont {Knight}},\ }\bibfield  {title} {\bibinfo {title} {Superpositions of coherent states: Squeezing and dissipation},\ }\href {https://doi.org/10.1103/PhysRevA.45.6570} {\bibfield  {journal} {\bibinfo  {journal} {Phys. Rev. A}\ }\textbf {\bibinfo {volume} {45}},\ \bibinfo {pages} {6570} (\bibinfo {year} {1992})}\BibitemShut {NoStop}%
\bibitem [{\citenamefont {Veitch}\ \emph {et~al.}(2012)\citenamefont {Veitch}, \citenamefont {Ferrie}, \citenamefont {Gross},\ and\ \citenamefont {Emerson}}]{veitch2012negative}%
  \BibitemOpen
  \bibfield  {author} {\bibinfo {author} {\bibfnamefont {V.}~\bibnamefont {Veitch}}, \bibinfo {author} {\bibfnamefont {C.}~\bibnamefont {Ferrie}}, \bibinfo {author} {\bibfnamefont {D.}~\bibnamefont {Gross}},\ and\ \bibinfo {author} {\bibfnamefont {J.}~\bibnamefont {Emerson}},\ }\bibfield  {title} {\bibinfo {title} {Negative quasi-probability as a resource for quantum computation},\ }\href {https://iopscience.iop.org/article/10.1088/1367-2630/14/11/113011/meta} {\bibfield  {journal} {\bibinfo  {journal} {New Journal of Physics}\ }\textbf {\bibinfo {volume} {14}},\ \bibinfo {pages} {113011} (\bibinfo {year} {2012})}\BibitemShut {NoStop}%
\bibitem [{\citenamefont {Bravyi}\ and\ \citenamefont {Gosset}(2016)}]{PhysRevLett.116.250501}%
  \BibitemOpen
  \bibfield  {author} {\bibinfo {author} {\bibfnamefont {S.}~\bibnamefont {Bravyi}}\ and\ \bibinfo {author} {\bibfnamefont {D.}~\bibnamefont {Gosset}},\ }\bibfield  {title} {\bibinfo {title} {Improved classical simulation of quantum circuits dominated by clifford gates},\ }\href {https://doi.org/10.1103/PhysRevLett.116.250501} {\bibfield  {journal} {\bibinfo  {journal} {Phys. Rev. Lett.}\ }\textbf {\bibinfo {volume} {116}},\ \bibinfo {pages} {250501} (\bibinfo {year} {2016})}\BibitemShut {NoStop}%
\bibitem [{\citenamefont {Bravyi}\ \emph {et~al.}(2016)\citenamefont {Bravyi}, \citenamefont {Smith},\ and\ \citenamefont {Smolin}}]{bravyi2016trading}%
  \BibitemOpen
  \bibfield  {author} {\bibinfo {author} {\bibfnamefont {S.}~\bibnamefont {Bravyi}}, \bibinfo {author} {\bibfnamefont {G.}~\bibnamefont {Smith}},\ and\ \bibinfo {author} {\bibfnamefont {J.~A.}\ \bibnamefont {Smolin}},\ }\bibfield  {title} {\bibinfo {title} {Trading classical and quantum computational resources},\ }\href {https://doi.org/10.1103/PhysRevX.6.021043} {\bibfield  {journal} {\bibinfo  {journal} {Phys. Rev. X}\ }\textbf {\bibinfo {volume} {6}},\ \bibinfo {pages} {021043} (\bibinfo {year} {2016})}\BibitemShut {NoStop}%
\bibitem [{\citenamefont {Howard}\ and\ \citenamefont {Campbell}(2017)}]{PhysRevLett.118.090501}%
  \BibitemOpen
  \bibfield  {author} {\bibinfo {author} {\bibfnamefont {M.}~\bibnamefont {Howard}}\ and\ \bibinfo {author} {\bibfnamefont {E.}~\bibnamefont {Campbell}},\ }\bibfield  {title} {\bibinfo {title} {Application of a resource theory for magic states to fault-tolerant quantum computing},\ }\href {https://doi.org/10.1103/PhysRevLett.118.090501} {\bibfield  {journal} {\bibinfo  {journal} {Phys. Rev. Lett.}\ }\textbf {\bibinfo {volume} {118}},\ \bibinfo {pages} {090501} (\bibinfo {year} {2017})}\BibitemShut {NoStop}%
\bibitem [{\citenamefont {Heinrich}\ and\ \citenamefont {Gross}(2019)}]{heinrich2019robustness}%
  \BibitemOpen
  \bibfield  {author} {\bibinfo {author} {\bibfnamefont {M.}~\bibnamefont {Heinrich}}\ and\ \bibinfo {author} {\bibfnamefont {D.}~\bibnamefont {Gross}},\ }\bibfield  {title} {\bibinfo {title} {Robustness of magic and symmetries of the stabiliser polytope},\ }\href {https://quantum-journal.org/papers/q-2019-04-08-132/} {\bibfield  {journal} {\bibinfo  {journal} {Quantum}\ }\textbf {\bibinfo {volume} {3}},\ \bibinfo {pages} {132} (\bibinfo {year} {2019})}\BibitemShut {NoStop}%
\bibitem [{\citenamefont {Wang}\ \emph {et~al.}(2020)\citenamefont {Wang}, \citenamefont {Wilde},\ and\ \citenamefont {Su}}]{PhysRevLett.124.090505}%
  \BibitemOpen
  \bibfield  {author} {\bibinfo {author} {\bibfnamefont {X.}~\bibnamefont {Wang}}, \bibinfo {author} {\bibfnamefont {M.~M.}\ \bibnamefont {Wilde}},\ and\ \bibinfo {author} {\bibfnamefont {Y.}~\bibnamefont {Su}},\ }\bibfield  {title} {\bibinfo {title} {Efficiently computable bounds for magic state distillation},\ }\href {https://doi.org/10.1103/PhysRevLett.124.090505} {\bibfield  {journal} {\bibinfo  {journal} {Phys. Rev. Lett.}\ }\textbf {\bibinfo {volume} {124}},\ \bibinfo {pages} {090505} (\bibinfo {year} {2020})}\BibitemShut {NoStop}%
\bibitem [{\citenamefont {Heimendahl}\ \emph {et~al.}(2021)\citenamefont {Heimendahl}, \citenamefont {Montealegre-Mora}, \citenamefont {Vallentin},\ and\ \citenamefont {Gross}}]{heimendahl2021stabilizer}%
  \BibitemOpen
  \bibfield  {author} {\bibinfo {author} {\bibfnamefont {A.}~\bibnamefont {Heimendahl}}, \bibinfo {author} {\bibfnamefont {F.}~\bibnamefont {Montealegre-Mora}}, \bibinfo {author} {\bibfnamefont {F.}~\bibnamefont {Vallentin}},\ and\ \bibinfo {author} {\bibfnamefont {D.}~\bibnamefont {Gross}},\ }\bibfield  {title} {\bibinfo {title} {Stabilizer extent is not multiplicative},\ }\href {https://quantum-journal.org/papers/q-2021-02-24-400/} {\bibfield  {journal} {\bibinfo  {journal} {Quantum}\ }\textbf {\bibinfo {volume} {5}},\ \bibinfo {pages} {400} (\bibinfo {year} {2021})}\BibitemShut {NoStop}%
\bibitem [{\citenamefont {Veitch}\ \emph {et~al.}(2014)\citenamefont {Veitch}, \citenamefont {Mousavian}, \citenamefont {Gottesman},\ and\ \citenamefont {Emerson}}]{veitch2014resource}%
  \BibitemOpen
  \bibfield  {author} {\bibinfo {author} {\bibfnamefont {V.}~\bibnamefont {Veitch}}, \bibinfo {author} {\bibfnamefont {S.~A.~H.}\ \bibnamefont {Mousavian}}, \bibinfo {author} {\bibfnamefont {D.}~\bibnamefont {Gottesman}},\ and\ \bibinfo {author} {\bibfnamefont {J.}~\bibnamefont {Emerson}},\ }\bibfield  {title} {\bibinfo {title} {The resource theory of stabilizer quantum computation},\ }\href {https://doi.org/10.1088/1367-2630/16/1/013009} {\bibfield  {journal} {\bibinfo  {journal} {New Journal of Physics}\ }\textbf {\bibinfo {volume} {16}},\ \bibinfo {pages} {013009} (\bibinfo {year} {2014})}\BibitemShut {NoStop}%
\bibitem [{\citenamefont {Leone}\ \emph {et~al.}(2022)\citenamefont {Leone}, \citenamefont {Oliviero},\ and\ \citenamefont {Hamma}}]{leone2022stabilizer}%
  \BibitemOpen
  \bibfield  {author} {\bibinfo {author} {\bibfnamefont {L.}~\bibnamefont {Leone}}, \bibinfo {author} {\bibfnamefont {S.~F.~E.}\ \bibnamefont {Oliviero}},\ and\ \bibinfo {author} {\bibfnamefont {A.}~\bibnamefont {Hamma}},\ }\bibfield  {title} {\bibinfo {title} {Stabilizer r\'enyi entropy},\ }\href {https://doi.org/10.1103/PhysRevLett.128.050402} {\bibfield  {journal} {\bibinfo  {journal} {Phys. Rev. Lett.}\ }\textbf {\bibinfo {volume} {128}},\ \bibinfo {pages} {050402} (\bibinfo {year} {2022})}\BibitemShut {NoStop}%
\bibitem [{\citenamefont {Haug}\ and\ \citenamefont {Piroli}(2023)}]{haug2023stabilizer}%
  \BibitemOpen
  \bibfield  {author} {\bibinfo {author} {\bibfnamefont {T.}~\bibnamefont {Haug}}\ and\ \bibinfo {author} {\bibfnamefont {L.}~\bibnamefont {Piroli}},\ }\bibfield  {title} {\bibinfo {title} {Stabilizer entropies and nonstabilizerness monotones},\ }\href {https://doi.org/10.22331/q-2023-08-28-1092} {\bibfield  {journal} {\bibinfo  {journal} {Quantum}\ }\textbf {\bibinfo {volume} {7}},\ \bibinfo {pages} {1092} (\bibinfo {year} {2023})}\BibitemShut {NoStop}%
\bibitem [{\citenamefont {Leone}\ and\ \citenamefont {Bittel}(2024)}]{PhysRevA.110.L040403}%
  \BibitemOpen
  \bibfield  {author} {\bibinfo {author} {\bibfnamefont {L.}~\bibnamefont {Leone}}\ and\ \bibinfo {author} {\bibfnamefont {L.}~\bibnamefont {Bittel}},\ }\bibfield  {title} {\bibinfo {title} {Stabilizer entropies are monotones for magic-state resource theory},\ }\href {https://doi.org/10.1103/PhysRevA.110.L040403} {\bibfield  {journal} {\bibinfo  {journal} {Phys. Rev. A}\ }\textbf {\bibinfo {volume} {110}},\ \bibinfo {pages} {L040403} (\bibinfo {year} {2024})}\BibitemShut {NoStop}%
\bibitem [{\citenamefont {Sarkar}\ \emph {et~al.}(2020)\citenamefont {Sarkar}, \citenamefont {Mukhopadhyay},\ and\ \citenamefont {Bayat}}]{sarkar2020characterization}%
  \BibitemOpen
  \bibfield  {author} {\bibinfo {author} {\bibfnamefont {S.}~\bibnamefont {Sarkar}}, \bibinfo {author} {\bibfnamefont {C.}~\bibnamefont {Mukhopadhyay}},\ and\ \bibinfo {author} {\bibfnamefont {A.}~\bibnamefont {Bayat}},\ }\bibfield  {title} {\bibinfo {title} {Characterization of an operational quantum resource in a critical many-body system},\ }\href {https://iopscience.iop.org/article/10.1088/1367-2630/aba919} {\bibfield  {journal} {\bibinfo  {journal} {New Journal of Physics}\ }\textbf {\bibinfo {volume} {22}},\ \bibinfo {pages} {083077} (\bibinfo {year} {2020})}\BibitemShut {NoStop}%
\bibitem [{\citenamefont {Oliviero}\ \emph {et~al.}(2022{\natexlab{a}})\citenamefont {Oliviero}, \citenamefont {Leone},\ and\ \citenamefont {Hamma}}]{oliviero2022magic}%
  \BibitemOpen
  \bibfield  {author} {\bibinfo {author} {\bibfnamefont {S.~F.~E.}\ \bibnamefont {Oliviero}}, \bibinfo {author} {\bibfnamefont {L.}~\bibnamefont {Leone}},\ and\ \bibinfo {author} {\bibfnamefont {A.}~\bibnamefont {Hamma}},\ }\bibfield  {title} {\bibinfo {title} {Magic-state resource theory for the ground state of the transverse-field ising model},\ }\href {https://doi.org/10.1103/PhysRevA.106.042426} {\bibfield  {journal} {\bibinfo  {journal} {Phys. Rev. A}\ }\textbf {\bibinfo {volume} {106}},\ \bibinfo {pages} {042426} (\bibinfo {year} {2022}{\natexlab{a}})}\BibitemShut {NoStop}%
\bibitem [{\citenamefont {Liu}\ and\ \citenamefont {Winter}(2022)}]{liu2022manybody}%
  \BibitemOpen
  \bibfield  {author} {\bibinfo {author} {\bibfnamefont {Z.-W.}\ \bibnamefont {Liu}}\ and\ \bibinfo {author} {\bibfnamefont {A.}~\bibnamefont {Winter}},\ }\bibfield  {title} {\bibinfo {title} {Many-body quantum magic},\ }\href {https://doi.org/10.1103/PRXQuantum.3.020333} {\bibfield  {journal} {\bibinfo  {journal} {PRX Quantum}\ }\textbf {\bibinfo {volume} {3}},\ \bibinfo {pages} {020333} (\bibinfo {year} {2022})}\BibitemShut {NoStop}%
\bibitem [{\citenamefont {Viscardi}\ \emph {et~al.}(2025)\citenamefont {Viscardi}, \citenamefont {Dalmonte}, \citenamefont {Hamma},\ and\ \citenamefont {Tirrito}}]{viscardi2025interplay}%
  \BibitemOpen
  \bibfield  {author} {\bibinfo {author} {\bibfnamefont {M.}~\bibnamefont {Viscardi}}, \bibinfo {author} {\bibfnamefont {M.}~\bibnamefont {Dalmonte}}, \bibinfo {author} {\bibfnamefont {A.}~\bibnamefont {Hamma}},\ and\ \bibinfo {author} {\bibfnamefont {E.}~\bibnamefont {Tirrito}},\ }\bibfield  {title} {\bibinfo {title} {Interplay of entanglement structures and stabilizer entropy in spin models},\ }\href {https://arxiv.org/abs/2503.08620} {\bibfield  {journal} {\bibinfo  {journal} {arXiv:2503.08620}\ } (\bibinfo {year} {2025})}\BibitemShut {NoStop}%
\bibitem [{\citenamefont {Falc{\~a}o}\ \emph {et~al.}(2025)\citenamefont {Falc{\~a}o}, \citenamefont {Sierant}, \citenamefont {Zakrzewski},\ and\ \citenamefont {Tirrito}}]{falcao2025magic}%
  \BibitemOpen
  \bibfield  {author} {\bibinfo {author} {\bibfnamefont {P.~R.~N.}\ \bibnamefont {Falc{\~a}o}}, \bibinfo {author} {\bibfnamefont {P.}~\bibnamefont {Sierant}}, \bibinfo {author} {\bibfnamefont {J.}~\bibnamefont {Zakrzewski}},\ and\ \bibinfo {author} {\bibfnamefont {E.}~\bibnamefont {Tirrito}},\ }\bibfield  {title} {\bibinfo {title} {Magic dynamics in many-body localized systems},\ }\href {https://arxiv.org/abs/2503.07468} {\bibfield  {journal} {\bibinfo  {journal} {arXiv:2503.07468}\ } (\bibinfo {year} {2025})}\BibitemShut {NoStop}%
\bibitem [{\citenamefont {Santra}\ \emph {et~al.}(2025{\natexlab{b}})\citenamefont {Santra}, \citenamefont {Windey}, \citenamefont {Bandyopadhyay}, \citenamefont {Legramandi},\ and\ \citenamefont {Hauke}}]{santra2025complexity}%
  \BibitemOpen
  \bibfield  {author} {\bibinfo {author} {\bibfnamefont {G.~C.}\ \bibnamefont {Santra}}, \bibinfo {author} {\bibfnamefont {A.}~\bibnamefont {Windey}}, \bibinfo {author} {\bibfnamefont {S.}~\bibnamefont {Bandyopadhyay}}, \bibinfo {author} {\bibfnamefont {A.}~\bibnamefont {Legramandi}},\ and\ \bibinfo {author} {\bibfnamefont {P.}~\bibnamefont {Hauke}},\ }\bibfield  {title} {\bibinfo {title} {Complexity transitions in chaotic quantum systems},\ }\href {https://arxiv.org/abs/2505.09707} {\bibfield  {journal} {\bibinfo  {journal} {arXiv:2505.09707}\ } (\bibinfo {year} {2025}{\natexlab{b}})}\BibitemShut {NoStop}%
\bibitem [{\citenamefont {Korbany}\ \emph {et~al.}(2025)\citenamefont {Korbany}, \citenamefont {Gullans},\ and\ \citenamefont {Piroli}}]{korbany2025long}%
  \BibitemOpen
  \bibfield  {author} {\bibinfo {author} {\bibfnamefont {D.~A.}\ \bibnamefont {Korbany}}, \bibinfo {author} {\bibfnamefont {M.~J.}\ \bibnamefont {Gullans}},\ and\ \bibinfo {author} {\bibfnamefont {L.}~\bibnamefont {Piroli}},\ }\bibfield  {title} {\bibinfo {title} {Long-range nonstabilizerness and phases of matter},\ }\href {https://arxiv.org/abs/2502.19504} {\bibfield  {journal} {\bibinfo  {journal} {arXiv:2502.19504}\ } (\bibinfo {year} {2025})}\BibitemShut {NoStop}%
\bibitem [{\citenamefont {Turkeshi}\ \emph {et~al.}(2025{\natexlab{a}})\citenamefont {Turkeshi}, \citenamefont {Tirrito},\ and\ \citenamefont {Sierant}}]{turkeshi2025magic}%
  \BibitemOpen
  \bibfield  {author} {\bibinfo {author} {\bibfnamefont {X.}~\bibnamefont {Turkeshi}}, \bibinfo {author} {\bibfnamefont {E.}~\bibnamefont {Tirrito}},\ and\ \bibinfo {author} {\bibfnamefont {P.}~\bibnamefont {Sierant}},\ }\bibfield  {title} {\bibinfo {title} {Magic spreading in random quantum circuits},\ }\href {https://www.nature.com/articles/s41467-025-57704-x} {\bibfield  {journal} {\bibinfo  {journal} {Nature Communications}\ }\textbf {\bibinfo {volume} {16}},\ \bibinfo {pages} {2575} (\bibinfo {year} {2025}{\natexlab{a}})}\BibitemShut {NoStop}%
\bibitem [{\citenamefont {Haug}\ \emph {et~al.}(2024)\citenamefont {Haug}, \citenamefont {Aolita},\ and\ \citenamefont {Kim}}]{haug2024probing}%
  \BibitemOpen
  \bibfield  {author} {\bibinfo {author} {\bibfnamefont {T.}~\bibnamefont {Haug}}, \bibinfo {author} {\bibfnamefont {L.}~\bibnamefont {Aolita}},\ and\ \bibinfo {author} {\bibfnamefont {M.}~\bibnamefont {Kim}},\ }\bibfield  {title} {\bibinfo {title} {Probing quantum complexity via universal saturation of stabilizer entropies},\ }\href {https://arxiv.org/abs/2406.04190} {\bibfield  {journal} {\bibinfo  {journal} {arXiv:2406.04190}\ } (\bibinfo {year} {2024})}\BibitemShut {NoStop}%
\bibitem [{\citenamefont {Varikuti}\ \emph {et~al.}(2025)\citenamefont {Varikuti}, \citenamefont {Bandyopadhyay},\ and\ \citenamefont {Hauke}}]{dileepNonstab2025}%
  \BibitemOpen
  \bibfield  {author} {\bibinfo {author} {\bibfnamefont {N.~D.}\ \bibnamefont {Varikuti}}, \bibinfo {author} {\bibfnamefont {S.}~\bibnamefont {Bandyopadhyay}},\ and\ \bibinfo {author} {\bibfnamefont {P.}~\bibnamefont {Hauke}},\ }\bibfield  {title} {\bibinfo {title} {Impact of clifford operations on non-stabilizing power and quantum chaos},\ }\href {https://arxiv.org/abs/2505.14793} {\bibfield  {journal} {\bibinfo  {journal} {arXiv:2505.14793}\ } (\bibinfo {year} {2025})}\BibitemShut {NoStop}%
\bibitem [{\citenamefont {White}\ \emph {et~al.}(2021)\citenamefont {White}, \citenamefont {Cao},\ and\ \citenamefont {Swingle}}]{white2021conformal}%
  \BibitemOpen
  \bibfield  {author} {\bibinfo {author} {\bibfnamefont {C.~D.}\ \bibnamefont {White}}, \bibinfo {author} {\bibfnamefont {C.}~\bibnamefont {Cao}},\ and\ \bibinfo {author} {\bibfnamefont {B.}~\bibnamefont {Swingle}},\ }\bibfield  {title} {\bibinfo {title} {Conformal field theories are magical},\ }\href {https://doi.org/10.1103/PhysRevB.103.075145} {\bibfield  {journal} {\bibinfo  {journal} {Phys. Rev. B}\ }\textbf {\bibinfo {volume} {103}},\ \bibinfo {pages} {075145} (\bibinfo {year} {2021})}\BibitemShut {NoStop}%
\bibitem [{\citenamefont {Tarabunga}(2024)}]{tarabunga2024critical}%
  \BibitemOpen
  \bibfield  {author} {\bibinfo {author} {\bibfnamefont {P.~S.}\ \bibnamefont {Tarabunga}},\ }\bibfield  {title} {\bibinfo {title} {Critical behaviors of non-stabilizerness in quantum spin chains},\ }\href {https://quantum-journal.org/papers/q-2024-07-17-1413/} {\bibfield  {journal} {\bibinfo  {journal} {Quantum}\ }\textbf {\bibinfo {volume} {8}},\ \bibinfo {pages} {1413} (\bibinfo {year} {2024})}\BibitemShut {NoStop}%
\bibitem [{\citenamefont {Hoshino}\ \emph {et~al.}(2025)\citenamefont {Hoshino}, \citenamefont {Oshikawa},\ and\ \citenamefont {Ashida}}]{hoshino2025stabilizer}%
  \BibitemOpen
  \bibfield  {author} {\bibinfo {author} {\bibfnamefont {M.}~\bibnamefont {Hoshino}}, \bibinfo {author} {\bibfnamefont {M.}~\bibnamefont {Oshikawa}},\ and\ \bibinfo {author} {\bibfnamefont {Y.}~\bibnamefont {Ashida}},\ }\bibfield  {title} {\bibinfo {title} {Stabilizer r$\backslash$'enyi entropy and conformal field theory},\ }\href {https://arxiv.org/abs/2503.13599} {\bibfield  {journal} {\bibinfo  {journal} {arXiv:2503.13599}\ } (\bibinfo {year} {2025})}\BibitemShut {NoStop}%
\bibitem [{\citenamefont {Spriggs}\ \emph {et~al.}(2025)\citenamefont {Spriggs}, \citenamefont {Ahmadi}, \citenamefont {Chen},\ and\ \citenamefont {Greplova}}]{Spriggs_2025}%
  \BibitemOpen
  \bibfield  {author} {\bibinfo {author} {\bibfnamefont {T.}~\bibnamefont {Spriggs}}, \bibinfo {author} {\bibfnamefont {A.}~\bibnamefont {Ahmadi}}, \bibinfo {author} {\bibfnamefont {B.}~\bibnamefont {Chen}},\ and\ \bibinfo {author} {\bibfnamefont {E.}~\bibnamefont {Greplova}},\ }\bibfield  {title} {\bibinfo {title} {Quantum resources of quantum and classical variational methods},\ }\href {https://doi.org/10.1088/2632-2153/adaca2} {\bibfield  {journal} {\bibinfo  {journal} {Machine Learning: Science and Technology}\ }\textbf {\bibinfo {volume} {6}},\ \bibinfo {pages} {015042} (\bibinfo {year} {2025})}\BibitemShut {NoStop}%
\bibitem [{\citenamefont {Sinibaldi}\ \emph {et~al.}(2025)\citenamefont {Sinibaldi}, \citenamefont {Mello}, \citenamefont {Collura},\ and\ \citenamefont {Carleo}}]{sinibaldi2025}%
  \BibitemOpen
  \bibfield  {author} {\bibinfo {author} {\bibfnamefont {A.}~\bibnamefont {Sinibaldi}}, \bibinfo {author} {\bibfnamefont {A.~F.}\ \bibnamefont {Mello}}, \bibinfo {author} {\bibfnamefont {M.}~\bibnamefont {Collura}},\ and\ \bibinfo {author} {\bibfnamefont {G.}~\bibnamefont {Carleo}},\ }\bibfield  {title} {\bibinfo {title} {Non-stabilizerness of neural quantum states},\ }\href {https://arxiv.org/abs/2502.09725} {\bibfield  {journal} {\bibinfo  {journal} {arXiv:2502.09725}\ } (\bibinfo {year} {2025})}\BibitemShut {NoStop}%
\bibitem [{\citenamefont {Panchenko}(2013)}]{panchenko2013sherrington}%
  \BibitemOpen
  \bibfield  {author} {\bibinfo {author} {\bibfnamefont {D.}~\bibnamefont {Panchenko}},\ }\href {https://link.springer.com/book/10.1007/978-1-4614-6289-7} {\emph {\bibinfo {title} {The sherrington-kirkpatrick model}}}\ (\bibinfo  {publisher} {Springer Science \& Business Media},\ \bibinfo {year} {2013})\BibitemShut {NoStop}%
\bibitem [{\citenamefont {Dupont}\ \emph {et~al.}(2022{\natexlab{b}})\citenamefont {Dupont}, \citenamefont {Didier}, \citenamefont {Hodson}, \citenamefont {Moore},\ and\ \citenamefont {Reagor}}]{dupont2022entanglement}%
  \BibitemOpen
  \bibfield  {author} {\bibinfo {author} {\bibfnamefont {M.}~\bibnamefont {Dupont}}, \bibinfo {author} {\bibfnamefont {N.}~\bibnamefont {Didier}}, \bibinfo {author} {\bibfnamefont {M.~J.}\ \bibnamefont {Hodson}}, \bibinfo {author} {\bibfnamefont {J.~E.}\ \bibnamefont {Moore}},\ and\ \bibinfo {author} {\bibfnamefont {M.~J.}\ \bibnamefont {Reagor}},\ }\bibfield  {title} {\bibinfo {title} {Entanglement perspective on the quantum approximate optimization algorithm},\ }\href {https://link.aps.org/doi/10.1103/PhysRevA.106.022423} {\bibfield  {journal} {\bibinfo  {journal} {Phys. Rev. A}\ }\textbf {\bibinfo {volume} {106}},\ \bibinfo {pages} {022423} (\bibinfo {year} {2022}{\natexlab{b}})}\BibitemShut {NoStop}%
\bibitem [{\citenamefont {Oliviero}\ \emph {et~al.}(2022{\natexlab{b}})\citenamefont {Oliviero}, \citenamefont {Leone}, \citenamefont {Hamma},\ and\ \citenamefont {Lloyd}}]{oliviero2022measuring}%
  \BibitemOpen
  \bibfield  {author} {\bibinfo {author} {\bibfnamefont {S.~F.}\ \bibnamefont {Oliviero}}, \bibinfo {author} {\bibfnamefont {L.}~\bibnamefont {Leone}}, \bibinfo {author} {\bibfnamefont {A.}~\bibnamefont {Hamma}},\ and\ \bibinfo {author} {\bibfnamefont {S.}~\bibnamefont {Lloyd}},\ }\bibfield  {title} {\bibinfo {title} {Measuring magic on a quantum processor},\ }\href {https://www.nature.com/articles/s41534-022-00666-5} {\bibfield  {journal} {\bibinfo  {journal} {npj Quantum Information}\ }\textbf {\bibinfo {volume} {8}},\ \bibinfo {pages} {148} (\bibinfo {year} {2022}{\natexlab{b}})}\BibitemShut {NoStop}%
\bibitem [{\citenamefont {Niroula}\ \emph {et~al.}(2024)\citenamefont {Niroula}, \citenamefont {Wang}, \citenamefont {Johri}, \citenamefont {Zhu}, \citenamefont {Monroe}, \citenamefont {Noel},\ and\ \citenamefont {Gullans}}]{Niroula2024}%
  \BibitemOpen
  \bibfield  {author} {\bibinfo {author} {\bibfnamefont {C.~D.}\ \bibnamefont {Niroula}, \bibfnamefont {Pradeep~andWhite}}, \bibinfo {author} {\bibfnamefont {Q.}~\bibnamefont {Wang}}, \bibinfo {author} {\bibfnamefont {S.}~\bibnamefont {Johri}}, \bibinfo {author} {\bibfnamefont {D.}~\bibnamefont {Zhu}}, \bibinfo {author} {\bibfnamefont {C.}~\bibnamefont {Monroe}}, \bibinfo {author} {\bibfnamefont {C.}~\bibnamefont {Noel}},\ and\ \bibinfo {author} {\bibfnamefont {M.~J.}\ \bibnamefont {Gullans}},\ }\bibfield  {title} {\bibinfo {title} {Phase transition in magic with random quantum circuits},\ }\href {https://doi.org/10.1038/s41567-024-02637-3} {\bibfield  {journal} {\bibinfo  {journal} {Nature Physics}\ }\textbf {\bibinfo {volume} {20}},\ \bibinfo {pages} {1786–1792} (\bibinfo {year} {2024})}\BibitemShut {NoStop}%
\bibitem [{\citenamefont {Bluvstein}\ \emph {et~al.}(2024)\citenamefont {Bluvstein}, \citenamefont {Evered}, \citenamefont {Geim}, \citenamefont {Li}, \citenamefont {Zhou}, \citenamefont {Manovitz}, \citenamefont {Ebadi}, \citenamefont {Cain}, \citenamefont {Kalinowski}, \citenamefont {Hangleiter} \emph {et~al.}}]{bluvstein2024logical}%
  \BibitemOpen
  \bibfield  {author} {\bibinfo {author} {\bibfnamefont {D.}~\bibnamefont {Bluvstein}}, \bibinfo {author} {\bibfnamefont {S.~J.}\ \bibnamefont {Evered}}, \bibinfo {author} {\bibfnamefont {A.~A.}\ \bibnamefont {Geim}}, \bibinfo {author} {\bibfnamefont {S.~H.}\ \bibnamefont {Li}}, \bibinfo {author} {\bibfnamefont {H.}~\bibnamefont {Zhou}}, \bibinfo {author} {\bibfnamefont {T.}~\bibnamefont {Manovitz}}, \bibinfo {author} {\bibfnamefont {S.}~\bibnamefont {Ebadi}}, \bibinfo {author} {\bibfnamefont {M.}~\bibnamefont {Cain}}, \bibinfo {author} {\bibfnamefont {M.}~\bibnamefont {Kalinowski}}, \bibinfo {author} {\bibfnamefont {D.}~\bibnamefont {Hangleiter}}, \emph {et~al.},\ }\bibfield  {title} {\bibinfo {title} {Logical quantum processor based on reconfigurable atom arrays},\ }\href {https://www.nature.com/articles/s41586-023-06927-3} {\bibfield  {journal} {\bibinfo  {journal} {Nature}\ }\textbf {\bibinfo {volume} {626}},\ \bibinfo {pages} {58} (\bibinfo {year} {2024})}\BibitemShut {NoStop}%
\bibitem [{\citenamefont {Gross}\ \emph {et~al.}(2021)\citenamefont {Gross}, \citenamefont {Nezami},\ and\ \citenamefont {Walter}}]{gross2021schurweylduality}%
  \BibitemOpen
  \bibfield  {author} {\bibinfo {author} {\bibfnamefont {D.}~\bibnamefont {Gross}}, \bibinfo {author} {\bibfnamefont {S.}~\bibnamefont {Nezami}},\ and\ \bibinfo {author} {\bibfnamefont {M.}~\bibnamefont {Walter}},\ }\bibfield  {title} {\bibinfo {title} {Schur–weyl duality for the clifford group with applications: Property testing, a robust hudson theorem, and de finetti representations},\ }\href {https://doi.org/10.1007/s00220-021-04118-7} {\bibfield  {journal} {\bibinfo  {journal} {Commun. Math. Phys.}\ }\textbf {\bibinfo {volume} {385}},\ \bibinfo {pages} {1325} (\bibinfo {year} {2021})}\BibitemShut {NoStop}%
\bibitem [{\citenamefont {Wootters}(1987{\natexlab{b}})}]{WOOTTERS19871}%
  \BibitemOpen
  \bibfield  {author} {\bibinfo {author} {\bibfnamefont {W.~K.}\ \bibnamefont {Wootters}},\ }\bibfield  {title} {\bibinfo {title} {A wigner-function formulation of finite-state quantum mechanics},\ }\href {https://doi.org/https://doi.org/10.1016/0003-4916(87)90176-X} {\bibfield  {journal} {\bibinfo  {journal} {Annals of Physics}\ }\textbf {\bibinfo {volume} {176}},\ \bibinfo {pages} {1} (\bibinfo {year} {1987}{\natexlab{b}})}\BibitemShut {NoStop}%
\bibitem [{\citenamefont {Streif}\ and\ \citenamefont {Leib}(2019)}]{streif2019comparison}%
  \BibitemOpen
  \bibfield  {author} {\bibinfo {author} {\bibfnamefont {M.}~\bibnamefont {Streif}}\ and\ \bibinfo {author} {\bibfnamefont {M.}~\bibnamefont {Leib}},\ }\bibfield  {title} {\bibinfo {title} {Comparison of qaoa with quantum and simulated annealing},\ }\href {https://arxiv.org/abs/1901.01903} {\bibfield  {journal} {\bibinfo  {journal} {arXiv:1901.01903}\ } (\bibinfo {year} {2019})}\BibitemShut {NoStop}%
\bibitem [{\citenamefont {Bottrill}\ \emph {et~al.}(2023)\citenamefont {Bottrill}, \citenamefont {Pandey},\ and\ \citenamefont {Di~Matteo}}]{bottrillQutrit2023}%
  \BibitemOpen
  \bibfield  {author} {\bibinfo {author} {\bibfnamefont {G.}~\bibnamefont {Bottrill}}, \bibinfo {author} {\bibfnamefont {M.}~\bibnamefont {Pandey}},\ and\ \bibinfo {author} {\bibfnamefont {O.}~\bibnamefont {Di~Matteo}},\ }\bibfield  {title} {\bibinfo {title} {Exploring the potential of qutrits for quantum optimization of graph coloring},\ }\href {https://doi.org/10.1109/QCE57702.2023.00028} {\bibfield  {journal} {\bibinfo  {journal} {2023 IEEE International Conference on Quantum Computing and Engineering (QCE)}\ }\textbf {\bibinfo {volume} {01}},\ \bibinfo {pages} {177} (\bibinfo {year} {2023})}\BibitemShut {NoStop}%
\bibitem [{\citenamefont {Kar{\'a}csony}\ \emph {et~al.}(2024)\citenamefont {Kar{\'a}csony}, \citenamefont {Oroszl{\'a}ny},\ and\ \citenamefont {Zimbor{\'a}s}}]{karacsony2024efficient}%
  \BibitemOpen
  \bibfield  {author} {\bibinfo {author} {\bibfnamefont {M.}~\bibnamefont {Kar{\'a}csony}}, \bibinfo {author} {\bibfnamefont {L.}~\bibnamefont {Oroszl{\'a}ny}},\ and\ \bibinfo {author} {\bibfnamefont {Z.}~\bibnamefont {Zimbor{\'a}s}},\ }\bibfield  {title} {\bibinfo {title} {Efficient qudit based scheme for photonic quantum computing},\ }\href {https://scipost.org/preprints/scipost_202304_00012v2/} {\bibfield  {journal} {\bibinfo  {journal} {SciPost Physics Core}\ }\textbf {\bibinfo {volume} {7}},\ \bibinfo {pages} {032} (\bibinfo {year} {2024})}\BibitemShut {NoStop}%
\bibitem [{Note1()}]{Note1}%
  \BibitemOpen
  \bibinfo {note} {The $\mathinner {|{+}\rangle }=\protect \frac {1}{\protect \sqrt {D}}\DOTSB \sum@ \slimits@ _{i=0}^{D-1}\mathinner {|{i}\rangle }$ qudit state is a stabilizer state because it is a $(+1)$ eigenstate of a $d$-element subgroup of the Pauli group; for example, it is $\{I, X\}$ for $d=2$. Using the additive property, $\protect \mathcal {M}(\mathinner {|{+}\rangle }^{\otimes N}) = N\protect \mathcal {M}(\mathinner {|{+}\rangle }) = 0$.}\BibitemShut {Stop}%
\bibitem [{\citenamefont {Sherrington}\ and\ \citenamefont {Kirkpatrick}(1975)}]{sherrington1975solvable}%
  \BibitemOpen
  \bibfield  {author} {\bibinfo {author} {\bibfnamefont {D.}~\bibnamefont {Sherrington}}\ and\ \bibinfo {author} {\bibfnamefont {S.}~\bibnamefont {Kirkpatrick}},\ }\bibfield  {title} {\bibinfo {title} {Solvable model of a spin-glass},\ }\href {https://doi.org/10.1103/PhysRevLett.35.1792} {\bibfield  {journal} {\bibinfo  {journal} {Phys. Rev. Lett.}\ }\textbf {\bibinfo {volume} {35}},\ \bibinfo {pages} {1792} (\bibinfo {year} {1975})}\BibitemShut {NoStop}%
\bibitem [{\citenamefont {Venturelli}\ \emph {et~al.}(2015)\citenamefont {Venturelli}, \citenamefont {Mandr\`a}, \citenamefont {Knysh}, \citenamefont {O'Gorman}, \citenamefont {Biswas},\ and\ \citenamefont {Smelyanskiy}}]{venturelli2015quantum}%
  \BibitemOpen
  \bibfield  {author} {\bibinfo {author} {\bibfnamefont {D.}~\bibnamefont {Venturelli}}, \bibinfo {author} {\bibfnamefont {S.}~\bibnamefont {Mandr\`a}}, \bibinfo {author} {\bibfnamefont {S.}~\bibnamefont {Knysh}}, \bibinfo {author} {\bibfnamefont {B.}~\bibnamefont {O'Gorman}}, \bibinfo {author} {\bibfnamefont {R.}~\bibnamefont {Biswas}},\ and\ \bibinfo {author} {\bibfnamefont {V.}~\bibnamefont {Smelyanskiy}},\ }\bibfield  {title} {\bibinfo {title} {Quantum optimization of fully connected spin glasses},\ }\href {https://doi.org/10.1103/PhysRevX.5.031040} {\bibfield  {journal} {\bibinfo  {journal} {Phys. Rev. X}\ }\textbf {\bibinfo {volume} {5}},\ \bibinfo {pages} {031040} (\bibinfo {year} {2015})}\BibitemShut {NoStop}%
\bibitem [{\citenamefont {Mugel}\ \emph {et~al.}(2022)\citenamefont {Mugel}, \citenamefont {Kuchkovsky}, \citenamefont {S{\'a}nchez}, \citenamefont {Fern{\'a}ndez-Lorenzo}, \citenamefont {Luis-Hita}, \citenamefont {Lizaso},\ and\ \citenamefont {Or{\'u}s}}]{mugel2022dynamic}%
  \BibitemOpen
  \bibfield  {author} {\bibinfo {author} {\bibfnamefont {S.}~\bibnamefont {Mugel}}, \bibinfo {author} {\bibfnamefont {C.}~\bibnamefont {Kuchkovsky}}, \bibinfo {author} {\bibfnamefont {E.}~\bibnamefont {S{\'a}nchez}}, \bibinfo {author} {\bibfnamefont {S.}~\bibnamefont {Fern{\'a}ndez-Lorenzo}}, \bibinfo {author} {\bibfnamefont {J.}~\bibnamefont {Luis-Hita}}, \bibinfo {author} {\bibfnamefont {E.}~\bibnamefont {Lizaso}},\ and\ \bibinfo {author} {\bibfnamefont {R.}~\bibnamefont {Or{\'u}s}},\ }\bibfield  {title} {\bibinfo {title} {Dynamic portfolio optimization with real datasets using quantum processors and quantum-inspired tensor networks},\ }\href {https://journals.aps.org/prresearch/abstract/10.1103/PhysRevResearch.4.013006} {\bibfield  {journal} {\bibinfo  {journal} {Phys. Rev. R}\ }\textbf {\bibinfo {volume} {4}},\ \bibinfo {pages} {013006} (\bibinfo {year} {2022})}\BibitemShut {NoStop}%
\bibitem [{\citenamefont {Sack}\ and\ \citenamefont {Serbyn}(2021)}]{sack2021quantum}%
  \BibitemOpen
  \bibfield  {author} {\bibinfo {author} {\bibfnamefont {S.~H.}\ \bibnamefont {Sack}}\ and\ \bibinfo {author} {\bibfnamefont {M.}~\bibnamefont {Serbyn}},\ }\bibfield  {title} {\bibinfo {title} {Quantum annealing initialization of the quantum approximate optimization algorithm},\ }\href {https://quantum-journal.org/papers/q-2021-07-01-491/} {\bibfield  {journal} {\bibinfo  {journal} {Quantum}\ }\textbf {\bibinfo {volume} {5}},\ \bibinfo {pages} {491} (\bibinfo {year} {2021})}\BibitemShut {NoStop}%
\bibitem [{\citenamefont {Turkeshi}\ \emph {et~al.}(2025{\natexlab{b}})\citenamefont {Turkeshi}, \citenamefont {Dymarsky},\ and\ \citenamefont {Sierant}}]{turkeshi2025pauli}%
  \BibitemOpen
  \bibfield  {author} {\bibinfo {author} {\bibfnamefont {X.}~\bibnamefont {Turkeshi}}, \bibinfo {author} {\bibfnamefont {A.}~\bibnamefont {Dymarsky}},\ and\ \bibinfo {author} {\bibfnamefont {P.}~\bibnamefont {Sierant}},\ }\bibfield  {title} {\bibinfo {title} {Pauli spectrum and nonstabilizerness of typical quantum many-body states},\ }\href {https://doi.org/10.1103/PhysRevB.111.054301} {\bibfield  {journal} {\bibinfo  {journal} {Phys. Rev. B}\ }\textbf {\bibinfo {volume} {111}},\ \bibinfo {pages} {054301} (\bibinfo {year} {2025}{\natexlab{b}})}\BibitemShut {NoStop}%
\bibitem [{\citenamefont {Hauke}\ \emph {et~al.}(2020)\citenamefont {Hauke}, \citenamefont {Katzgraber}, \citenamefont {Lechner}, \citenamefont {Nishimori},\ and\ \citenamefont {Oliver}}]{hauke2020perspectives}%
  \BibitemOpen
  \bibfield  {author} {\bibinfo {author} {\bibfnamefont {P.}~\bibnamefont {Hauke}}, \bibinfo {author} {\bibfnamefont {H.~G.}\ \bibnamefont {Katzgraber}}, \bibinfo {author} {\bibfnamefont {W.}~\bibnamefont {Lechner}}, \bibinfo {author} {\bibfnamefont {H.}~\bibnamefont {Nishimori}},\ and\ \bibinfo {author} {\bibfnamefont {W.~D.}\ \bibnamefont {Oliver}},\ }\bibfield  {title} {\bibinfo {title} {Perspectives of quantum annealing: methods and implementations},\ }\href {https://doi.org/10.1088/1361-6633/ab85b8} {\bibfield  {journal} {\bibinfo  {journal} {Reports on Progress in Physics}\ }\textbf {\bibinfo {volume} {83}},\ \bibinfo {pages} {054401} (\bibinfo {year} {2020})}\BibitemShut {NoStop}%
\bibitem [{\citenamefont {Santoro}\ \emph {et~al.}(2002)\citenamefont {Santoro}, \citenamefont {Martoňák}, \citenamefont {Tosatti},\ and\ \citenamefont {Car}}]{Santoro_2002_QA}%
  \BibitemOpen
  \bibfield  {author} {\bibinfo {author} {\bibfnamefont {G.~E.}\ \bibnamefont {Santoro}}, \bibinfo {author} {\bibfnamefont {R.}~\bibnamefont {Martoňák}}, \bibinfo {author} {\bibfnamefont {E.}~\bibnamefont {Tosatti}},\ and\ \bibinfo {author} {\bibfnamefont {R.}~\bibnamefont {Car}},\ }\bibfield  {title} {\bibinfo {title} {Theory of quantum annealing of an ising spin glass},\ }\href {https://doi.org/10.1126/science.1068774} {\bibfield  {journal} {\bibinfo  {journal} {Science}\ }\textbf {\bibinfo {volume} {295}},\ \bibinfo {pages} {2427} (\bibinfo {year} {2002})}\BibitemShut {NoStop}%
\bibitem [{\citenamefont {Morita}\ and\ \citenamefont {Nishimori}(2008)}]{morita2008mathematical}%
  \BibitemOpen
  \bibfield  {author} {\bibinfo {author} {\bibfnamefont {S.}~\bibnamefont {Morita}}\ and\ \bibinfo {author} {\bibfnamefont {H.}~\bibnamefont {Nishimori}},\ }\bibfield  {title} {\bibinfo {title} {Mathematical foundation of quantum annealing},\ }\href {https://pubs.aip.org/aip/jmp/article/49/12/125210/231148} {\bibfield  {journal} {\bibinfo  {journal} {Journal of Mathematical Physics}\ }\textbf {\bibinfo {volume} {49}} (\bibinfo {year} {2008})}\BibitemShut {NoStop}%
\bibitem [{\citenamefont {Rajak}\ \emph {et~al.}(2023)\citenamefont {Rajak}, \citenamefont {Suzuki}, \citenamefont {Dutta},\ and\ \citenamefont {Chakrabarti}}]{rajak2023quantum}%
  \BibitemOpen
  \bibfield  {author} {\bibinfo {author} {\bibfnamefont {A.}~\bibnamefont {Rajak}}, \bibinfo {author} {\bibfnamefont {S.}~\bibnamefont {Suzuki}}, \bibinfo {author} {\bibfnamefont {A.}~\bibnamefont {Dutta}},\ and\ \bibinfo {author} {\bibfnamefont {B.~K.}\ \bibnamefont {Chakrabarti}},\ }\bibfield  {title} {\bibinfo {title} {Quantum annealing: An overview},\ }\href {https://royalsocietypublishing.org/doi/10.1098/rsta.2021.0417} {\bibfield  {journal} {\bibinfo  {journal} {Philosophical Transactions of the Royal Society A}\ }\textbf {\bibinfo {volume} {381}},\ \bibinfo {pages} {20210417} (\bibinfo {year} {2023})}\BibitemShut {NoStop}%
\bibitem [{\citenamefont {Schollw{\"o}ck}(2011)}]{schollwock2011density}%
  \BibitemOpen
  \bibfield  {author} {\bibinfo {author} {\bibfnamefont {U.}~\bibnamefont {Schollw{\"o}ck}},\ }\bibfield  {title} {\bibinfo {title} {The density-matrix renormalization group in the age of matrix product states},\ }\href {https://www.sciencedirect.com/science/article/pii/S0003491610001752} {\bibfield  {journal} {\bibinfo  {journal} {Annals of physics}\ }\textbf {\bibinfo {volume} {326}},\ \bibinfo {pages} {96} (\bibinfo {year} {2011})}\BibitemShut {NoStop}%
\bibitem [{\citenamefont {Or{\'u}s}(2014)}]{orus2014practical}%
  \BibitemOpen
  \bibfield  {author} {\bibinfo {author} {\bibfnamefont {R.}~\bibnamefont {Or{\'u}s}},\ }\bibfield  {title} {\bibinfo {title} {A practical introduction to tensor networks: Matrix product states and projected entangled pair states},\ }\href {https://www.sciencedirect.com/science/article/pii/S0003491614001596} {\bibfield  {journal} {\bibinfo  {journal} {Annals of physics}\ }\textbf {\bibinfo {volume} {349}},\ \bibinfo {pages} {117} (\bibinfo {year} {2014})}\BibitemShut {NoStop}%
\bibitem [{\citenamefont {Or{\'u}s}(2019)}]{orus2019tensor}%
  \BibitemOpen
  \bibfield  {author} {\bibinfo {author} {\bibfnamefont {R.}~\bibnamefont {Or{\'u}s}},\ }\bibfield  {title} {\bibinfo {title} {Tensor networks for complex quantum systems},\ }\href {https://www.nature.com/articles/s42254-019-0086-7} {\bibfield  {journal} {\bibinfo  {journal} {Nature Reviews Physics}\ }\textbf {\bibinfo {volume} {1}},\ \bibinfo {pages} {538} (\bibinfo {year} {2019})}\BibitemShut {NoStop}%
\bibitem [{\citenamefont {Ran}\ \emph {et~al.}(2020)\citenamefont {Ran}, \citenamefont {Tirrito}, \citenamefont {Peng}, \citenamefont {Chen}, \citenamefont {Tagliacozzo}, \citenamefont {Su},\ and\ \citenamefont {Lewenstein}}]{ran2020tensor}%
  \BibitemOpen
  \bibfield  {author} {\bibinfo {author} {\bibfnamefont {S.-J.}\ \bibnamefont {Ran}}, \bibinfo {author} {\bibfnamefont {E.}~\bibnamefont {Tirrito}}, \bibinfo {author} {\bibfnamefont {C.}~\bibnamefont {Peng}}, \bibinfo {author} {\bibfnamefont {X.}~\bibnamefont {Chen}}, \bibinfo {author} {\bibfnamefont {L.}~\bibnamefont {Tagliacozzo}}, \bibinfo {author} {\bibfnamefont {G.}~\bibnamefont {Su}},\ and\ \bibinfo {author} {\bibfnamefont {M.}~\bibnamefont {Lewenstein}},\ }\href {https://link.springer.com/book/10.1007/978-3-030-34489-4} {\emph {\bibinfo {title} {Tensor network contractions: methods and applications to quantum many-body systems}}}\ (\bibinfo  {publisher} {Springer Nature},\ \bibinfo {year} {2020})\BibitemShut {NoStop}%
\bibitem [{\citenamefont {Tarabunga}\ \emph {et~al.}(2024)\citenamefont {Tarabunga}, \citenamefont {Tirrito}, \citenamefont {Ba{\~n}uls},\ and\ \citenamefont {Dalmonte}}]{tarabunga2024nonstabilizerness}%
  \BibitemOpen
  \bibfield  {author} {\bibinfo {author} {\bibfnamefont {P.~S.}\ \bibnamefont {Tarabunga}}, \bibinfo {author} {\bibfnamefont {E.}~\bibnamefont {Tirrito}}, \bibinfo {author} {\bibfnamefont {M.~C.}\ \bibnamefont {Ba{\~n}uls}},\ and\ \bibinfo {author} {\bibfnamefont {M.}~\bibnamefont {Dalmonte}},\ }\bibfield  {title} {\bibinfo {title} {Nonstabilizerness via matrix product states in the pauli basis},\ }\href {https://journals.aps.org/prl/abstract/10.1103/PhysRevLett.131.180401} {\bibfield  {journal} {\bibinfo  {journal} {Physical Review Letters}\ }\textbf {\bibinfo {volume} {133}},\ \bibinfo {pages} {010601} (\bibinfo {year} {2024})}\BibitemShut {NoStop}%
\bibitem [{\citenamefont {O'Gorman}\ and\ \citenamefont {Campbell}(2017)}]{ogorman2017quantum}%
  \BibitemOpen
  \bibfield  {author} {\bibinfo {author} {\bibfnamefont {J.}~\bibnamefont {O'Gorman}}\ and\ \bibinfo {author} {\bibfnamefont {E.~T.}\ \bibnamefont {Campbell}},\ }\bibfield  {title} {\bibinfo {title} {Quantum computation with realistic magic-state factories},\ }\href {https://doi.org/10.1103/PhysRevA.95.032338} {\bibfield  {journal} {\bibinfo  {journal} {Phys. Rev. A}\ }\textbf {\bibinfo {volume} {95}},\ \bibinfo {pages} {032338} (\bibinfo {year} {2017})}\BibitemShut {NoStop}%
\bibitem [{\citenamefont {Souza}\ \emph {et~al.}(2011)\citenamefont {Souza}, \citenamefont {Zhang}, \citenamefont {Ryan},\ and\ \citenamefont {Laflamme}}]{souza2011experimental}%
  \BibitemOpen
  \bibfield  {author} {\bibinfo {author} {\bibfnamefont {A.~M.}\ \bibnamefont {Souza}}, \bibinfo {author} {\bibfnamefont {J.}~\bibnamefont {Zhang}}, \bibinfo {author} {\bibfnamefont {C.~A.}\ \bibnamefont {Ryan}},\ and\ \bibinfo {author} {\bibfnamefont {R.}~\bibnamefont {Laflamme}},\ }\bibfield  {title} {\bibinfo {title} {Experimental magic state distillation for fault-tolerant quantum computing},\ }\href {https://www.nature.com/articles/ncomms1166} {\bibfield  {journal} {\bibinfo  {journal} {Nature communications}\ }\textbf {\bibinfo {volume} {2}},\ \bibinfo {pages} {169} (\bibinfo {year} {2011})}\BibitemShut {NoStop}%
\bibitem [{\citenamefont {Beverland}\ \emph {et~al.}(2020)\citenamefont {Beverland}, \citenamefont {Campbell}, \citenamefont {Howard},\ and\ \citenamefont {Kliuchnikov}}]{beverland2020lower}%
  \BibitemOpen
  \bibfield  {author} {\bibinfo {author} {\bibfnamefont {M.}~\bibnamefont {Beverland}}, \bibinfo {author} {\bibfnamefont {E.}~\bibnamefont {Campbell}}, \bibinfo {author} {\bibfnamefont {M.}~\bibnamefont {Howard}},\ and\ \bibinfo {author} {\bibfnamefont {V.}~\bibnamefont {Kliuchnikov}},\ }\bibfield  {title} {\bibinfo {title} {Lower bounds on the non-clifford resources for quantum computations},\ }\href {https://iopscience.iop.org/article/10.1088/2058-9565/ab8963/meta?casa_token=J4VJf4bJu-8AAAAA:W8SlvH7BTutwCfLu8Iris9AonrTvDD8-C69UAB_xBSq2bli_GB85FOR2_N5z4y5dLIW7GNaLK3djVCOEk_P2ikLBFOqnMg&casa_token=k-P2s554PtkAAAAA:b3gFLU3bXk3O-0B91RlHZnG0D6rxLfXalY-YR59K_TYCdxUmyFVs9-g0wbfoIxLYZmXMGoxsLg4aQHtewOBM4teYqtdNKA} {\bibfield  {journal} {\bibinfo  {journal} {Quantum Science and Technology}\ }\textbf {\bibinfo {volume} {5}},\ \bibinfo {pages} {035009} (\bibinfo {year} {2020})}\BibitemShut {NoStop}%
\bibitem [{\citenamefont {Campbell}(2021)}]{campbell2021early}%
  \BibitemOpen
  \bibfield  {author} {\bibinfo {author} {\bibfnamefont {E.~T.}\ \bibnamefont {Campbell}},\ }\bibfield  {title} {\bibinfo {title} {Early fault-tolerant simulations of the hubbard model},\ }\href {https://iopscience.iop.org/article/10.1088/2058-9565/ac3110/meta} {\bibfield  {journal} {\bibinfo  {journal} {Quantum Science and Technology}\ }\textbf {\bibinfo {volume} {7}},\ \bibinfo {pages} {015007} (\bibinfo {year} {2021})}\BibitemShut {NoStop}%
\bibitem [{\citenamefont {Pagano}\ \emph {et~al.}(2020)\citenamefont {Pagano}, \citenamefont {Bapat}, \citenamefont {Becker}, \citenamefont {Collins}, \citenamefont {De}, \citenamefont {Hess}, \citenamefont {Kaplan}, \citenamefont {Kyprianidis}, \citenamefont {Tan}, \citenamefont {Baldwin} \emph {et~al.}}]{pagano2020}%
  \BibitemOpen
  \bibfield  {author} {\bibinfo {author} {\bibfnamefont {G.}~\bibnamefont {Pagano}}, \bibinfo {author} {\bibfnamefont {A.}~\bibnamefont {Bapat}}, \bibinfo {author} {\bibfnamefont {P.}~\bibnamefont {Becker}}, \bibinfo {author} {\bibfnamefont {K.~S.}\ \bibnamefont {Collins}}, \bibinfo {author} {\bibfnamefont {A.}~\bibnamefont {De}}, \bibinfo {author} {\bibfnamefont {P.~W.}\ \bibnamefont {Hess}}, \bibinfo {author} {\bibfnamefont {H.~B.}\ \bibnamefont {Kaplan}}, \bibinfo {author} {\bibfnamefont {A.}~\bibnamefont {Kyprianidis}}, \bibinfo {author} {\bibfnamefont {W.~L.}\ \bibnamefont {Tan}}, \bibinfo {author} {\bibfnamefont {C.}~\bibnamefont {Baldwin}}, \emph {et~al.},\ }\bibfield  {title} {\bibinfo {title} {Quantum approximate optimization of the long-range ising model with a trapped-ion quantum simulator},\ }\href {https://www.pnas.org/doi/10.1073/pnas.2006373117} {\bibfield  {journal} {\bibinfo  {journal} {Proceedings of the National Academy of Sciences}\ }\textbf {\bibinfo {volume} {117}},\ \bibinfo {pages}
  {25396} (\bibinfo {year} {2020})}\BibitemShut {NoStop}%
\bibitem [{\citenamefont {Huber}\ \emph {et~al.}(2021)\citenamefont {Huber}, \citenamefont {Haber}, \citenamefont {Barthel}, \citenamefont {Garc{\'\i}a-Ripoll}, \citenamefont {Torrontegui},\ and\ \citenamefont {Wunderlich}}]{Huber2021}%
  \BibitemOpen
  \bibfield  {author} {\bibinfo {author} {\bibfnamefont {P.}~\bibnamefont {Huber}}, \bibinfo {author} {\bibfnamefont {J.}~\bibnamefont {Haber}}, \bibinfo {author} {\bibfnamefont {P.}~\bibnamefont {Barthel}}, \bibinfo {author} {\bibfnamefont {J.}~\bibnamefont {Garc{\'\i}a-Ripoll}}, \bibinfo {author} {\bibfnamefont {E.}~\bibnamefont {Torrontegui}},\ and\ \bibinfo {author} {\bibfnamefont {C.}~\bibnamefont {Wunderlich}},\ }\bibfield  {title} {\bibinfo {title} {Realization of a quantum perceptron gate with trapped ions},\ }\href {https://arxiv.org/abs/2111.08977} {\bibfield  {journal} {\bibinfo  {journal} {arXiv:2111.08977}\ } (\bibinfo {year} {2021})}\BibitemShut {NoStop}%
\bibitem [{\citenamefont {Powell}(1994)}]{powell1994direct}%
  \BibitemOpen
  \bibfield  {author} {\bibinfo {author} {\bibfnamefont {M.~J.}\ \bibnamefont {Powell}},\ }\href {https://link.springer.com/chapter/10.1007/978-94-015-8330-5_4} {\emph {\bibinfo {title} {A direct search optimization method that models the objective and constraint functions by linear interpolation}}}\ (\bibinfo  {publisher} {Springer},\ \bibinfo {year} {1994})\BibitemShut {NoStop}%
\bibitem [{\citenamefont {Gheorghiu}(2014)}]{gheorghiu2014standard}%
  \BibitemOpen
  \bibfield  {author} {\bibinfo {author} {\bibfnamefont {V.}~\bibnamefont {Gheorghiu}},\ }\bibfield  {title} {\bibinfo {title} {Standard form of qudit stabilizer groups},\ }\href {https://www.sciencedirect.com/science/article/pii/S0375960113011080?via%3Dihub} {\bibfield  {journal} {\bibinfo  {journal} {Physics Letters A}\ }\textbf {\bibinfo {volume} {378}},\ \bibinfo {pages} {505} (\bibinfo {year} {2014})}\BibitemShut {NoStop}%
\bibitem [{\citenamefont {Liu}\ \emph {et~al.}(2025)\citenamefont {Liu}, \citenamefont {Low},\ and\ \citenamefont {Yin}}]{liu2025maximal}%
  \BibitemOpen
  \bibfield  {author} {\bibinfo {author} {\bibfnamefont {Q.}~\bibnamefont {Liu}}, \bibinfo {author} {\bibfnamefont {I.}~\bibnamefont {Low}},\ and\ \bibinfo {author} {\bibfnamefont {Z.}~\bibnamefont {Yin}},\ }\bibfield  {title} {\bibinfo {title} {Maximal magic for two-qubit states},\ }\href {https://arxiv.org/abs/2502.17550} {\bibfield  {journal} {\bibinfo  {journal} {arXiv:2502.17550}\ } (\bibinfo {year} {2025})}\BibitemShut {NoStop}%
\bibitem [{\citenamefont {Farhi}\ \emph {et~al.}(2022)\citenamefont {Farhi}, \citenamefont {Goldstone}, \citenamefont {Gutmann},\ and\ \citenamefont {Zhou}}]{farhi2022quantum}%
  \BibitemOpen
  \bibfield  {author} {\bibinfo {author} {\bibfnamefont {E.}~\bibnamefont {Farhi}}, \bibinfo {author} {\bibfnamefont {J.}~\bibnamefont {Goldstone}}, \bibinfo {author} {\bibfnamefont {S.}~\bibnamefont {Gutmann}},\ and\ \bibinfo {author} {\bibfnamefont {L.}~\bibnamefont {Zhou}},\ }\bibfield  {title} {\bibinfo {title} {The quantum approximate optimization algorithm and the sherrington-kirkpatrick model at infinite size},\ }\href {https://quantum-journal.org/papers/q-2022-07-07-759/} {\bibfield  {journal} {\bibinfo  {journal} {Quantum}\ }\textbf {\bibinfo {volume} {6}},\ \bibinfo {pages} {759} (\bibinfo {year} {2022})}\BibitemShut {NoStop}%
\end{thebibliography}
\end{document}